\documentclass[aps,preprint]{revtex4}%
\usepackage{amsfonts}
\usepackage{amsmath}
\usepackage{amssymb}
\usepackage{subfigure}
\usepackage{graphicx}
\usepackage{float}
\usepackage{color}%
\begin{document}
\title{Joule-Thomson expansion of d-dimensional charged AdS
black holes with cloud of strings and quintessence}

\author{Rui Yin$^{a,b}$}
\email{yrphysics@126.com}
\author{Jing Liang$^{a,b}$}
\email{ljphysics@163.com}
\author{Yue Song$^{a,c}$}
\email{syphysics@std.uestc.edu.cn}
\author{Yiqian He$^{a,c}$}
\email{heyiqian@std.uestc.edu.cn}
\author{Benrong Mu$^{a,b}$}
\email{benrongmu@cdutcm.edu.cn}
\affiliation{$^{a}$ Physics Teaching and Research section, College of Medical Technology,
Chengdu University of Traditional Chinese Medicine, Chengdu, 611137, PR China}
\affiliation{$^{b}$Center for Theoretical Physics, College of Physics, Sichuan University,
Chengdu, 610064, PR China}
\affiliation{$^{c}$School of Physics, University of Electronic Science and Technology of China, Chengdu 610054, China}

\begin{abstract}
Herein, we focus on the study of Joule-Thomson expansion corresponding to a d-dimensional charged AdS black hole with cloud of strings and quintessence. Then its relevant solution and some thermodynamic properties are investigated. Specifically, we evaluate its Joule-Thomson expansion from four important aspects, including the Joule-Thomson coefficient, inversion curve, isenthalpic curve, and ratio $\frac{T_{i}^{min}}{T_{c}}$. After analysis, different dimensions with strings of cloud and quintessence parameters have different effects on the Joule-Thomson coefficient (the same situation are found for the inversion curve, isenthalpic curve, and ratio $\frac{T_{i}^{min}}{T_{c}}$).

\end{abstract}
\maketitle
\tableofcontents

{}

\bigskip{}

%\affiliation{Center for Theoretical Physics, College of Physical Science and Technology,
%Sichuan University, Chengdu, 610064, PR China}

%\affiliation{Center for Theoretical Physics, College of Physical Science and Technology,
%Sichuan University, Chengdu, 610064, PR China}

\section{Introduction}
\label{sec:A}
In 1915, Einstein established the general theory of relativity, which describes the gravitational force of spacetime curvature due to the mass presence. The gravitational field equation satisfied by the spacetime gauge is the famous Einstein field equation. Since then, solving various exact solutions of Einstein's equation has become a hot research topic. Among the various gravitational solutions, it is generally believed that black holes are the best model to link gravitational theory and quantum field theory. The study of black hole physics contributes to the understanding of the nature of gravity and the establishment of quantum gravity theory. In 1916, Schwarzschild solved the first non-trivial exact solution of Einstein's equations except for flat spacetime, which is a spherically symmetric solution described by just one mass parameter in four-dimensional spacetime. Subsequent studies have mainly considered a black hole as a mathematical property. In 1973, black hole thermodynamics, which connects thermodynamics, classical gravity and quantum mechanics, became a new attractive topic of research. Early, Bekenstein pioneered the black hole entropy, and the black hole was treated as an interesting thermodynamic system \cite{Bekenstein:1972tm, Bardeen:1973gs}. In Ref. \cite{Bardeen:1973gs} the Four laws of black hole were established, then Hawking predicted the thermal emission of radiation (Hawking radiation) \cite{Hawking:1974rv, Hawking:1975vcx, Hawking:1976de}. After that, the complete theory of black hole thermodynamics was developed.

Generally, Einstein's equation contains a cosmological constant, which is used to describe the state of the universe. If a positive value is imposed on the cosmological constant, the associated negative pressure will be the cause of the accelerated expansion of the universe. Highly astronomical observations suggest that the universe is expanding at an accelerated rate \cite{SupernovaCosmologyProject:1998vns, SupernovaSearchTeam:1998cav}, which means there is the negative pressure. In addition to the cosmological constant, the quintessence hypothetical assumed form of dark energy is also an important factor affecting the negative pressure. In the origin of the universe associated with dark energy, the relationship between negative pressure and energy density can be described by an equation of state $p_{q}=\omega_{q}\rho_{q}$, where $q$ stands for quintessence. The origin of the quintessence is related to the dynamic scalar field \cite{Ratra:1987rm, Caldwell:1997ii}. The first solution to the Einstein equation associated with the quintessence was obtained by Kiselev in four dimensions \cite{Kiselev:2002dx}. Next, a solution of Einstein equations with quintessential matter in higher dimensional spacetime was presented \cite{Chen:2008ra}. After a series of corrections, the solutions of Einstein's equation about the quintessence were extended to different cases, such as charged black holes \cite{Azreg-Ainou:2014lua}, $5d$ black hole in Einstein Gauss-Bonnet gravity \cite{Ghosh:2016ddh}, $d$-dimensional Lovelock gravity \cite{Ghosh:2017cuq}.

On the other hand, the universe can also be described by a one-dimensional object namely strings. The rapidly accelerating expansion of the universe is also closely linked to the extension of such cosmic strings, which can penetrate anywhere in the universe that one can observe \cite{HenryTye:2006uv}. String theory treats particles and the fundamental forces of nature as vibrations of tiny supersymmetric strings, and it predicts the existence of quantum gravity. Then, studies on the gravitational effects of matter in the form of string clouds followed, one after another. The connection between the counting string state and the entropy of the black hole was investigated \cite{Letelier:1979ej}, in which Letelier first studied the general solutions of string clouds satisfying spherical symmetry. The relevant solutions are then generalized to the case in third-order Lovelock gravity \cite{Ghosh:2014pga} and the case for Einstein-Gauss-Bonnet theory in the Letelier spacetime \cite{Herscovich:2010vr}. In this context, many other extended solutions for a broad range of black holes have also been studied \cite{Richarte:2007bx, Yadav:2009zza, Ganguly:2014cqa, Bronnikov:2016dhz, Ghosh:2014dqa, Lee:2014dha}.

As the research proceeded, the thermodynamic properties of a variety of complicated black holes attracted attention and were studied \cite{Parikh:1999mf, Kerner:2007rr, Feng:2015jlj, Hod:2016hef, MunozdeNova:2018fxv, Robson:2018con, Feng:2018gqr, Moreno-Ruiz:2019lgn}. Among all studies, the thermodynamic behavior in the anti-de Sitter (AdS) space-time with a negative cosmological constant has become a major research topic. Hawking and Page first investigated thermodynamics of the AdS black holes, they also first pointed out the existence of a thermodynamic phase transition between a stable Schwarzschild-AdS black hole and the thermal gas in the AdS space \cite{Hawking:1982dh}. This work then inspired people to study the common properties between the charged AdS spacetime and general thermodynamic systems \cite{Chamblin:1999tk, Chamblin:1999hg, Wei:2018pnk, Banerjee:2011au}. It has been shown that there are similarities between RN-AdS black holes and van der Waals fluid phases in terms of phase transition.

In recent years, the cosmological constant has been treated as the thermodynamic pressure, and many studies have been carried out on this basis \cite{Caldarelli:1999xj, Kastor:2009wy, Kastor:2010gq, Dolan:2010ha, Dolan:2011xt, Cvetic:2010jb, Yin:2021fsg, Mu:2021xxx, Jing:2020sdf, Mu:2020szg, Liang:2020uul, Liang:2020hjz, Hong:2019yiz}. Among these studies, a typical example that deserves attention is the creative introduction of the concept of holographic heat engine, and another interesting example is the introduction of Joule-Thomson expansion (throttling process). \"O.~\"Okc\"u and E.~Ayd\i{}ner explored the Joule-Thomson expansion process in the charged AdS black holes and van der Waals system, in which the inversion curves and isenthalpic curves were obtained \cite{Okcu:2016tgt}.
In classical thermodynamics, Joule-Thomson expansion is the process of a high pressure gas passing through a porous plug into a low pressure, during which there is no change in enthalpy. Related studies were subsequently developed to different black holes, such as Kerr-AdS black holes \cite{Okcu:2017qgo}, AdS black holes with a global monopole \cite{RizwanCL:2018cyb}, AdS black holes in Lovelock gravity \cite{Mo:2018qkt}, charged Gauss-Bonnet black holes in AdS space \cite{Lan:2018nnp}, higher dimensional AdS black holes \cite{Mo:2018rgq}. These studies further confirm the basic findings in Ref. \cite{Okcu:2016tgt}. The recent progresses can be found in Refs. \cite{Ghaffarnejad:2018exz, DAlmeida:2018ldi, Chabab:2018zix, Liang:2021elg, Hegde:2020xlv, Wei:2017vqs, Kuang:2018goo,Yekta:2019wmt,Pu:2019bxf,Nam:2018sii,Zhao:2018kpz,Li:2019jcd,Hyun:2019gfz,Ghaffarnejad:2018tpr,Nam:2019zyk,Nam:2018ltb,Rostami:2019ivr,Haldar:2018cks,Guo:2019gkr,Lan:2019kak,Sadeghi:2020bon,Bi:2020vcg,Ranjbari:2019ktp,Guo:2019pzq,K.:2020rzl,Nam:2020gud,Meng:2020csd,Guo:2020qxy,Ghanaatian:2019xhi,Guo:2020zcr,Feng:2020swq,Debnath:2020zdv,Cao:2021dcq,Huang:2020xcs,Zhang:2021raw,Chen:2020igz,Jawad:2020mdc,Liang:2021xny,Debnath:2020inx,Mirza:2021kvi,Graca:2021izb} for various other black holes.

This paper is organized as follows: the thermodynamics of the quintessence surrounding d-dimensional Reissner-Nordstr\"om-Anti-de Sitter black hole with a
cloud of strings is reviewed in Sec. \ref{sec:B}. The Joule-Thomson expansion of a quintessence surrounding d-dimensional Reissner-Nordstr\"om-Anti-de Sitter black hole with a cloud of strings is discussed separately from four aspects in Sec. \ref{sec:C}. The Joule-Thomson coefficient is investigated in Sec. \ref{sec:CA}; the inversion curves are investigated in Sec. \ref{sec:CB}; the isenthalpic curves are investigated in Sec. \ref{sec:CC}; the ratio between $T_{i}^{min}$ and $T_{c}$ is investigated in Sec. \ref{sec:CD}. Finally, the results are discussed in Sec. \ref{sec:D}.

\section{Quintessence surrounding d-dimensional Reissner-Nordstr\"om-Anti-de Sitter black holes with a
cloud of strings}
\label{sec:B}

In this section, we focus on the relevant solution and some thermodynamic properties of a charged AdS black hole surrounded by quintessence with a cloud of strings in higher dimensional spacetime. Under the assumption, the quintessence and cloud of strings have no interaction \cite{Chabab:2020ejk,deMToledo:2018tjq}, the spacetime metric corresponding to a d-dimensional charged AdS black holes with cloud of strings and quintessence is given by the general form as follows
\begin{equation}
ds_{d}^{2}=-f(r)dt^{2}+f(r)^{-1}dr^{2}+r^{2}d\varOmega_{d-2}^{2},
\end{equation}
\label{eqn:two1}
 the metric function $f(r)$ is given by \cite{Chabab:2020ejk}
\begin{equation}
f(r)=1-\frac{m}{r^{d-3}}+\frac{q^{2}}{r^{2(d-3)}}-\frac{2\Lambda r^{2}%
}{(d-2)(d-1)}-\frac{\alpha}{r^{(d-1)\omega_{q}+d-3}}-\frac{2a}{(d-2)r^{d-4}%
},
\label{eqn:two2}
\end{equation}
where $d\varOmega_{d-2}^{2}$ denotes the metric on unit $(d-2)$-sphere, $\Omega_{d-2}$ is the volume of unit $(d-2)$-sphere, which takes the form \cite{deMToledo:2018tjq}
\begin{equation}\label{eqn:two3}
\varOmega_{d-2}=\frac{2\pi^{\frac{d-1}{2}}}{\Gamma(\frac{d-1}{2})},
\end{equation}
here $m$ is the integration constant proportional to the ADM mass, and $q$ is
charge of the black hole. There are two related expressions as follows \cite{Chamblin:1999tk,Gunasekaran:2012dq}
\begin{equation}
M=\frac{(d-2)}{16\pi}\Omega_{d-2}m,Q=\frac{\sqrt{2(d-2)(d-3)}\varOmega_{d-2}%
q}{8\pi},
\label{eqn:two4}
\end{equation}
moreover, $a$ in the above is an integration constant related to the presence of cloud of strings. And $\alpha$ is the quintessential parameter (a positive normalization factor), which has a relationship with the energy density for quintessence. The density of quintessence are defined by the equation $\rho_{q}=-\frac{\alpha\omega_{q}(d-1)(d-2)}{4r^{(d-1)(\omega_{q}+1)}}$, with $\omega_{q}$ is barotropic index such that $-1<\omega_{q}<-1/3$. We will set $\omega_{q}=-\frac{d-2}{d-1}$ in numerical analysis for only considering the asymptotically $dS$ behavior.
The definition of the cosmological constant is a key definition in the black hole thermodynamic framework. As usual, the cosmological constant is treated as the thermodynamic pressure \cite{Dolan:2011xt,Kubiznak:2012wp,Cvetic:2010jb,Caceres:2015vsa,Hendi:2012um,Pedraza:2018eey}, which is given by
\begin{equation}\label{eqn:two5}
  P=-\frac{\varLambda}{8\pi}=\frac{3}{8\pi l^{2}}.
\end{equation}

On the event horizon $r_{+}$ that corresponds to equation $f(r_{+})=0$, we can obtain the ADM mass
\begin{equation}
M=\frac{(d-2)\Omega_{d-2}r_{+}^{d-3}}{16\pi}+\frac{(d-2)\Omega_{d-2}q^{2}%
}{16\pi r_{+}^{d-3}}+\frac{\Omega_{d-2}Pr_{+}^{d-1}}{(d-1)}-\frac
{\alpha(d-2)\Omega_{d-2}}{16\pi r_{+}^{(d-1)\omega_{q}}}-\frac{ar_{+}%
\Omega_{d-2}}{8\pi}.\label{eqn:two6}%
\end{equation}
Therefore, the generalized first law of the black hole in the extended phase space is then expressed as
\begin{equation}
dM=TdS+VdP+\phi dQ+\mathcal{A}da+\mathcal{Q}d\alpha,\label{eqn:two7}%
\end{equation}
where
\begin{equation}
\begin{aligned}
&\mathcal{Q}=\frac{(2-d)\Omega_{d-2}}{16\pi r_{+}^{(d-1)\omega_{q}}},\mathcal{A}=\frac{-\varOmega_{d-2}r_{+}}{8\pi},\\
&\phi=\sqrt{\frac{d-2}{2(d-3)}}\frac{q}{r_{+}^{d-3}},V=\frac{\varOmega_{d-2}r_{+}^{(d-1)}}{d-1}.\\
\end{aligned}
\label{eqn:two8}
\end{equation}

Above quantities satisfy the generalized Smarr formula as follows
 \begin{equation}
M=\frac{d-2}{d-3}TS-\frac{2}{d-3}VP+\phi Q+(\frac{d-1}{d-3}\omega_{q}+1)\mathcal{Q\alpha+\textrm{\ensuremath{\frac{d-4}{d-3}\mathcal{A}a}}}.
\label{eqn:two9}%
\end{equation}

With the assistance of Bekenstein-Hawking formula \cite{Banerjee:2020xcn}, the entropy can be obtained as
\begin{equation}
S=\frac{A_{d-2}}{4}=\frac{\Omega_{d-2}}{4}r_{+}^{d-2}.\label{eqn:two10}
\end{equation}

Otherwise, the Hawking temperature of the black hole is given by
\begin{equation}\label{eqn:two11}
\begin{aligned}
 &T=(\frac{\partial M}{\partial S})_{P,Q}\\
 &=\frac{-2ar_{+}^{4-d}-8\pi^{3-d}Q^{2}r_{+}^{6-2d}\Gamma\left(\frac{d-1}{2}\right)^{2}-\alpha(d-2)^{2}r_{+}+(d-5)d+16\pi Pr_{+}^{2}+6}{4\pi(d-2)r_{+}}.
 \end{aligned}
\end{equation}

\section{Joule-Thomson expansion}
\label{sec:C}
\subsection{The Joule-Thomson coefficient}\label{sec:CA}
\begin{figure}[H]
  \centering
  \subfigure[{}]{
  \includegraphics[width=0.45\textwidth]{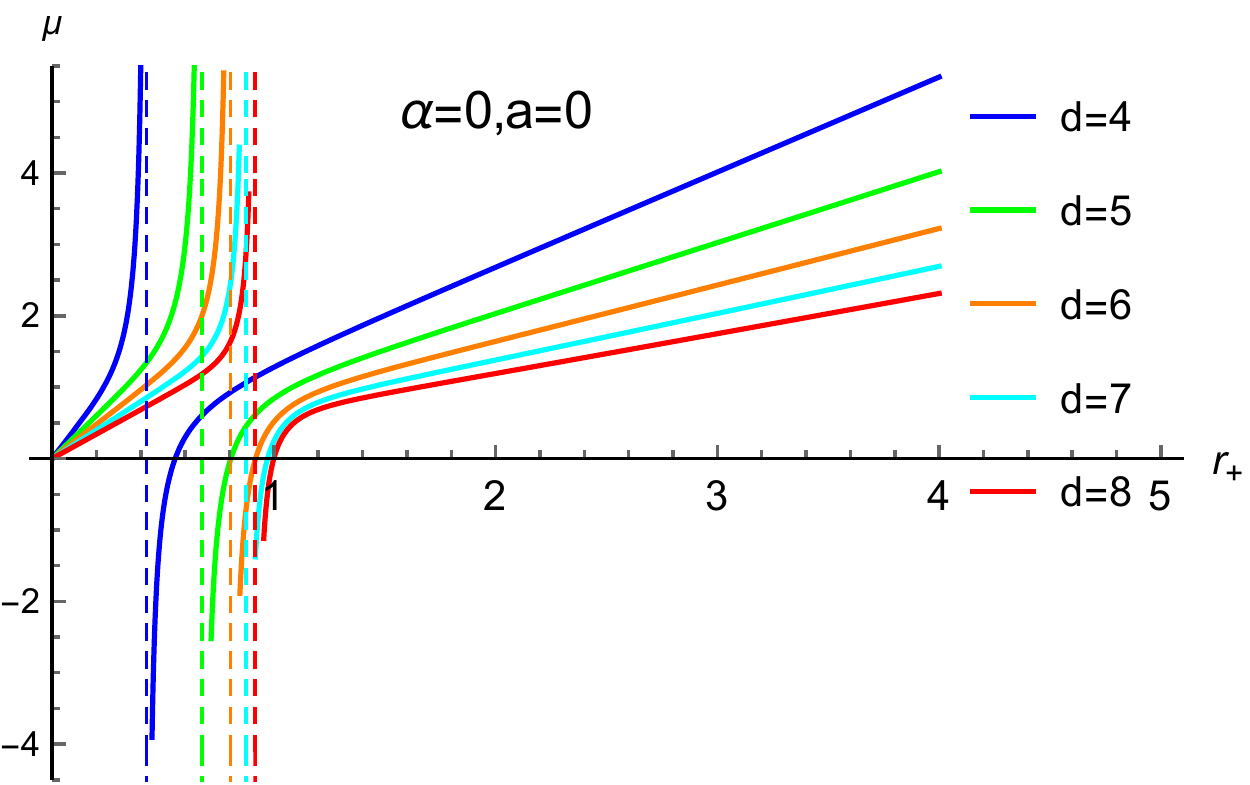}
 \label{fig:jt0}}
 \subfigure[{}]{
 \includegraphics[width=0.4\textwidth]{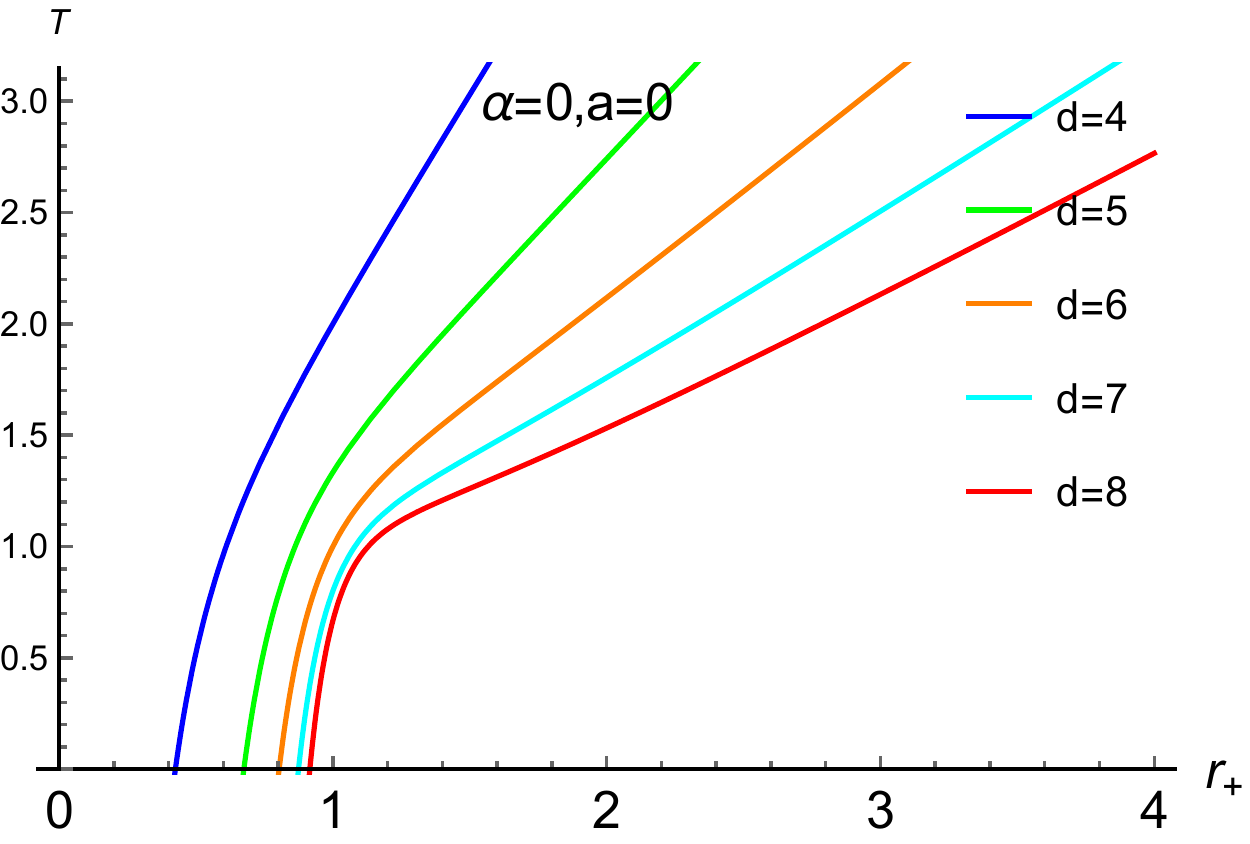}
 \label{fig:jt01}}
 \subfigure[{}]{
   \includegraphics[width=0.45\textwidth]{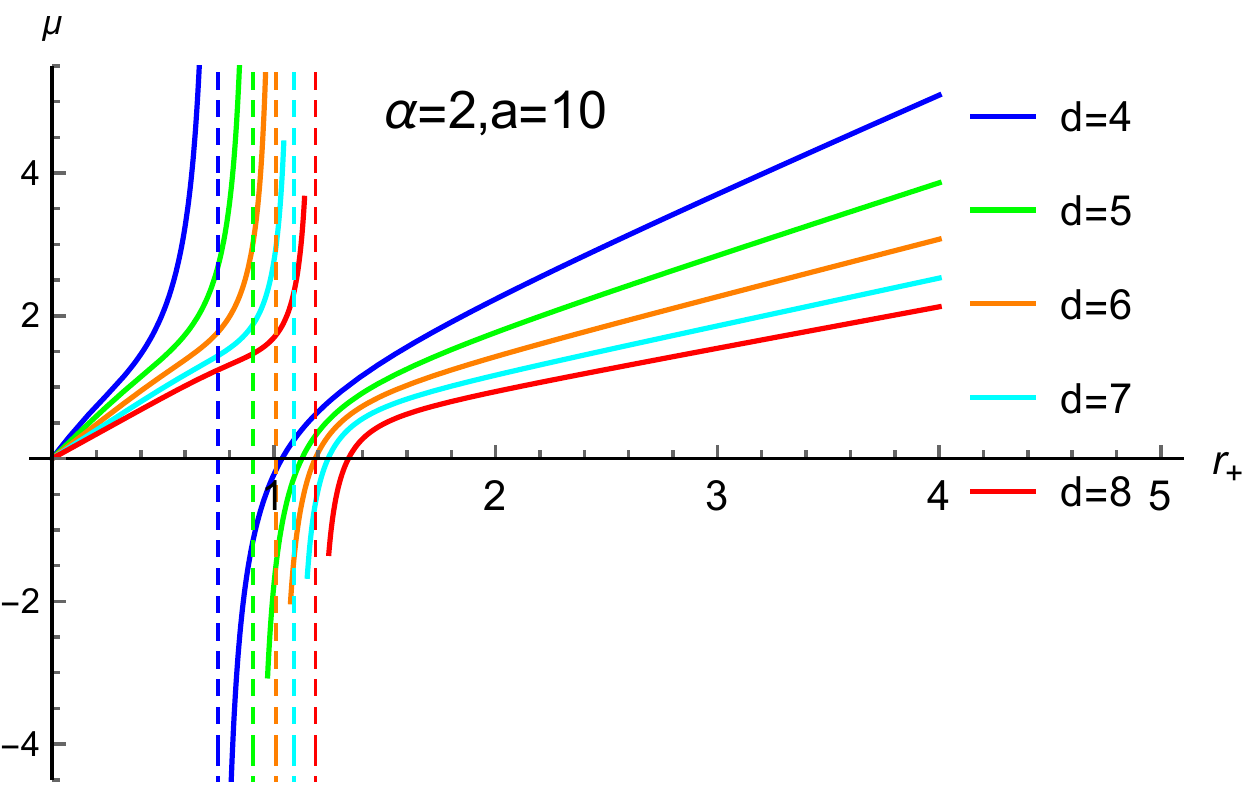}
 \label{fig:jt1}}
 \subfigure[{}]{
 \includegraphics[width=0.4\textwidth]{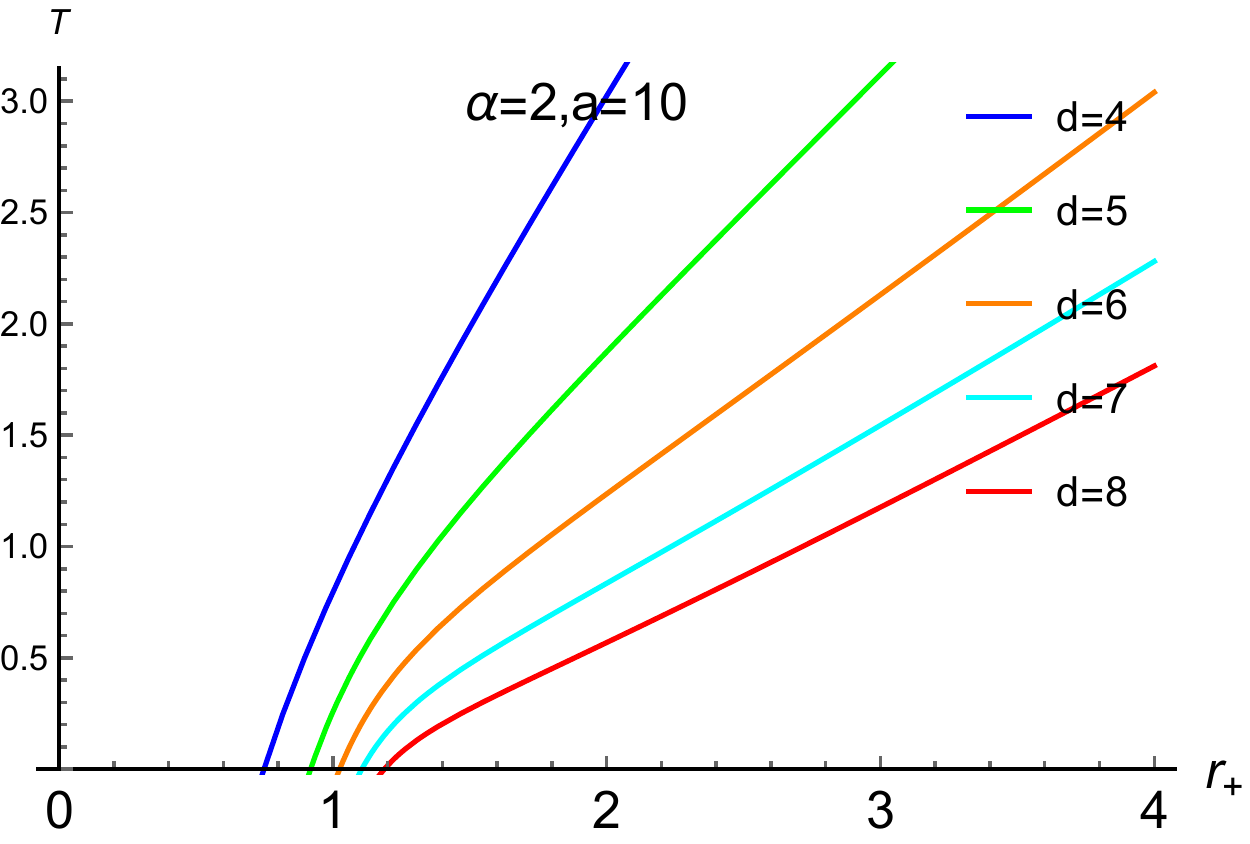}
 \label{fig:jt11}}
  \caption{Joule-Thomson coefficient $\mu$ and Hawking temperature $T$ versus event horizon $r_{+}$. Here, we use $P=1,q=1$.}\label{fig:JT1}
\end{figure}

\begin{figure}[H]
\begin{center}
\subfigure[{$d=5$}]{
\includegraphics[width=0.45\textwidth]{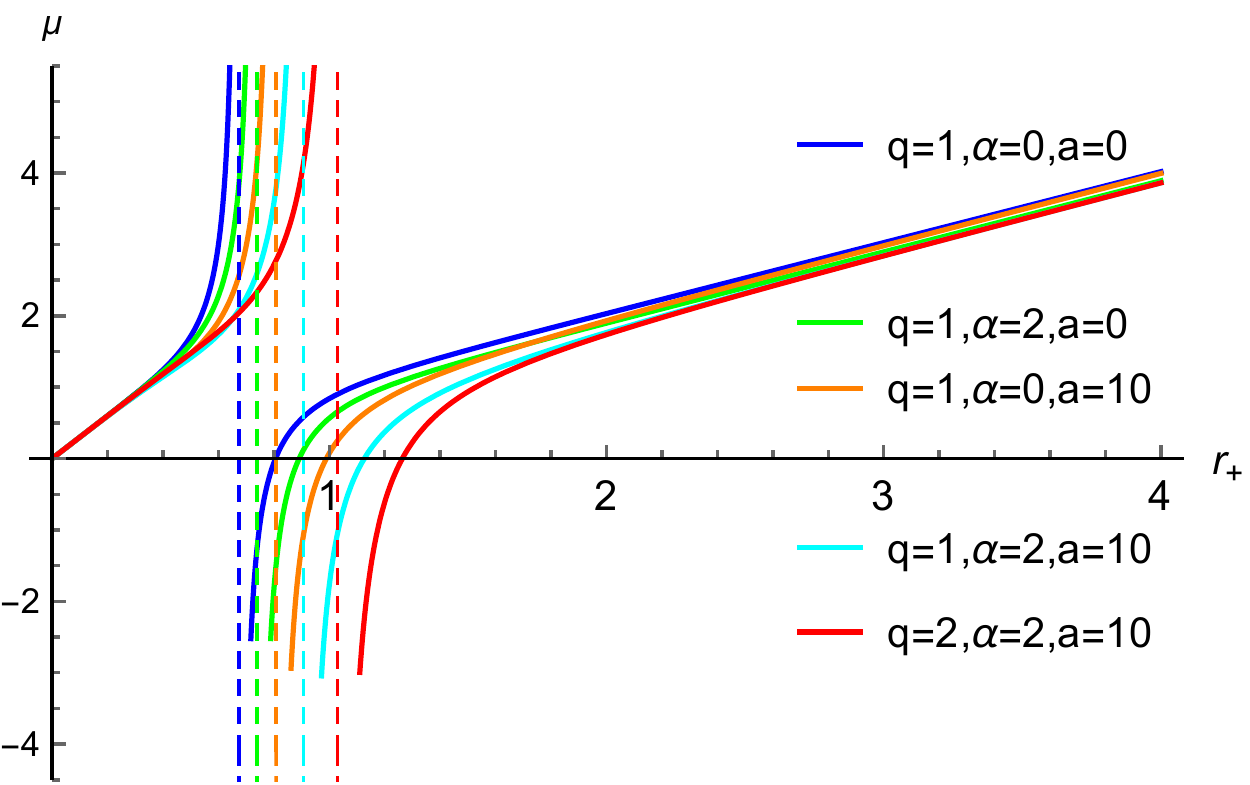}
 \label{fig:jt5}}
 \subfigure[{$d=5$}]{
 \includegraphics[width=0.4\textwidth]{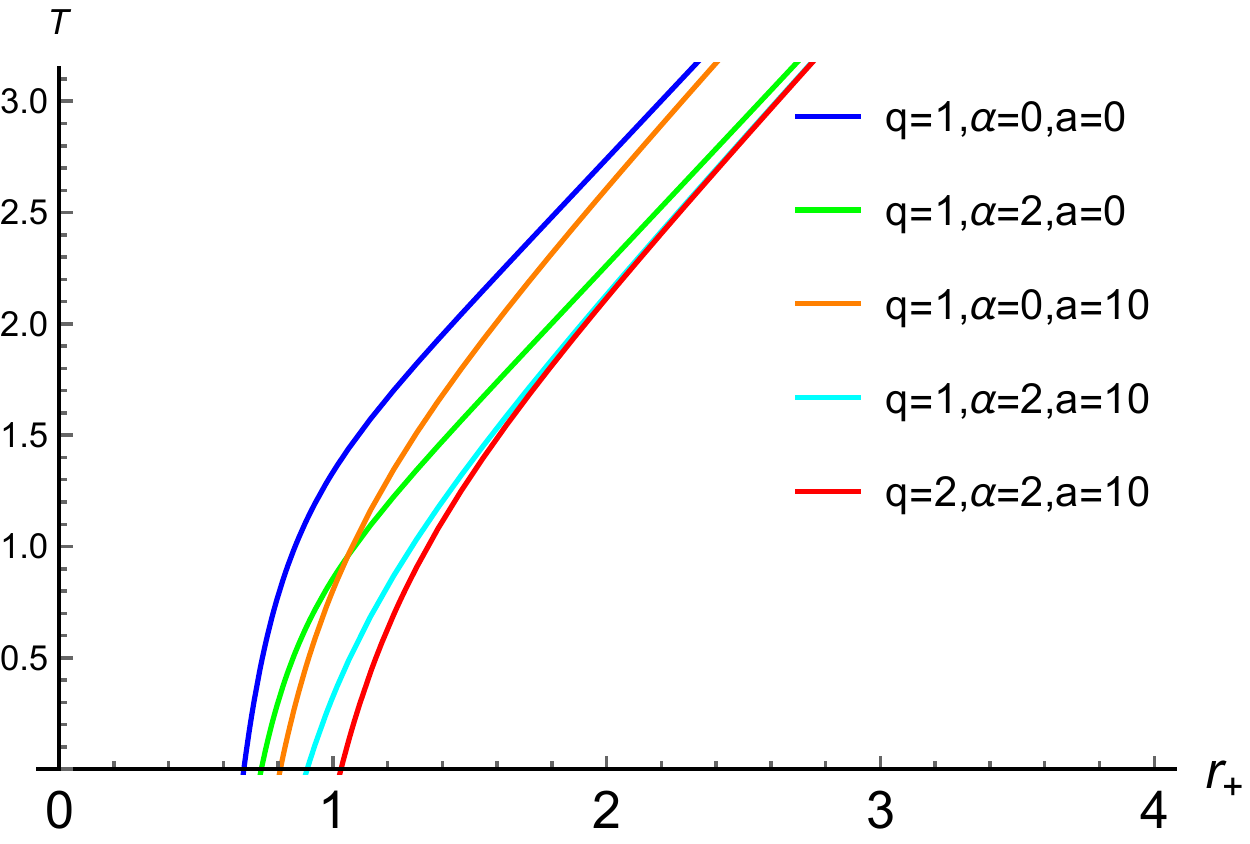}
 \label{fig:jt51}}
 \subfigure[{$d=6$}]{
   \includegraphics[width=0.45\textwidth]{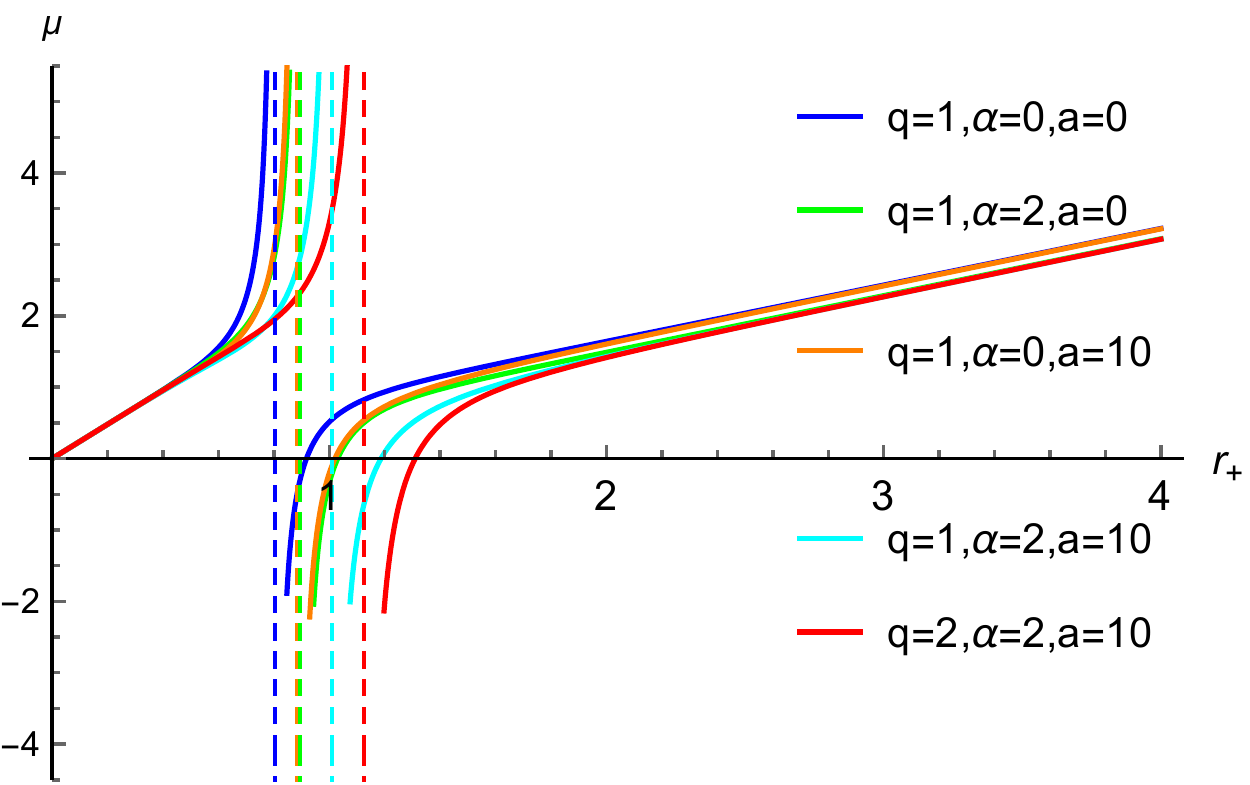}
 \label{fig:jt6}}
 \subfigure[{$d=6$}]{
 \includegraphics[width=0.4\textwidth]{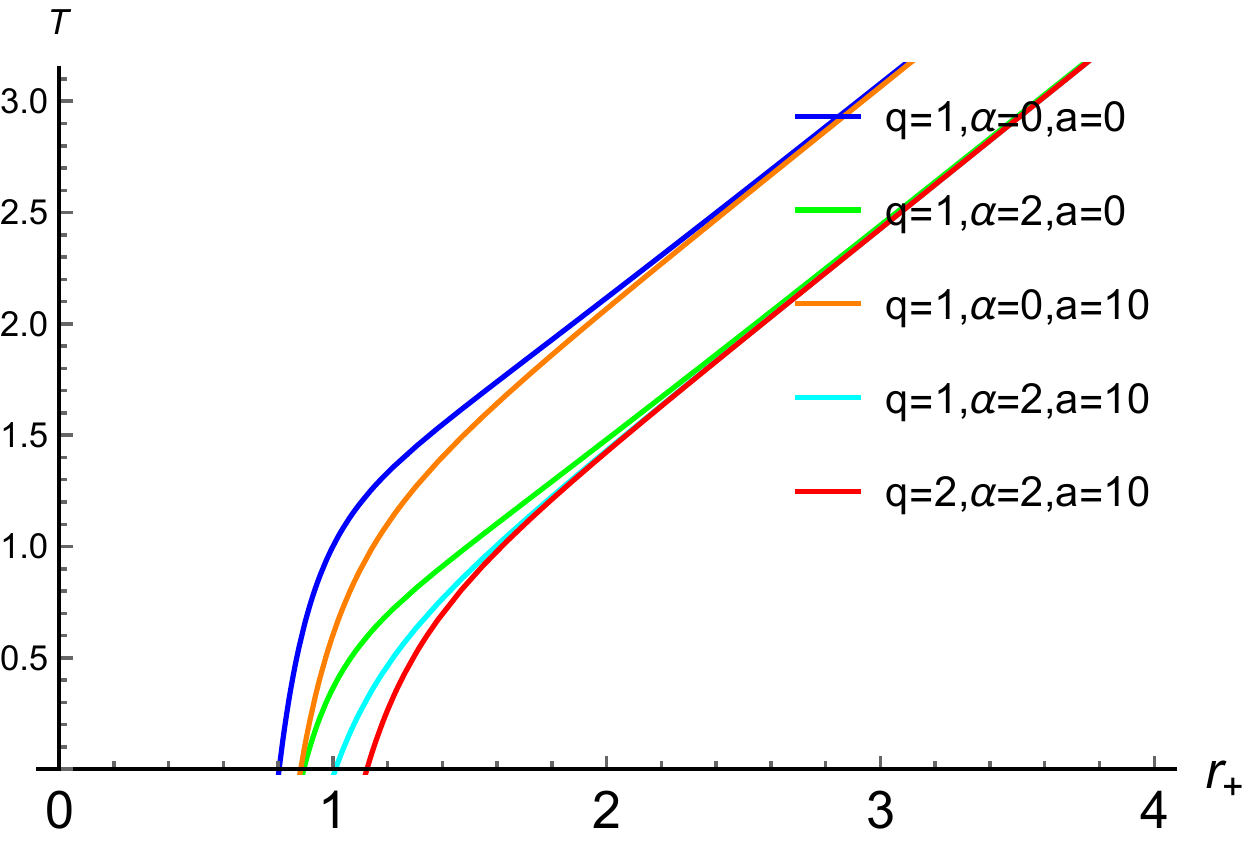}
 \label{fig:jt61}}
 \subfigure[{$d=7$}]{
  \includegraphics[width=0.45\textwidth]{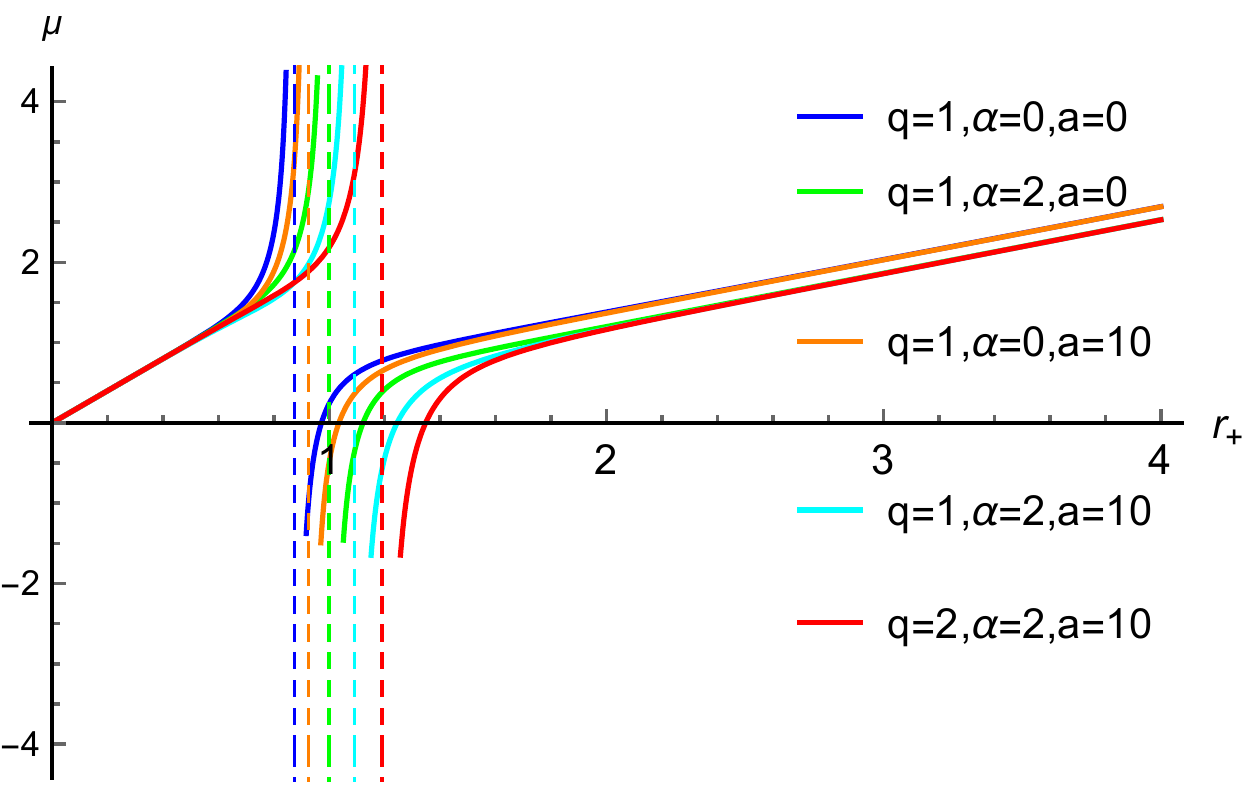}
 \label{fig:jt7}}
 \subfigure[{$d=7$}]{
 \includegraphics[width=0.4\textwidth]{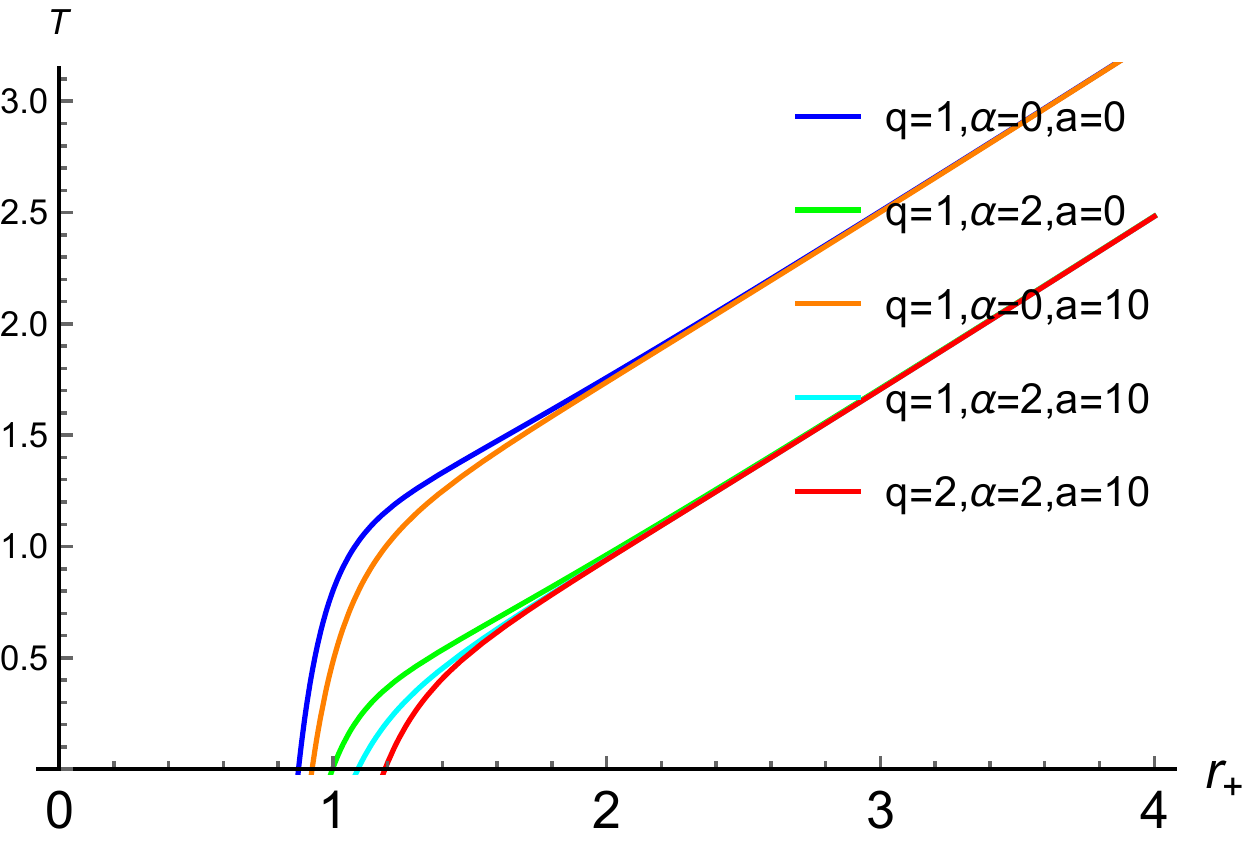}
 \label{fig:jt71}}
 \subfigure[{$d=8$}]{
   \includegraphics[width=0.45\textwidth]{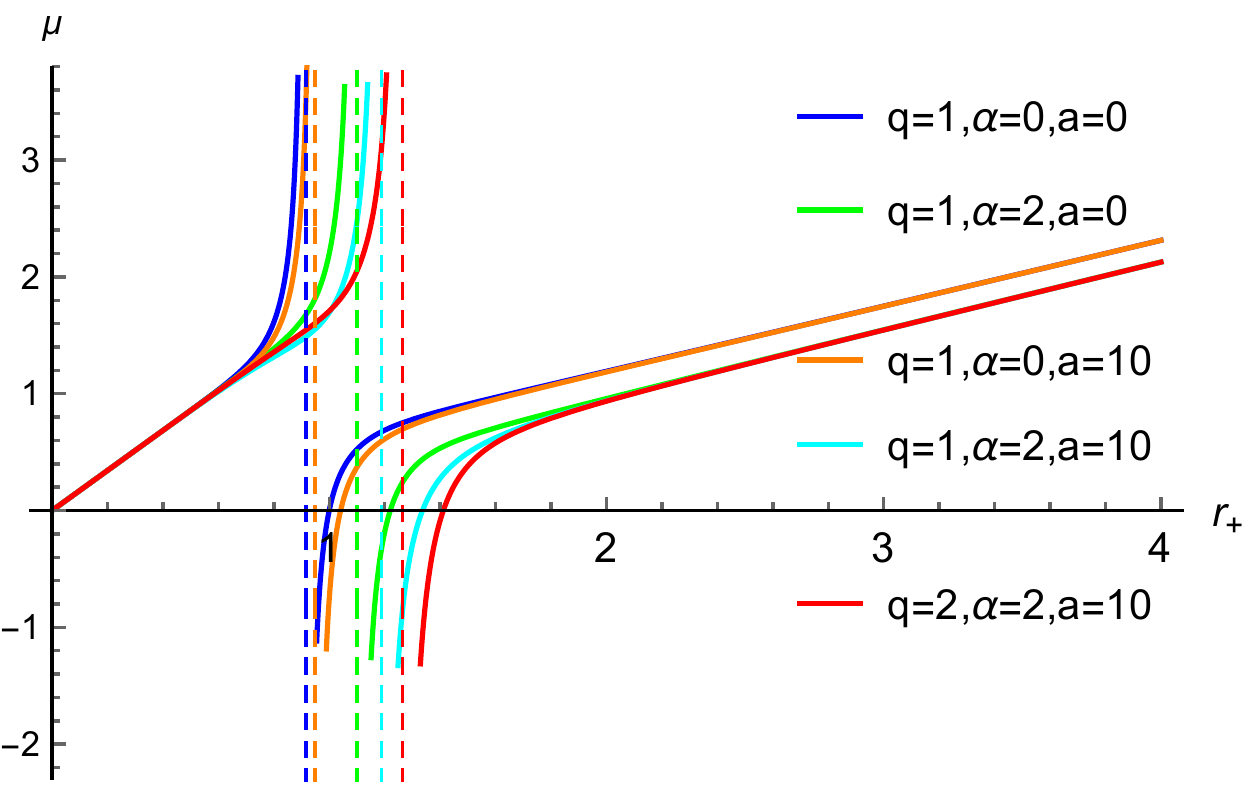}
 \label{fig:jt8}}
 \subfigure[{$d=8$}]{
 \includegraphics[width=0.4\textwidth]{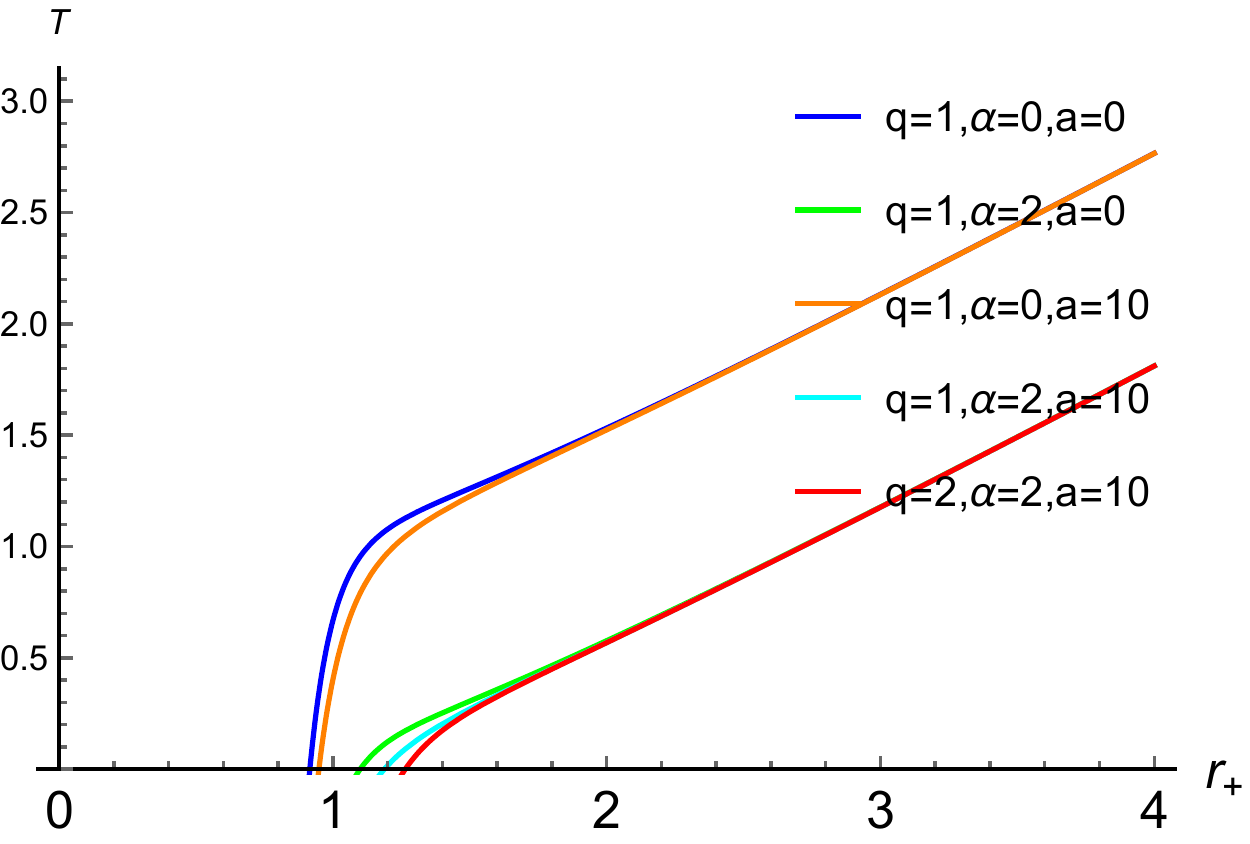}
 \label{fig:jt81}}
 \end{center}
  \caption{Joule-Thomson coefficient $\mu$ and Hawking temperature $T$ versus event horizon $r_{+}$ for fixed $P=1$.}\label{fig:JT2}
\end{figure}
In classical thermodynamics, Joule-Thomson expansion describes an adiabatic expansion, which is also called a throttling process. In this process, a nonideal fluid (gas or
liquid) expands from a high pressure section to a low pressure section through a porous plug. Besides, the temperature change of the fluid can lead to the cooling or liquefaction of the gas, so the throttling process is widely used in thermal machines including heat pumps, liquefiers, and air conditioners. This process can be considered as an isoenthalpy process, since the enthalpies of the initial and final states of the fluid do remain constant. The main characteristics of the throttling process are adiabatic and isoenthalpic. In the extended phase space, the Hawking radiation of the black hole is usually very tiny that can be regarded as satisfying the adiabatic condition. In addition, when radiation is not considered, the enthalpy of the black hole remains constant, so the process is also isoenthalpic. All above shows that the black hole satisfies the condition of the throttling process. In this process, the change in temperature with respect to pressure can be encoded in the Joule-Thomson coefficient, the sign of which can be used to determine whether heating occurs or cooling occurs. The Joule-Thomson coefficient is as follows
\begin{equation}\label{eqn:J1}
 \mu=(\frac{\partial T}{\partial P})_{H}.
\end{equation}

The negative coefficient $\mu$ corresponds to the heating process of the gas, and the positive coefficient $\mu$ corresponds to the cooling process of the gas.
Refs. \cite{Okcu:2016tgt,Okcu:2017qgo} derive the Joule-Thomson coefficients in two different ways, and Ref. \cite{Mo:2018rgq} demonstrates the same result for both methods in a d-dimensional black hole. In this paper we follow the approach proposed in Ref. \cite{Okcu:2016tgt}.
From the first law of thermodynamics and relating it to the differential form of enthalpy, we get
\begin{equation}\label{eqn:J2}
 dH=TdS+VdP+\phi dQ+\mathcal{A}da+\mathcal{Q}d\alpha.
\end{equation}

Using conditions $dH=0$ and $dQ=0$ yields
\begin{equation}\label{eqn:J3}
  T(\frac{\partial S}{\partial P})_{H}+V=0.
\end{equation}

Using the differential form of the entropy of the state function $dS=(\frac{\partial S}{\partial P})_{T,Q,a,\alpha}dP+(\frac{\partial S}{\partial T})_{P,Q,a,\alpha}dT+(\frac{\partial S}{\partial Q})_{T,P,a,\alpha}dQ+(\frac{\partial S}{\partial a})_{T,Q,P,\alpha}da+(\frac{\partial S}{\partial P})_{T,Q,a,P}d\alpha$ yields the following
\begin{equation}\label{eqn:J4}
 (\frac{\partial S}{\partial P})_{H}=(\frac{\partial S}{\partial P})_{T,Q,a,\alpha}+(\frac{\partial S}{\partial T})_{P,Q,a,\alpha}(\frac{\partial T}{\partial P})_{H}.
\end{equation}
Considering the Maxwell relation $(\frac{\partial S}{\partial P})_{T,Q,a,\alpha}=-(\frac{\partial V}{\partial T})_{P,Q,a,\alpha}$ and $C_{P}=T(\frac{\partial S}{\partial T})_{P,Q,a,\alpha}$, the above equation can be simplified as

\begin{equation}\label{eqn:J5}
 -T(\frac{\partial V}{\partial T})_{P,Q,a,\alpha}+C_{P}(\frac{\partial T}{\partial P})_{H}+V=0,
\end{equation}
thus we can obtain the Joule-Thomson coefficient as follows
\begin{equation}\label{eqn:J6}
  \mu=(\frac{\partial T}{\partial P})_{H}=\frac{1}{C_{P}}[T(\frac{\partial V}{\partial T})_{P,Q,a,\alpha}-V].
\end{equation}

 Next, substituting Eqs. $\left(\ref{eqn:two3}\right)$, $\left(\ref{eqn:two8}\right)$, $\left(\ref{eqn:two10}\right)$ and $\left(\ref{eqn:two11}\right)$ into Eq. $\left(\ref{eqn:J6}\right)$, yields
\begin{equation}\label{eqn:J7}
\begin{aligned}
 &C_{P}=-[(d-2)\pi^{\frac{d-1}{2}}r_{+}^{d-2}(-2ar_{+}^{d+4}+r_{+}^{2d}(d^{2}+\alpha(d-2)^{2}(-r_{+})\\
 &-5d+16\pi Pr_{+}^{2}+6)-(d-3)(d-2)q^{2}r_{+}^{6})]\\
 &\div[2\Gamma(\frac{d-1}{2})(-2a(d-3)r_{+}^{d+4}+r_{+}^{2d}((d-5)d-16\pi Pr_{+}^{2}+6)\\
 &-(d-3)(d-2)(2d-5)q^{2}r_{+}^{6})],\\
\end{aligned}
\end{equation}
and
\begin{equation}\label{eqn:J8}
\begin{aligned}
&\mu=[4r_{+}(4ar_{+}^{d+4}+r_{+}^{2d}((d-3)d(\alpha r_{+}-1)\\
&+2r_{+}(\alpha-8\pi Pr_{+}))+3(d-3)(d-2)q^{2}r_{+}^{6})]\\
&\div[(d-1)(2ar_{+}^{d+4}+r_{+}^{2d}(d(\alpha(d-4)r_{+}-d+5)\\
&+4r_{+}(\alpha-4\pi Pr_{+})-6)+(d-3)(d-2)q^{2}r_{+}^{6})].\\
\end{aligned}
\end{equation}

More further, a numerical analysis is performed for Joule-Thomson coefficient $\mu$. Without considering the effect of the parameters of cloud of strings and quintessence, the Fig. \ref{fig:jt0} is plotted at setting $P=1,q=1$. It is evident that the curves in the Fig. \ref{fig:jt0} and the one plotted in Ref. \cite{Mo:2018rgq} are almost identical in nature. Then, the effect of dimensionality on the Joule-Thomson coefficient is shown in the Fig. \ref{fig:jt1} for different cases of parameters $a$ and $\alpha$. From these two figures, the observation is that as the dimensionality $d$ increases the divergent point of the Joule-Thomson coefficient also moves to the right. The Hawking temperature is also plotted and shown in Fig. \ref{fig:jt01} and Fig. \ref{fig:jt11}. By comparison, the divergent point of the Joule-Thomson coefficient corresponds to the zero point of the Hawking temperature, which is strongly associated with extreme black holes. In addition, the phenomenon can also be explained with respect to the equation. Using Eqs. $\left(\ref{eqn:J5}\right)$ and $\left(\ref{eqn:J6}\right)$, we can derive the equation $\mu=(\frac{\partial V}{\partial S})_{P,Q,a,\alpha}-\frac{V}{T(\frac{\partial S}{\partial T})_{P,Q,a,\alpha}}$, which shows that the Joule-Thomson coefficient diverges as the Hawking temperature converges to zero. The dispersion of $\mu$ implies that the pressure vanishes, at which the microstate of the black hole has not been given any outward force. This happens because the total energy is used for the phase transition at these points, which leads to the vanishment of the outward pressure. Thus, the extreme black hole is obtained near the value of the dispersion $\mu$ \cite{Mandal:2016anc}.

The next focus is on studying the effect of parameters of cloud of strings and quintessence on the Joule-Thomson coefficient. By fixing the pressure $P$, the dimensionality $d$, and the charge $q$, one can see how the parameter of cloud of strings $a$ and parameter of quintessence $\alpha$ affect the
behaviors of the Joule-Thomson coefficient $\mu$, as well as how they affect Hawking temperature $T$. In Fig. \ref{fig:JT2}, the Joule-Thomson coefficient diverges where the Hawking temperature is zero. As $q$ increases, the Joule-Thomson coefficient $\mu$ shifts to the right and its divergence point also shifts to the right. Moreover, the same happens when $a$ and $\alpha$ increase. By comparing Figs. \ref{fig:jt5}, \ref{fig:jt6}, \ref{fig:jt7}, \ref{fig:jt8}, there is a novel phenomenon that the influence of $\alpha$ is greater than that of $a$, which is more obvious as the dimensionality $d$ increases.
\subsection{The inversion curve}
\label{sec:CB}
\begin{figure}[H]
  \centering
  \subfigure[{}]{
  \includegraphics[width=0.45\textwidth]{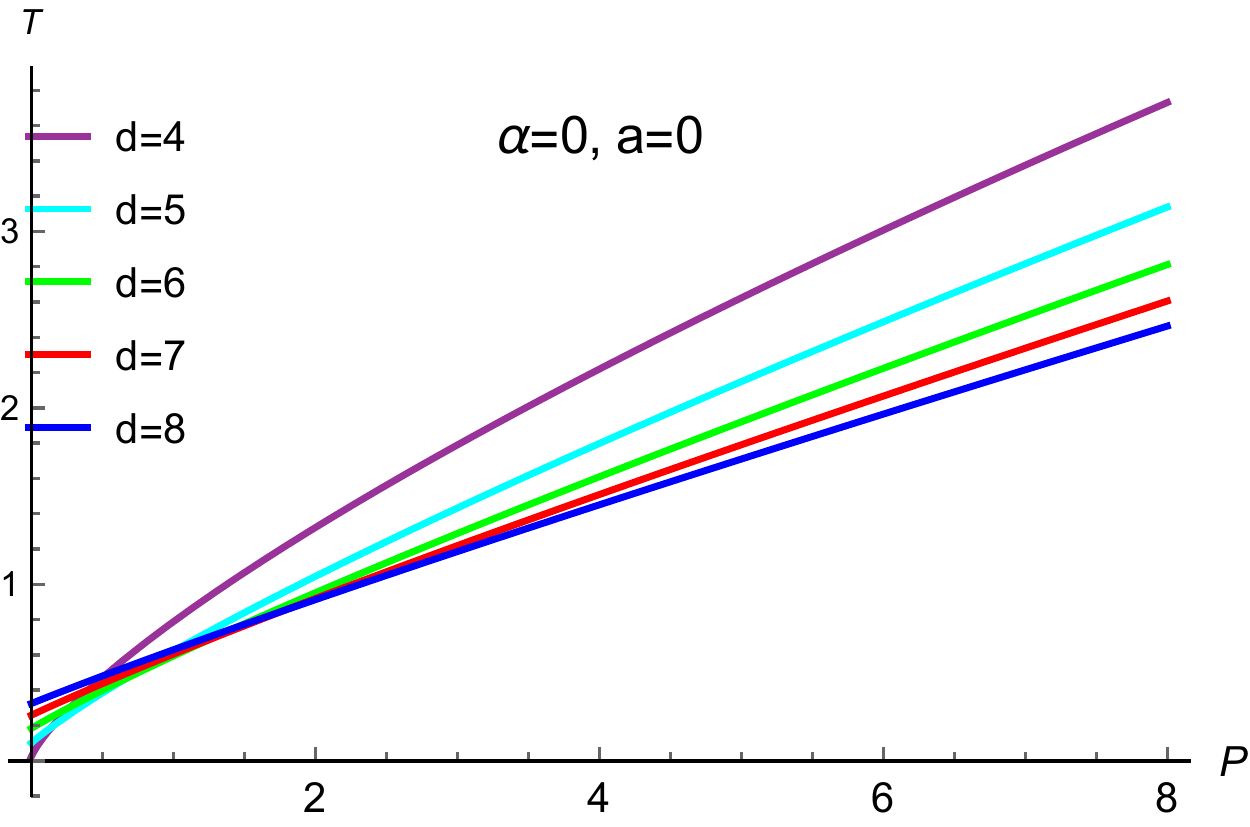}
 \label{fig:fz00}}
  \caption{The inversion curves versus $P$ for different dimensions. Here, we use $a=0$, $\alpha=0$ and $Q=1$.}\label{fig:TIC1}
\end{figure}

\begin{figure}[t]
  \centering
  \subfigure[{$d=5$}]{
  \includegraphics[width=0.31\textwidth]{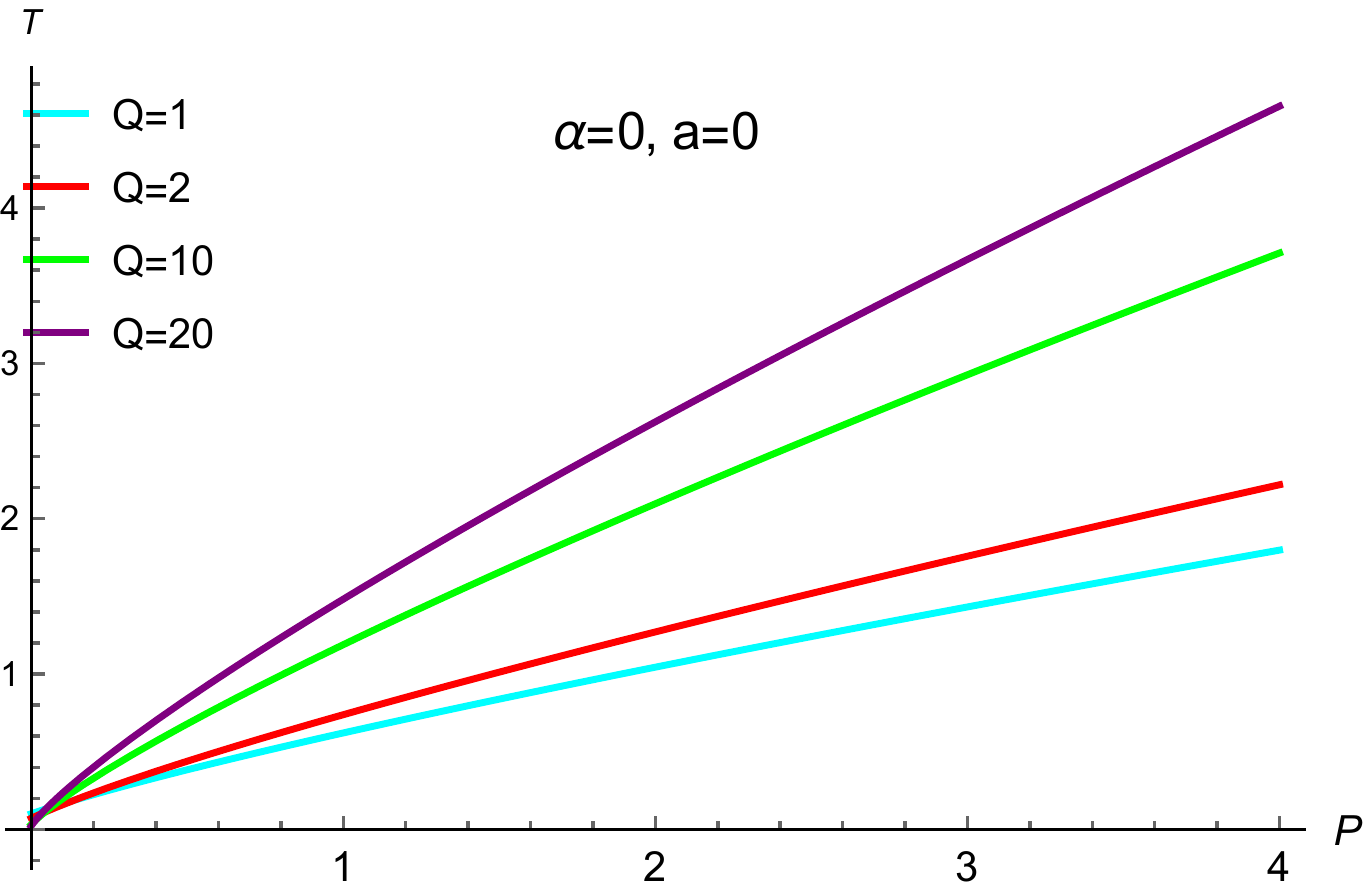}
 \label{fig:fz500}}
  \subfigure[{$d=5$}]{
  \includegraphics[width=0.31\textwidth]{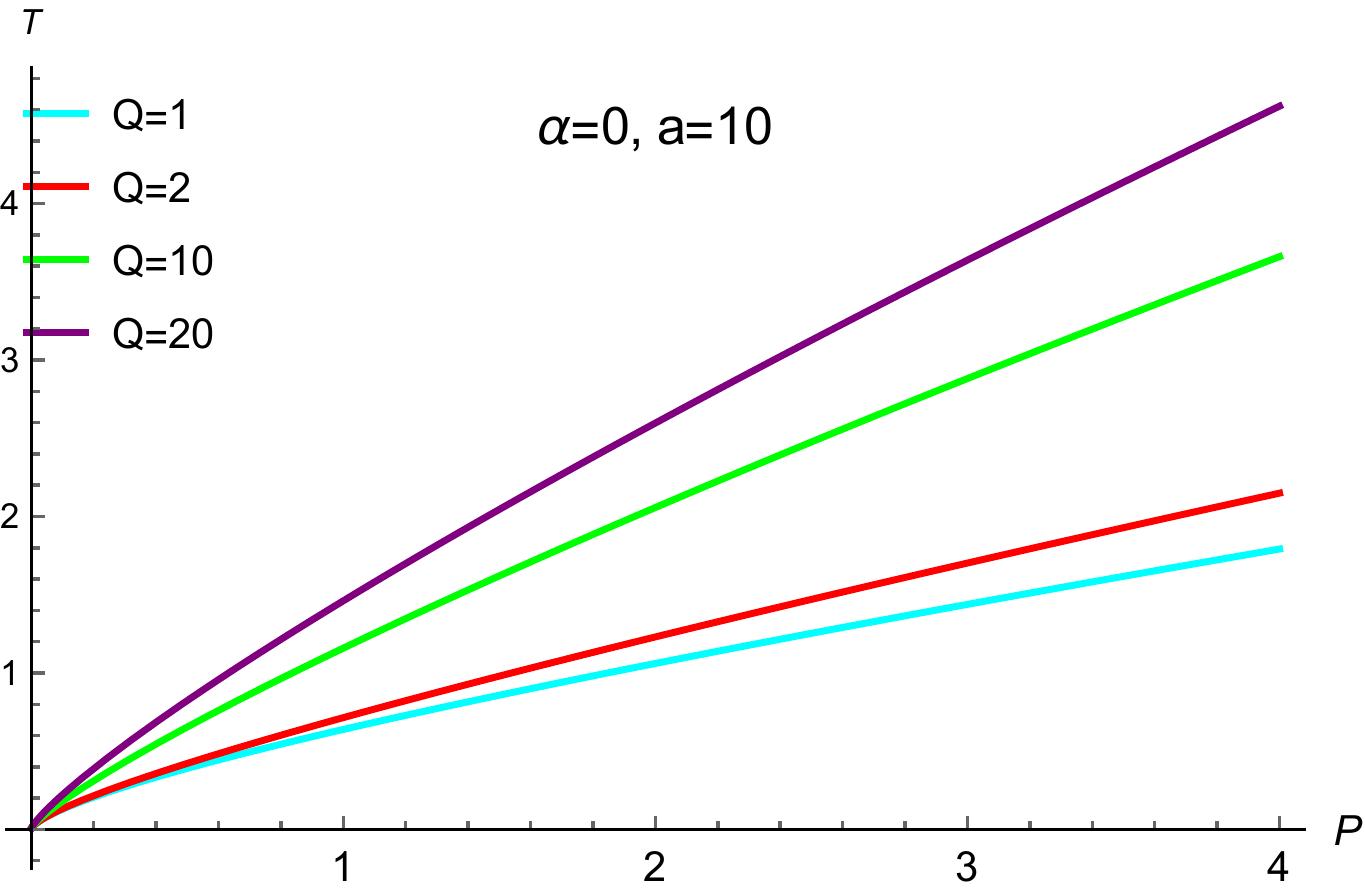}
 \label{fig:fz501}}
  \subfigure[{$d=5$}]{
  \includegraphics[width=0.31\textwidth]{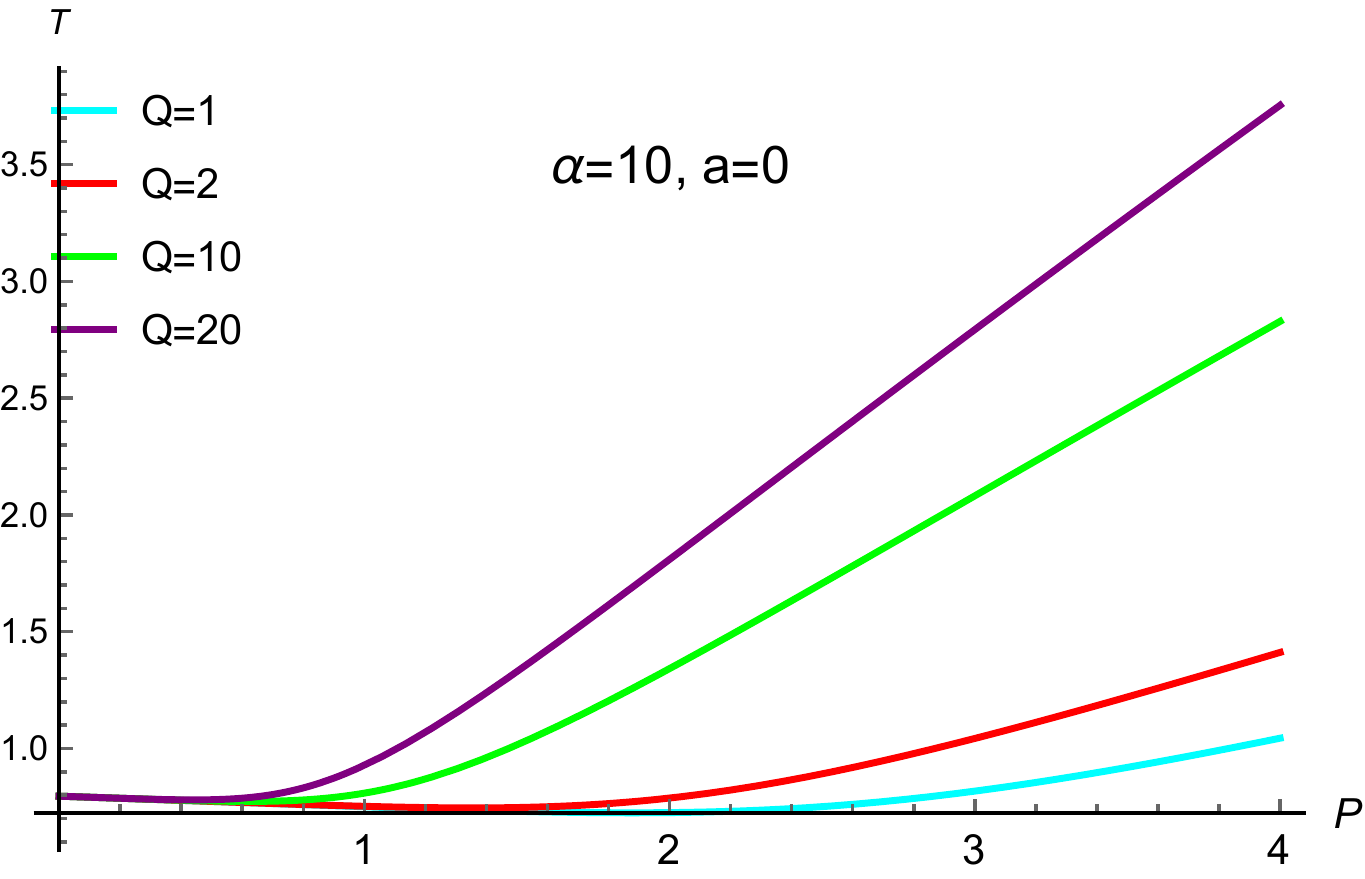}
 \label{fig:fz502}}
 \subfigure[{$d=6$}]{
  \includegraphics[width=0.31\textwidth]{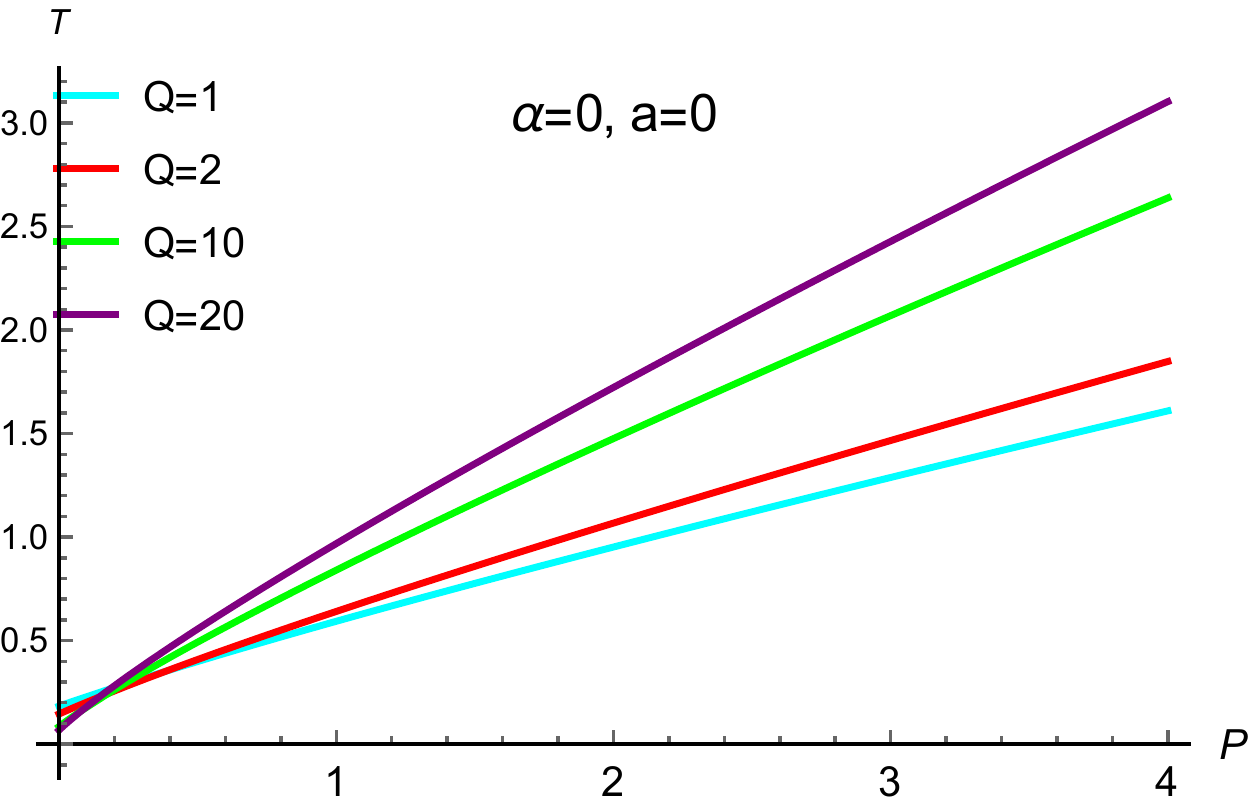}
 \label{fig:fz600}}
  \subfigure[{$d=6$}]{
  \includegraphics[width=0.31\textwidth]{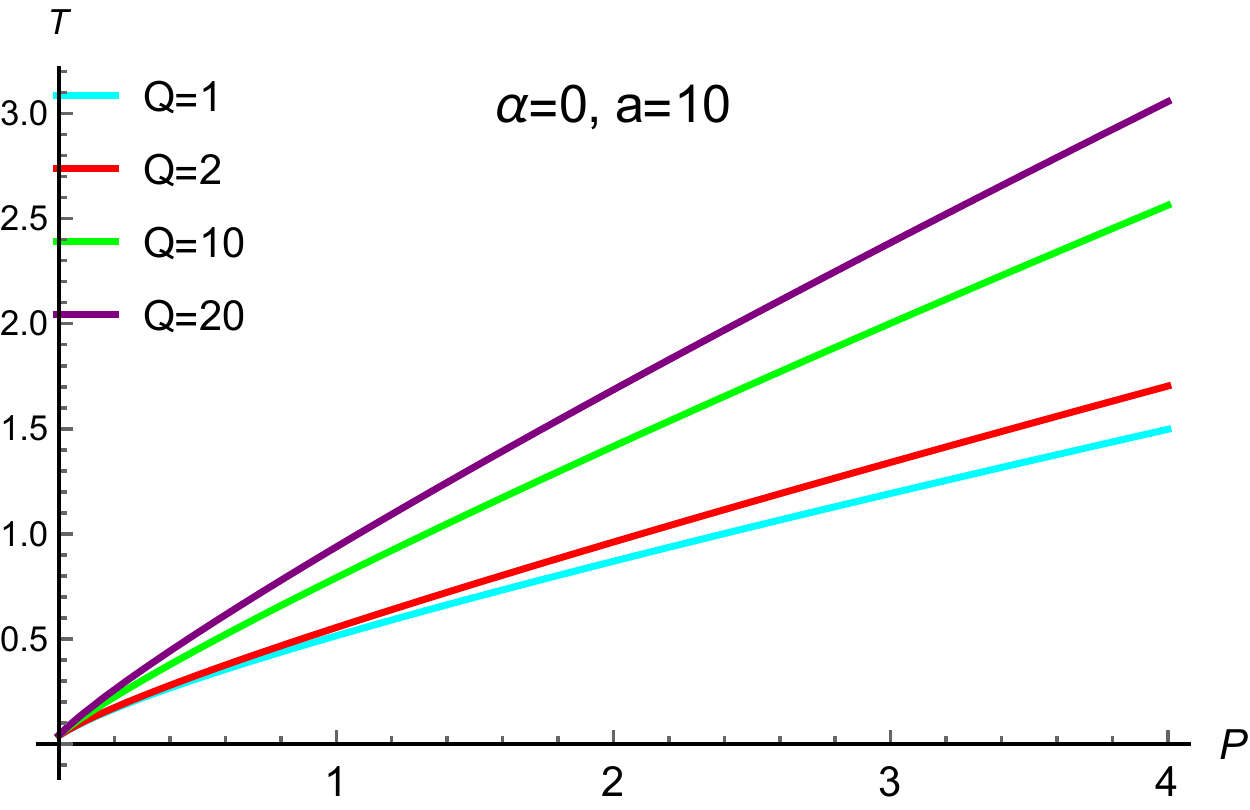}
 \label{fig:fz601}}
  \subfigure[{$d=6$}]{
  \includegraphics[width=0.31\textwidth]{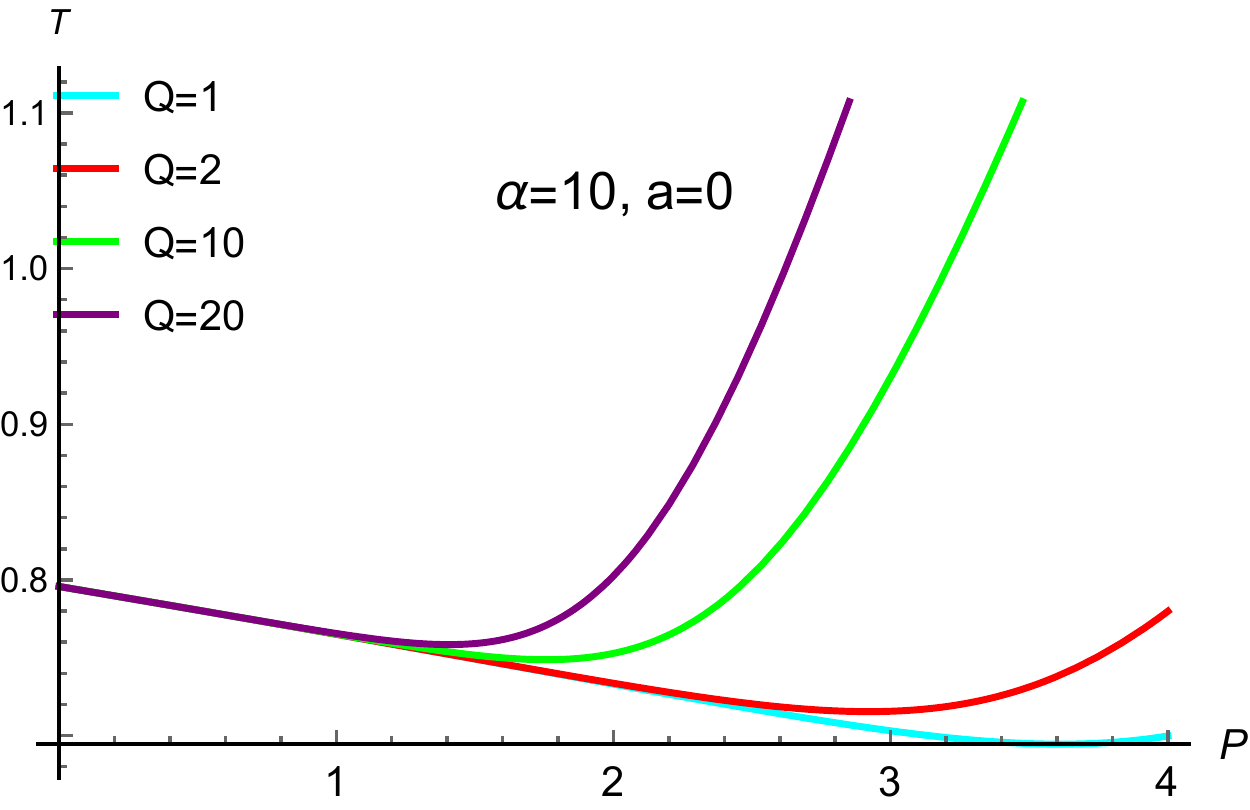}
 \label{fig:fz602}}
 \subfigure[{$d=7$}]{
  \includegraphics[width=0.31\textwidth]{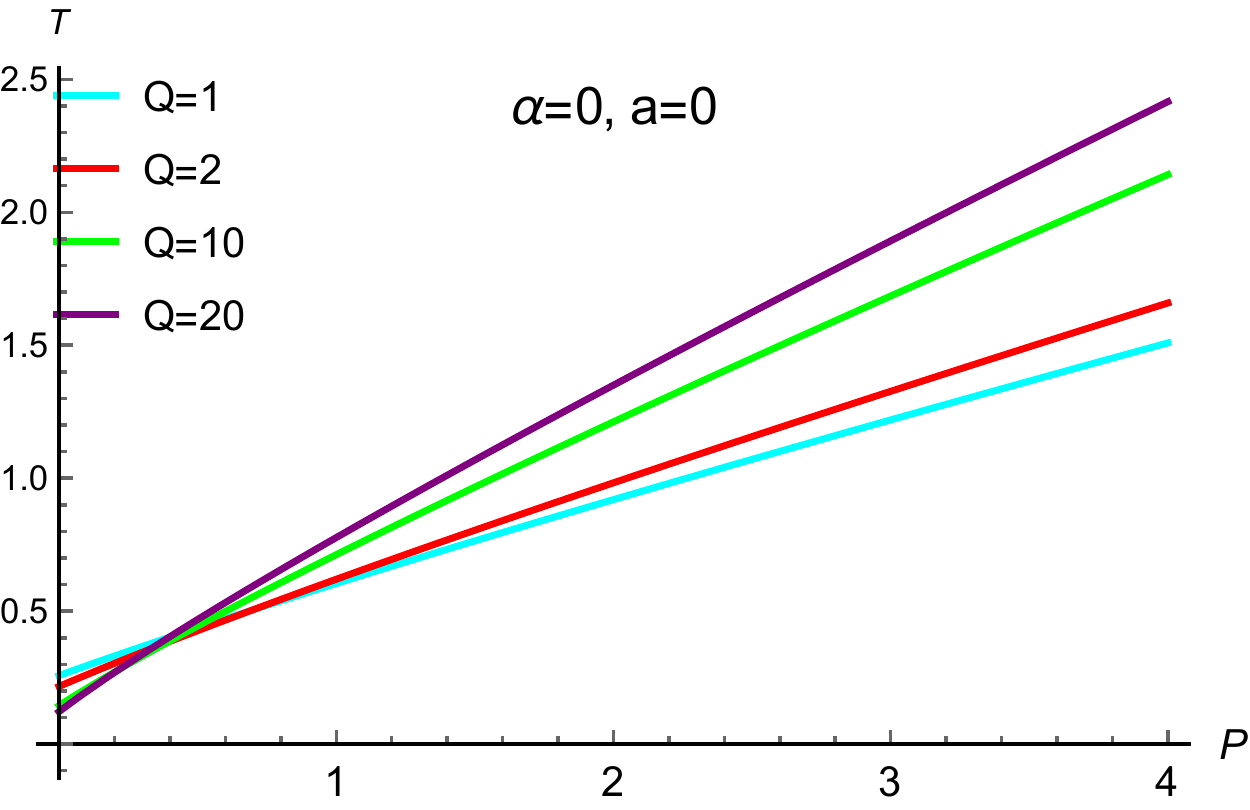}
 \label{fig:fz700}}
  \subfigure[{$d=7$}]{
  \includegraphics[width=0.31\textwidth]{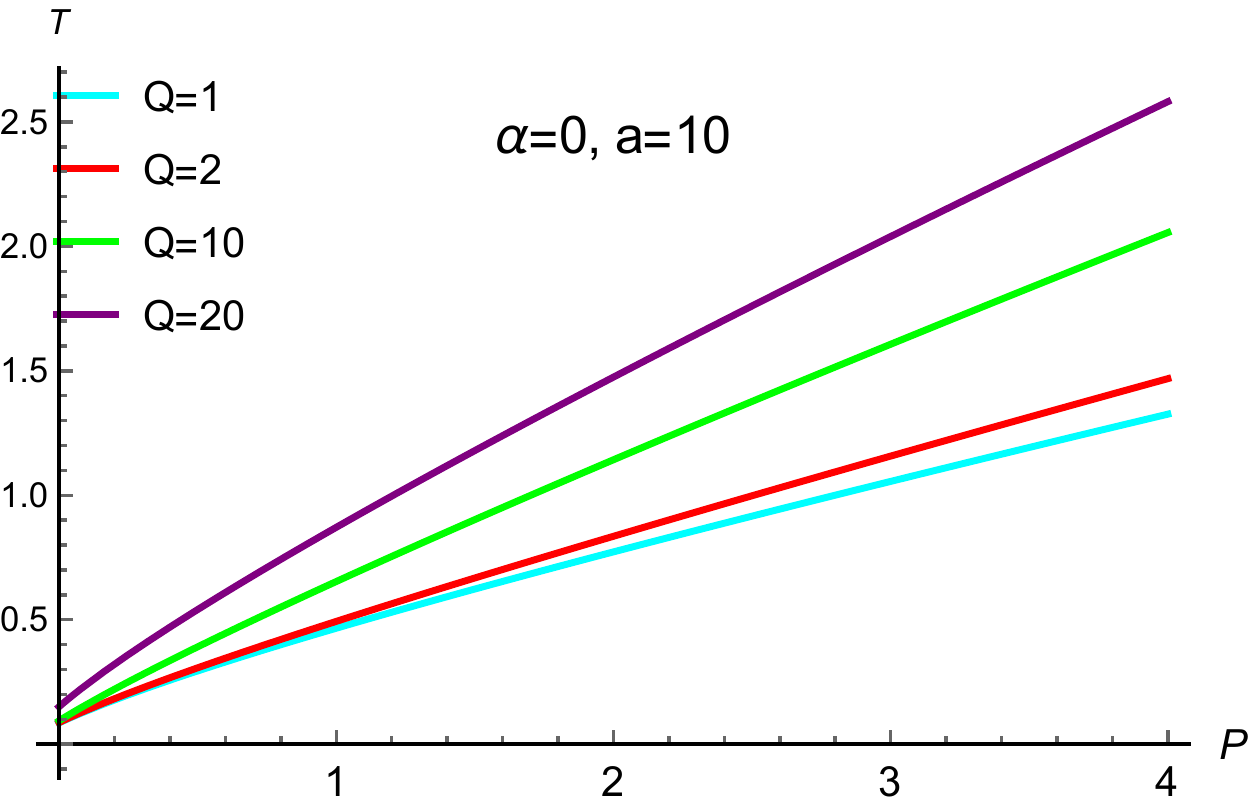}
 \label{fig:fz701}}
  \subfigure[{$d=7$}]{
  \includegraphics[width=0.31\textwidth]{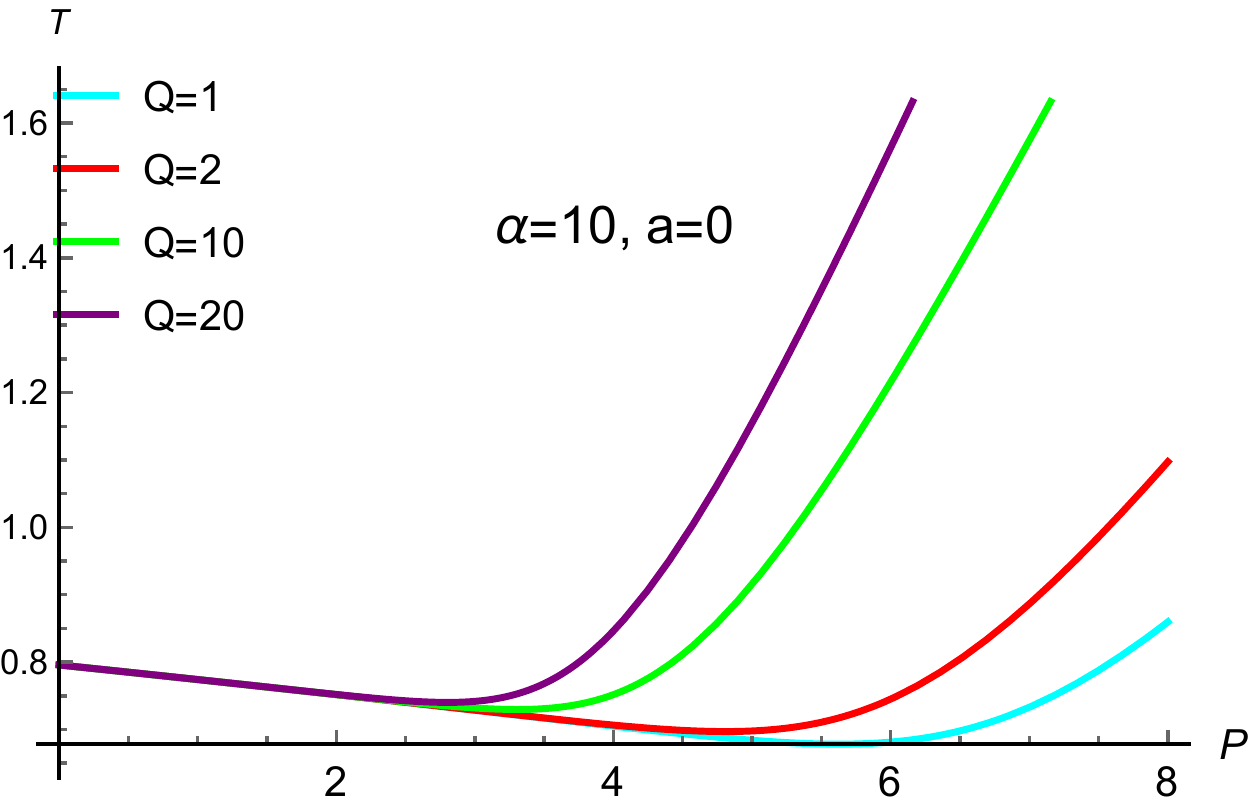}
 \label{fig:fz702}}
 \subfigure[{$d=8$}]{
  \includegraphics[width=0.31\textwidth]{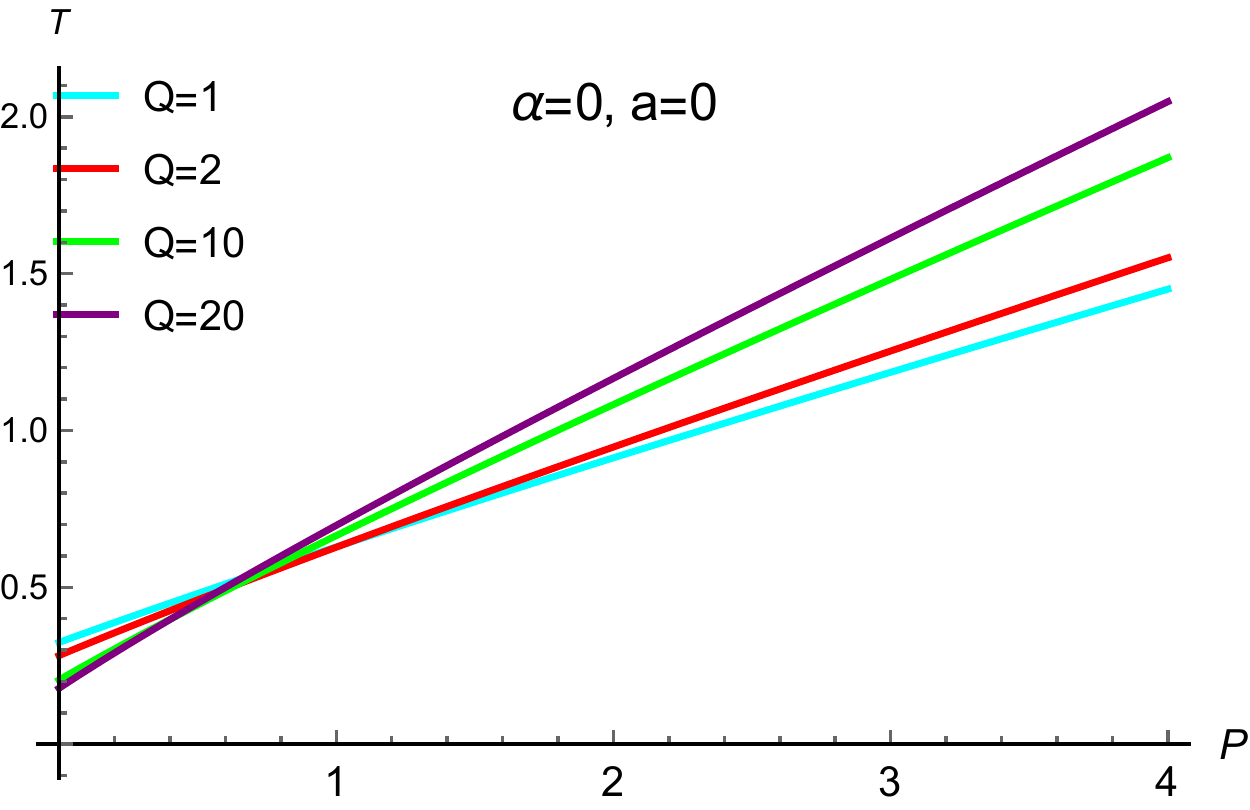}
 \label{fig:fz800}}
  \subfigure[{$d=8$}]{
  \includegraphics[width=0.31\textwidth]{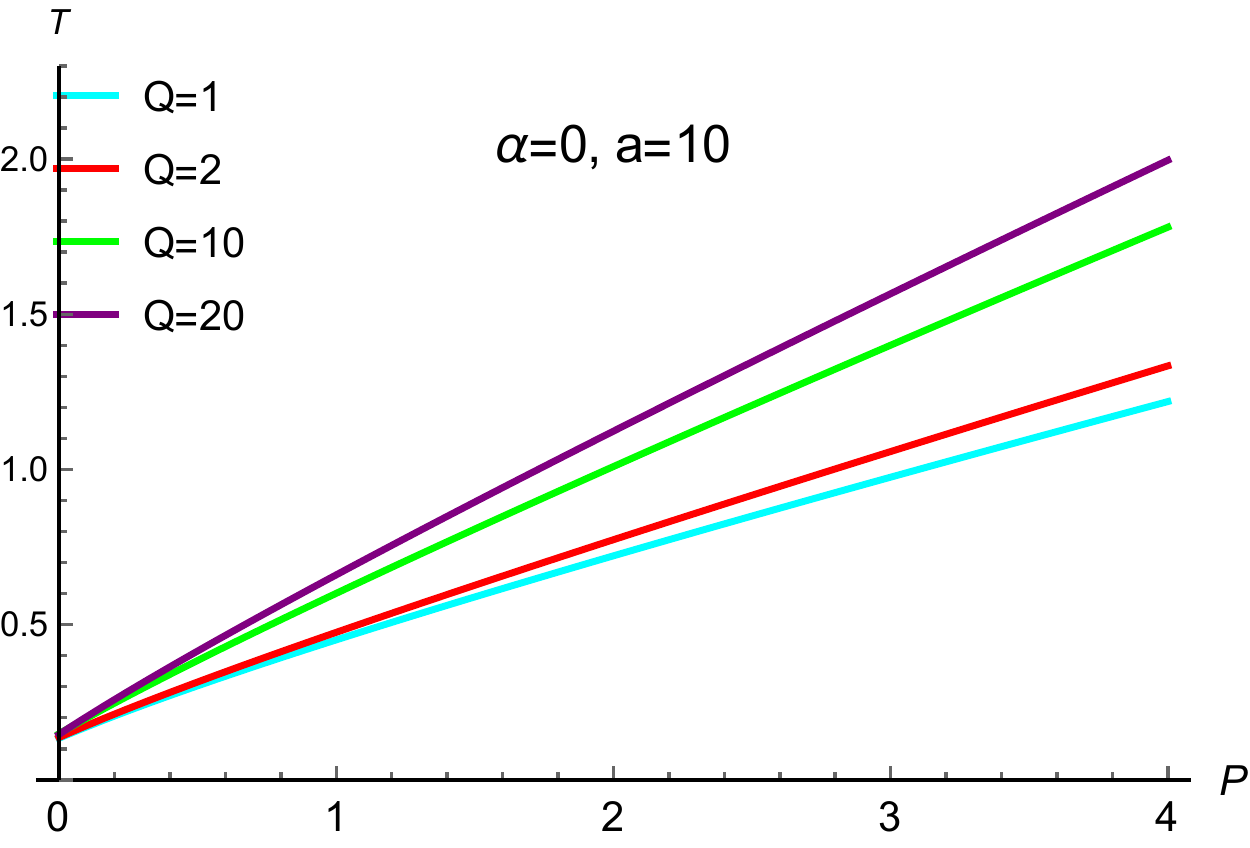}
 \label{fig:fz801}}
  \subfigure[{$d=8$}]{
  \includegraphics[width=0.31\textwidth]{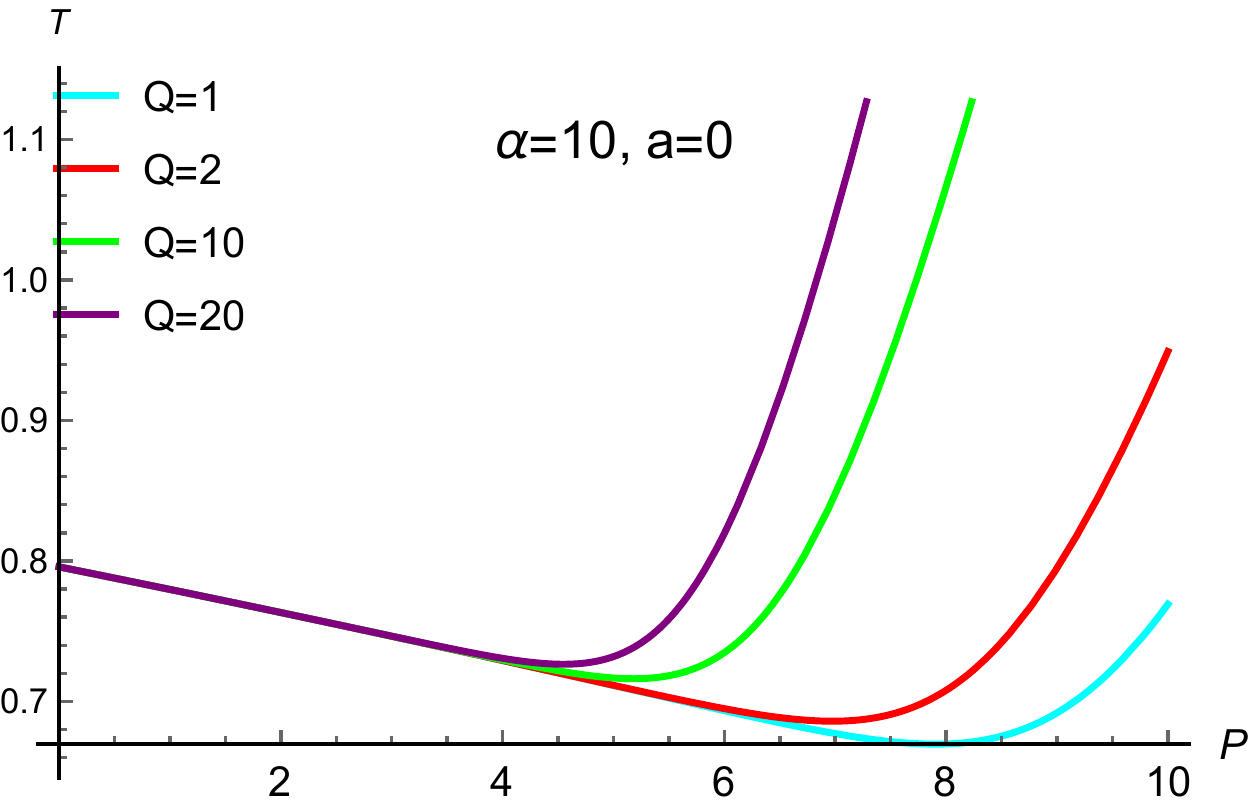}
 \label{fig:fz802}}
  \caption{The inversion curves versus $P$ for various combinations of $d$, $a$, $\alpha$ and $Q$.}\label{fig:TIC2}
\end{figure}

With the study of extending the black hole thermodynamics to the Joule-Thomson expansion regime, the inversion curves and the isenthalpic curves involved in the Joule-Thomson expansion of various black holes, have become major research interests. Besides, the results are often compared with van der Waals fluid. Here the investigation highlights the variations of the inversion curves concerning the parameters $d$, $a$, $\alpha$ and $Q$. As regards the derivation of the inversion temperature, we can obtain it from this approach by setting $\mu=0$, as follows
\begin{equation}\label{eqn:fz1}
  T_{i}=V(\frac{\partial T}{\partial V})_{P,Q,a,\alpha},
\end{equation}
then, the inversion temperature for a d-dimensional charged AdS black hole with cloud of strings and quintessence is
\begin{equation}\label{eqn:fz3}
 T_{i}=\frac{2a(d-3)r_{+}^{4-d}+8(2d-5)\pi^{3-d}Q^{2}r_{+}^{6-2d}\Gamma\left(\frac{d-1}{2}\right)^{2}-(d-5)d+16\pi Pr_{+}^{2}-6}{4\pi(d-2)(d-1)r_{+}}.
\end{equation}

In the case of $T_{i}=T=(\frac{\partial M}{\partial S})_{P,Q,a,\alpha}$, subtracting Eq. $\left(\ref{eqn:two11}\right)$ from Eq. $\left(\ref{eqn:fz3}\right)$ we can obtain
\begin{equation}\label{eqn:fz4}
\frac{4ar_{+}^{3-d}+24\pi^{3-d}Q^{2}r_{+}^{5-2d}\Gamma\left(\frac{d-1}{2}\right)^{2}+\frac{(d-3)d\left(\alpha r_{+}-1\right)}{r_{+}}+2\left(\alpha-8\pi Pr_{+}\right)}{4\pi(d-1)}=0.
\end{equation}

The explicit expressions for the inversion temperature can be obtained by substituting the roots into Eq. $\left(\ref{eqn:two11}\right)$, and the roots are obtained by solving Eq. $\left(\ref{eqn:fz4}\right)$. However, the explicit expressions for roots that satisfy the condition are so complicated that we do not represent them. In the following part of the paper, the numerical solution is the approach used to investigate.

In order to visualize how the inversion curve varies under the influence of $d$, Fig. \ref{fig:TIC1} is shown with $Q$ fixed and without considering parameters $a$ and $\alpha$. As a result, the same observations as in literature \cite{Mo:2018rgq} can be obtained, i.e., the effect of $d$ on the inversion curve is different in the high and low pressure cases. In the high-pressure case, the inversion curve decreases as $d$ increases, and the opposite is observed in the low-pressure case.

In Fig. \ref{fig:TIC2}, how parameters $Q$, $a$ and $\alpha$ affect the inversion curve are exhibited in the $T$-$P$ plane. By comparing the effect of Q on the inversion curve in different dimensions $d$ with $a$ and $\alpha$ fixed (e.g., comparing Figs. \ref{fig:fz500}, \ref{fig:fz600}, \ref{fig:fz700}, \ref{fig:fz800}), it is evident that the inversion temperature for a given pressure increases with the increase of $Q$ in the high-pressure region and decreases with the increase of $Q$ in the low-pressure region. Note that this result is independent of $d$. Further, by comparing the influence of $a$ on the inversion curve in different dimensions, we can compare the first and second columns of the Fig. \ref{fig:TIC2}, for example, the comparison between Fig. \ref{fig:fz800} and Fig. \ref{fig:fz801}. It can be seen that as $a$ changes the inversion curve also changes, whereby the high and low pressure dividing point becomes smaller as $a$ increases. Such a result holds in different dimensions $d$.

Observing the first and third columns of the figure, the change in $\alpha$ has changed the inversion curve significantly, i.e., the high and low pressure dividing points move to the right as $\alpha$ increases, moreover, the change is most pronounced in the low pressure region. Overall, the effect of $\alpha$ on the inversion curve is more profound than the effect of $a$ on the inversion curve. Combined with our findings in the literature \cite{Yin:2021akt} regarding the four dimensions, it is known that the above mentioned conclusion is independent of $d$.

We next extend the study to cover the case of the isenthalpic curve.

\subsection{The isenthalpic curve}
\label{sec:CC}

\begin{figure}[H]
  \centering
  \subfigure[{$d=5$}]{
  \includegraphics[width=0.31\textwidth]{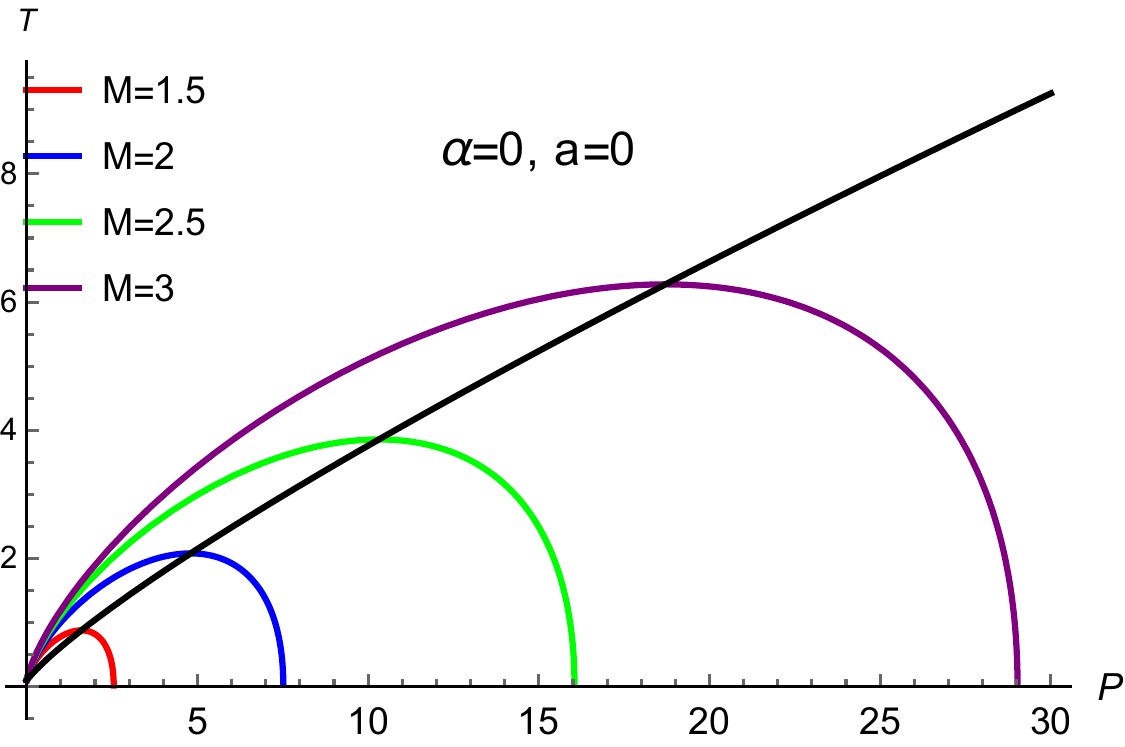}
 \label{fig:1dh500}}
   \subfigure[{$d=5$}]{
  \includegraphics[width=0.31\textwidth]{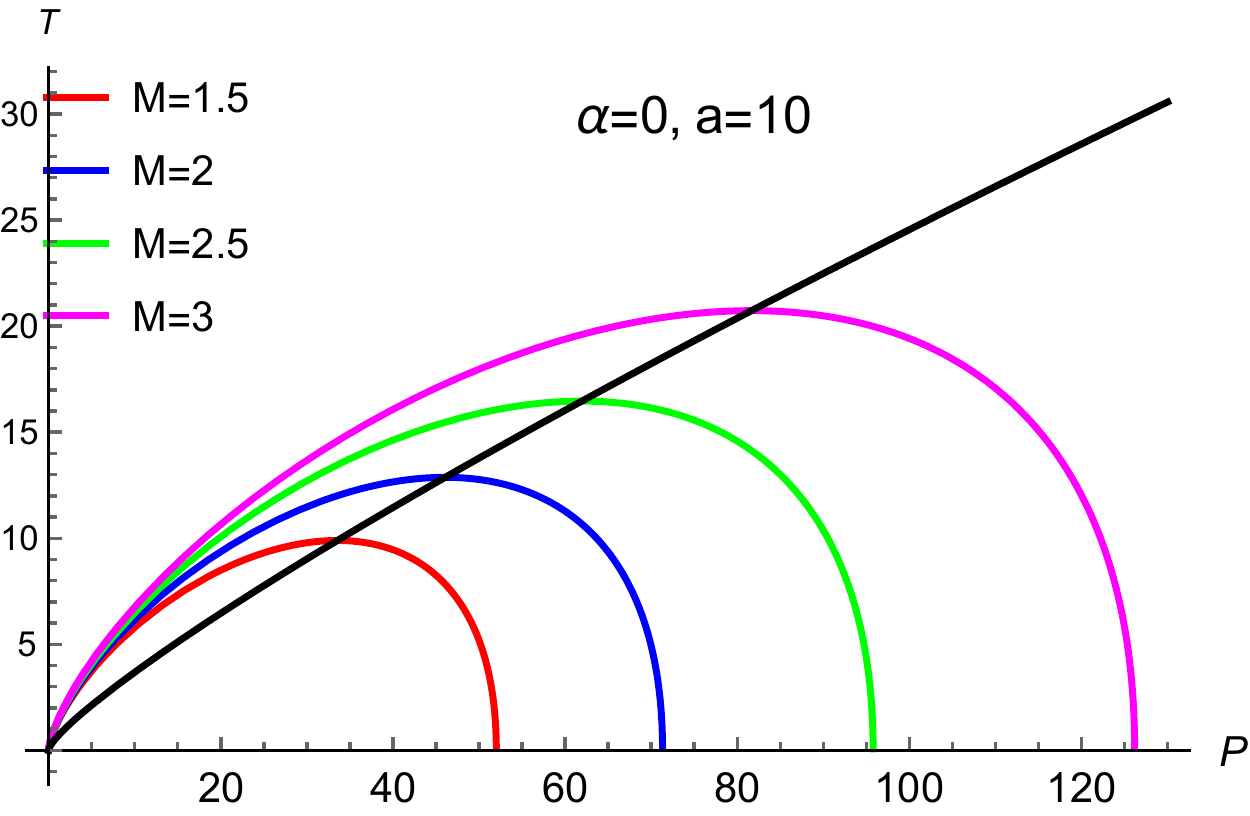}
 \label{fig:1dh5100}}
  \subfigure[{$d=5$}]{
  \includegraphics[width=0.31\textwidth]{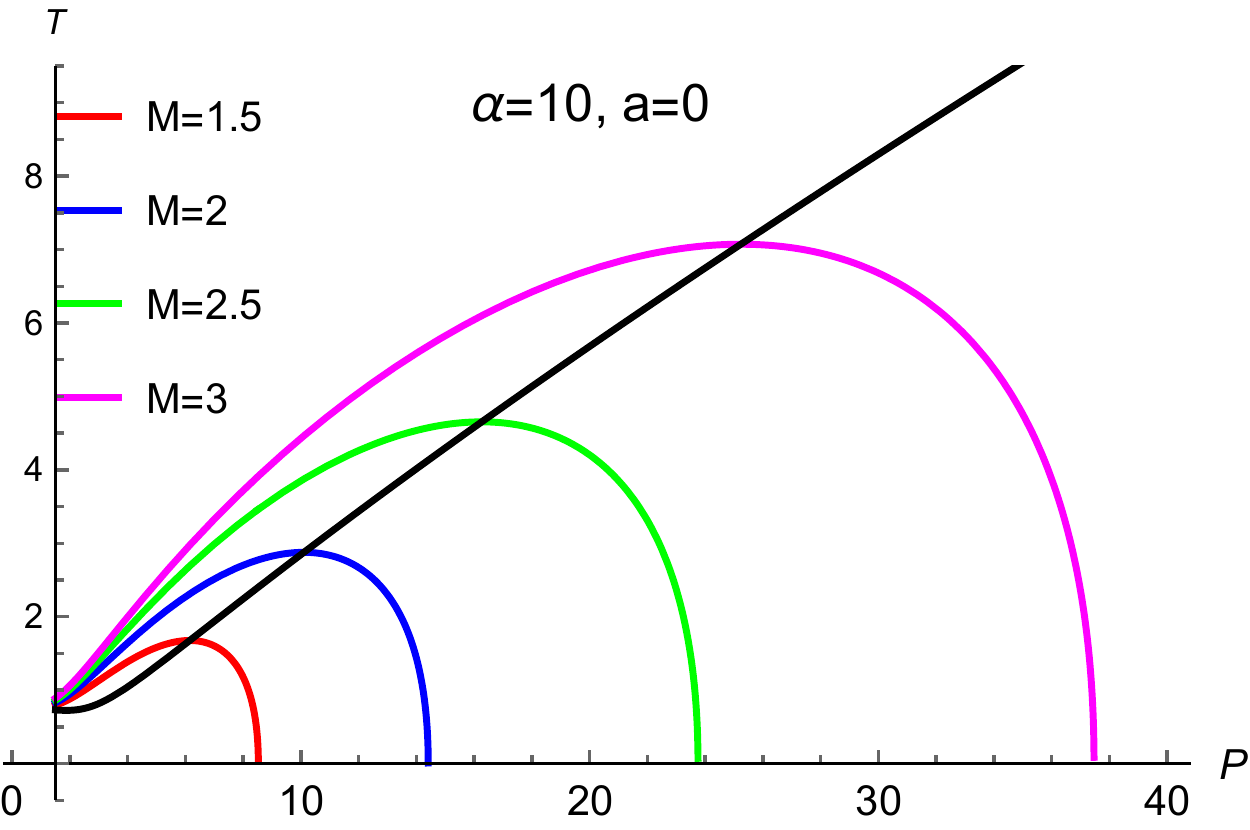}
 \label{fig:1dh5010}}
   \subfigure[{$d=6$}]{
  \includegraphics[width=0.31\textwidth]{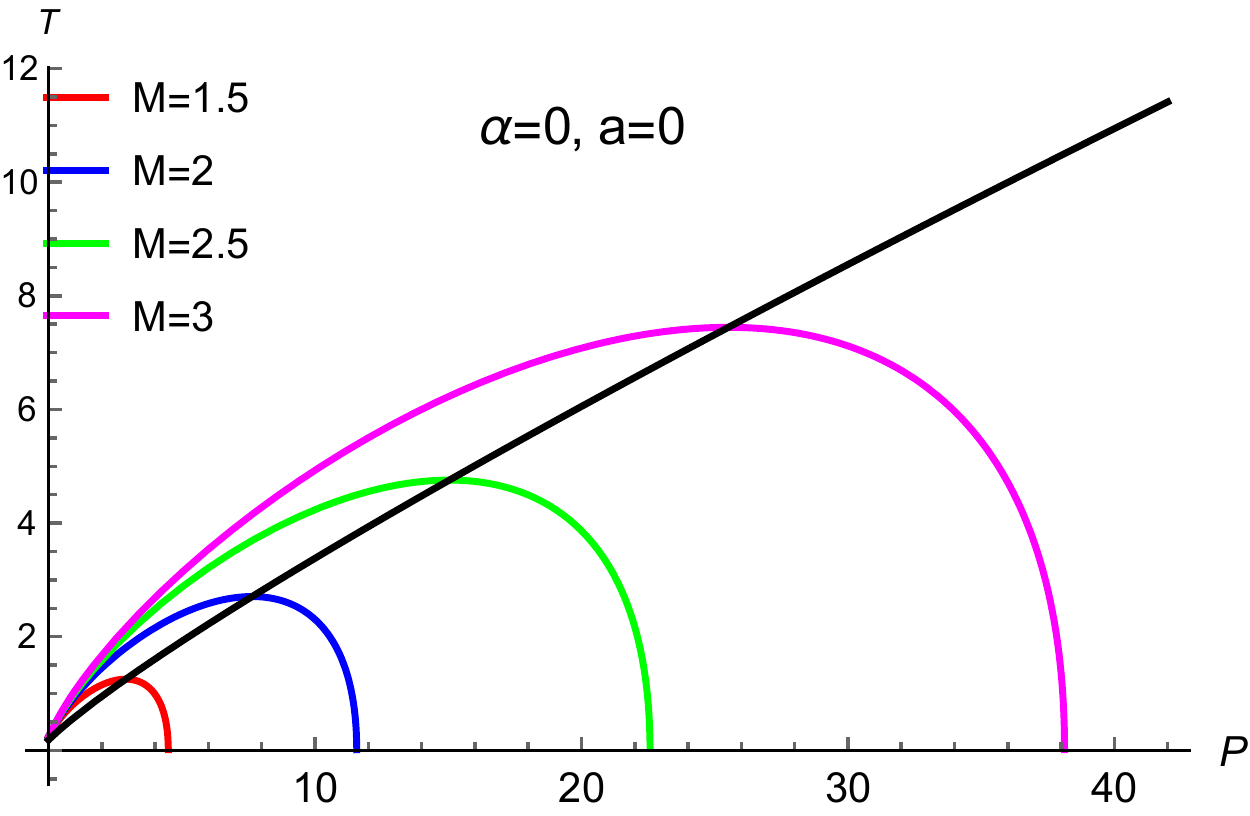}
 \label{fig:1dh600}}
   \subfigure[{$d=6$}]{
  \includegraphics[width=0.31\textwidth]{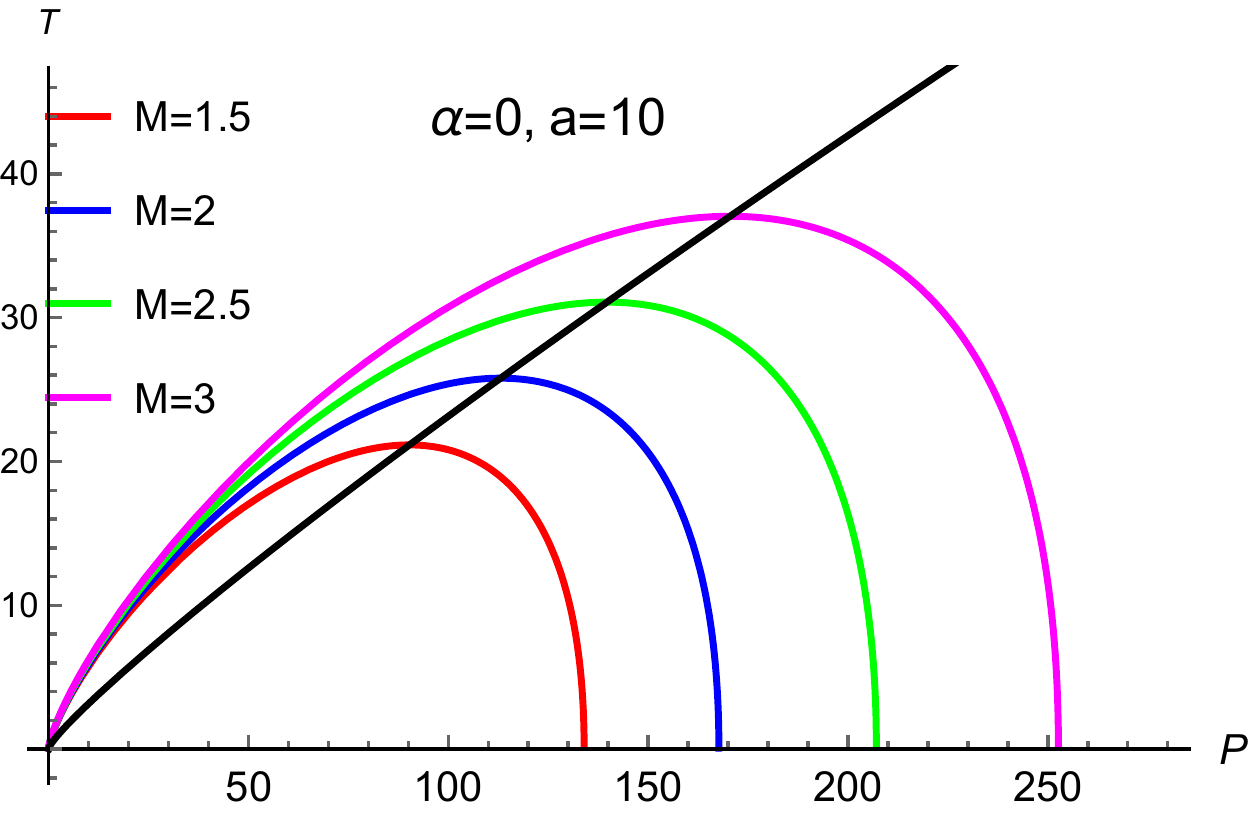}
 \label{fig:1dh6100}}
  \subfigure[{$d=6$}]{
  \includegraphics[width=0.31\textwidth]{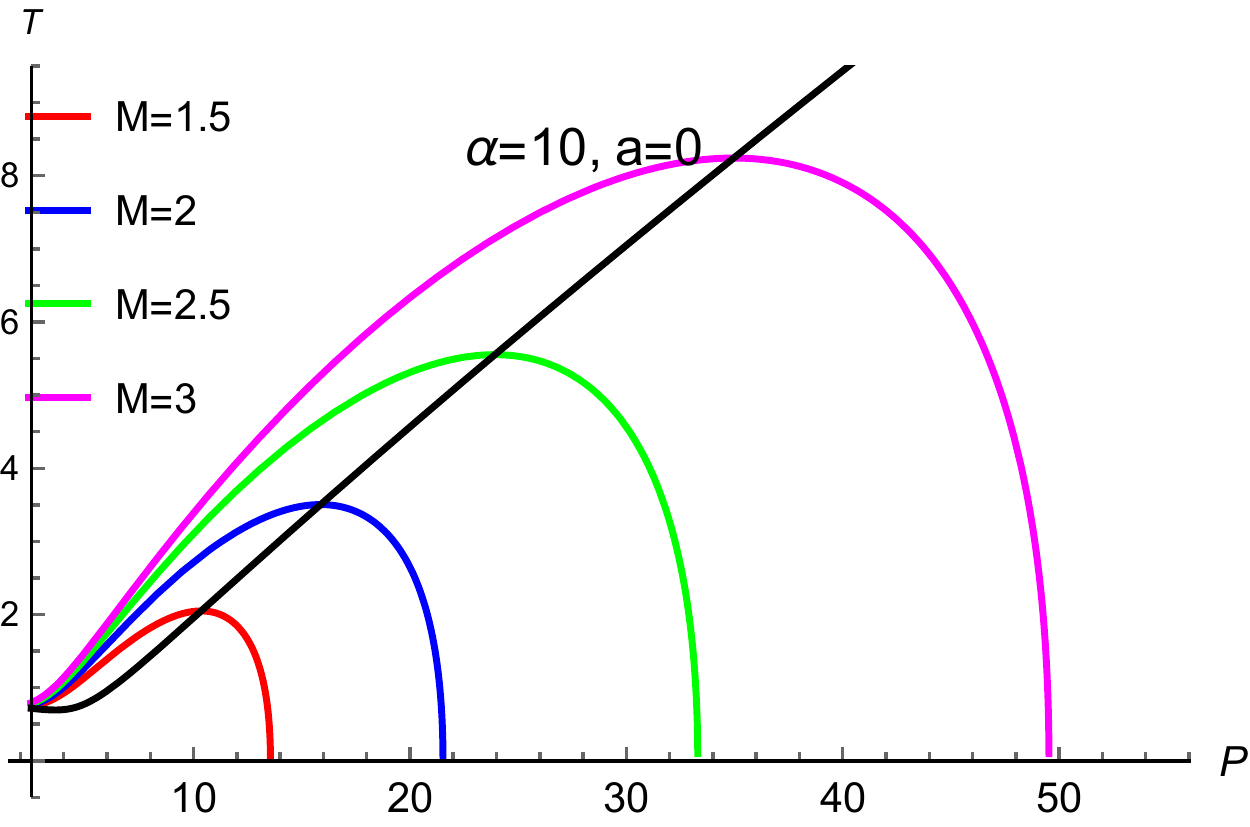}
 \label{fig:1dh6010}}
    \subfigure[{$d=7$}]{
  \includegraphics[width=0.31\textwidth]{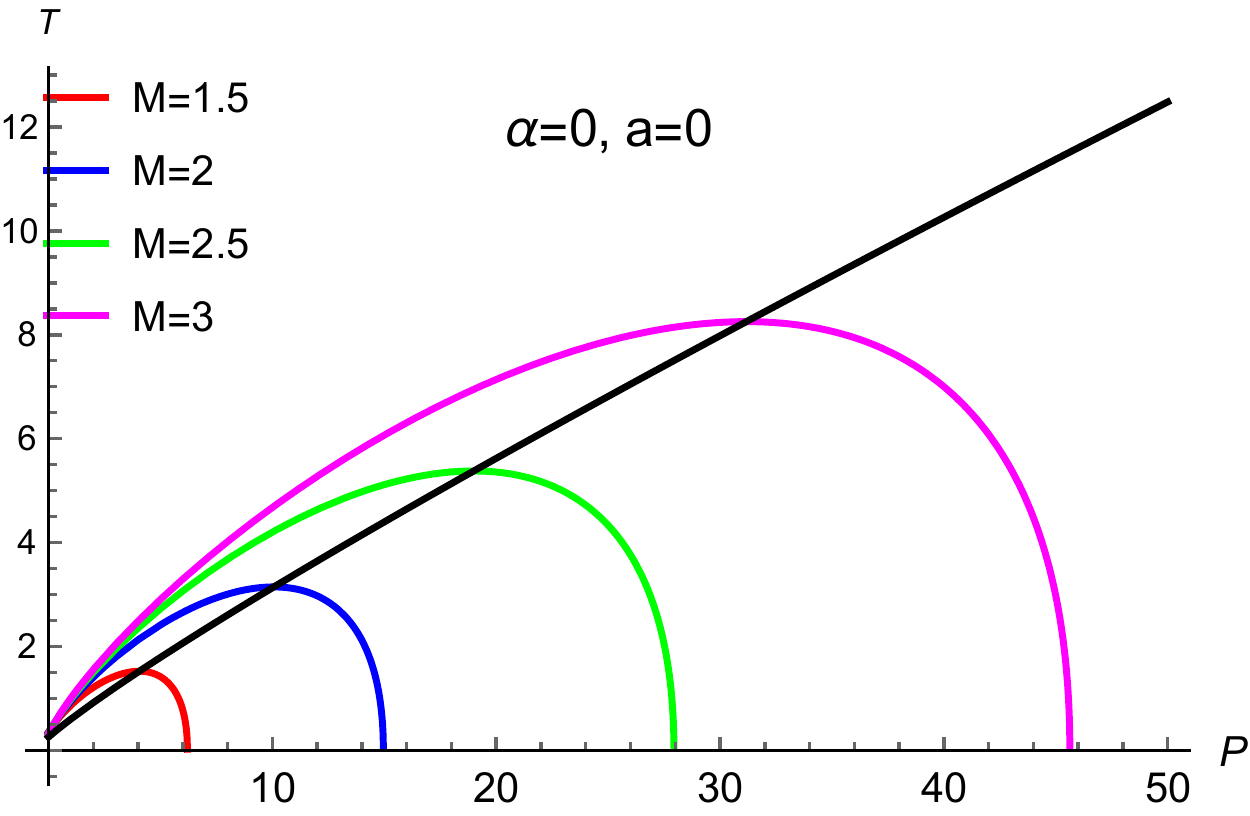}
 \label{fig:1dh700}}
   \subfigure[{$d=7$}]{
  \includegraphics[width=0.31\textwidth]{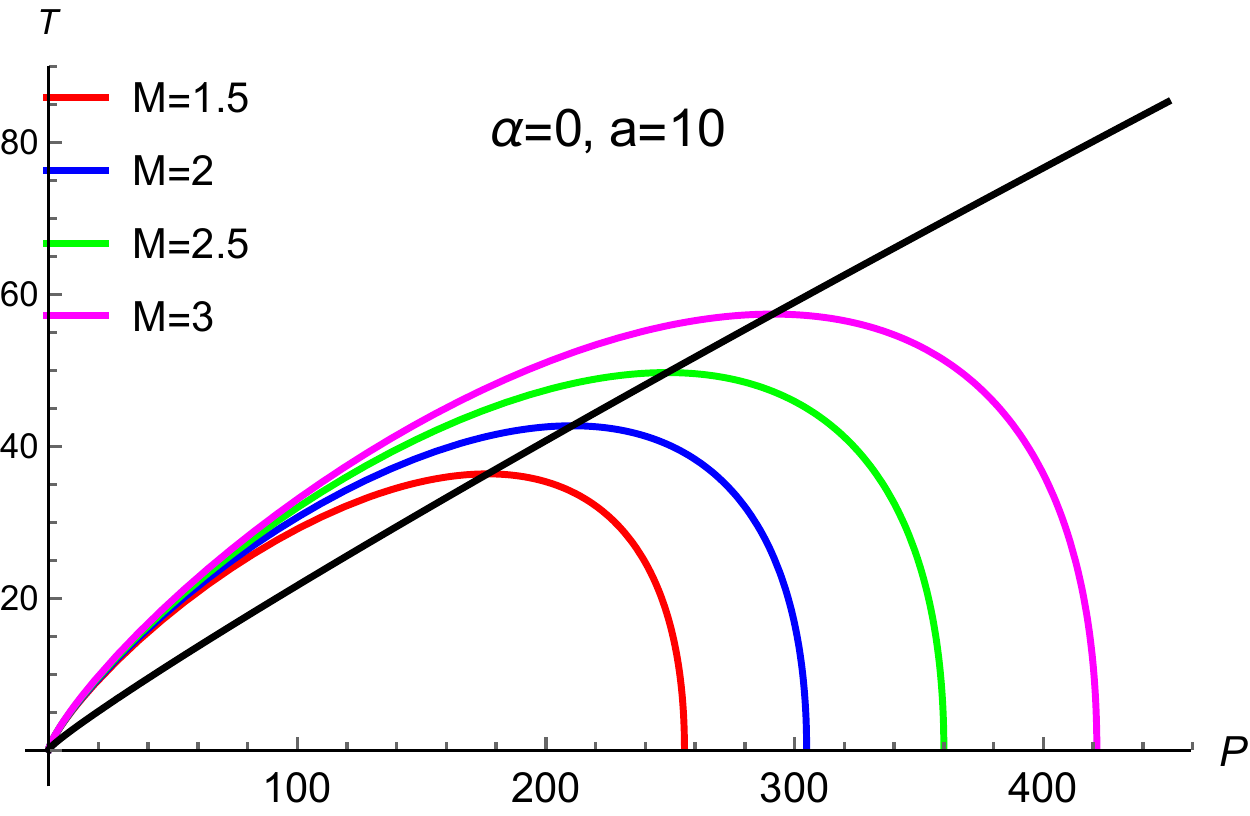}
 \label{fig:1dh7100}}
  \subfigure[{$d=7$}]{
  \includegraphics[width=0.31\textwidth]{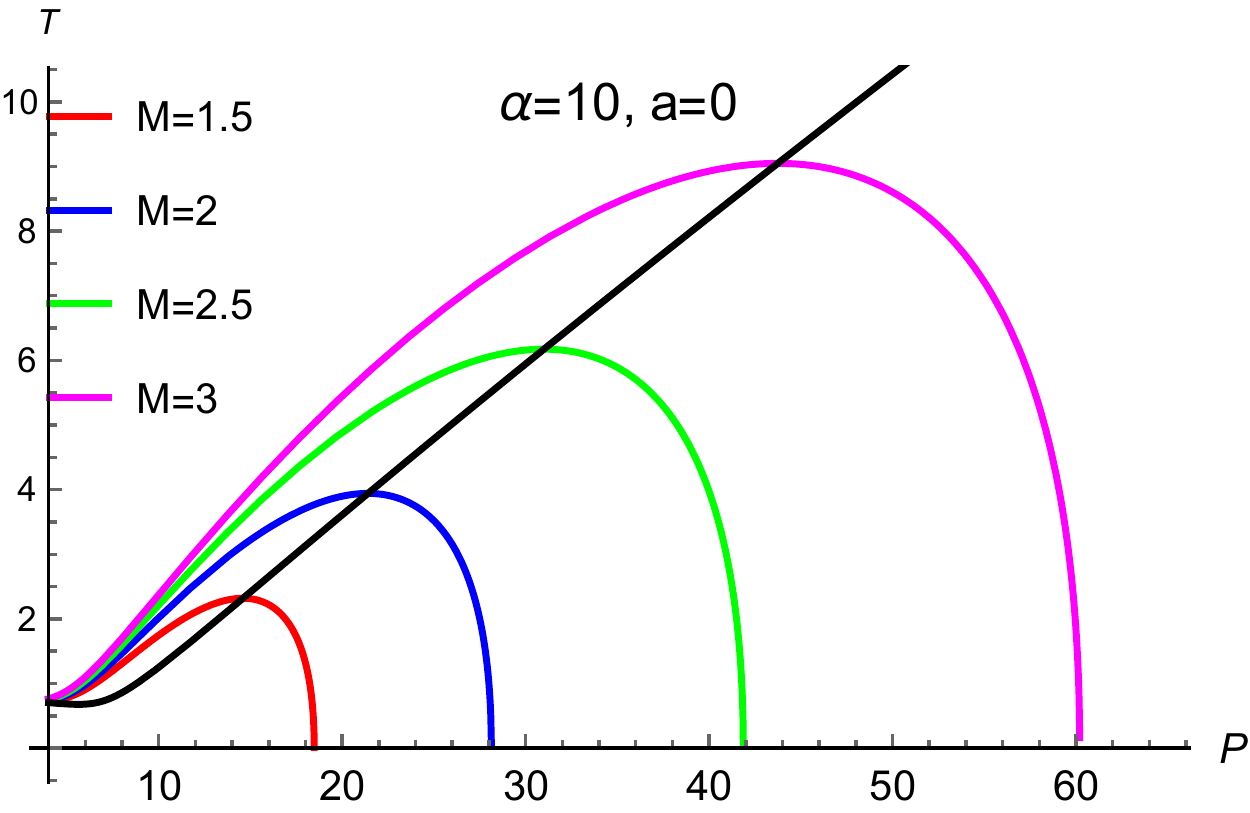}
 \label{fig:1dh7010}}
     \subfigure[{$d=8$}]{
  \includegraphics[width=0.31\textwidth]{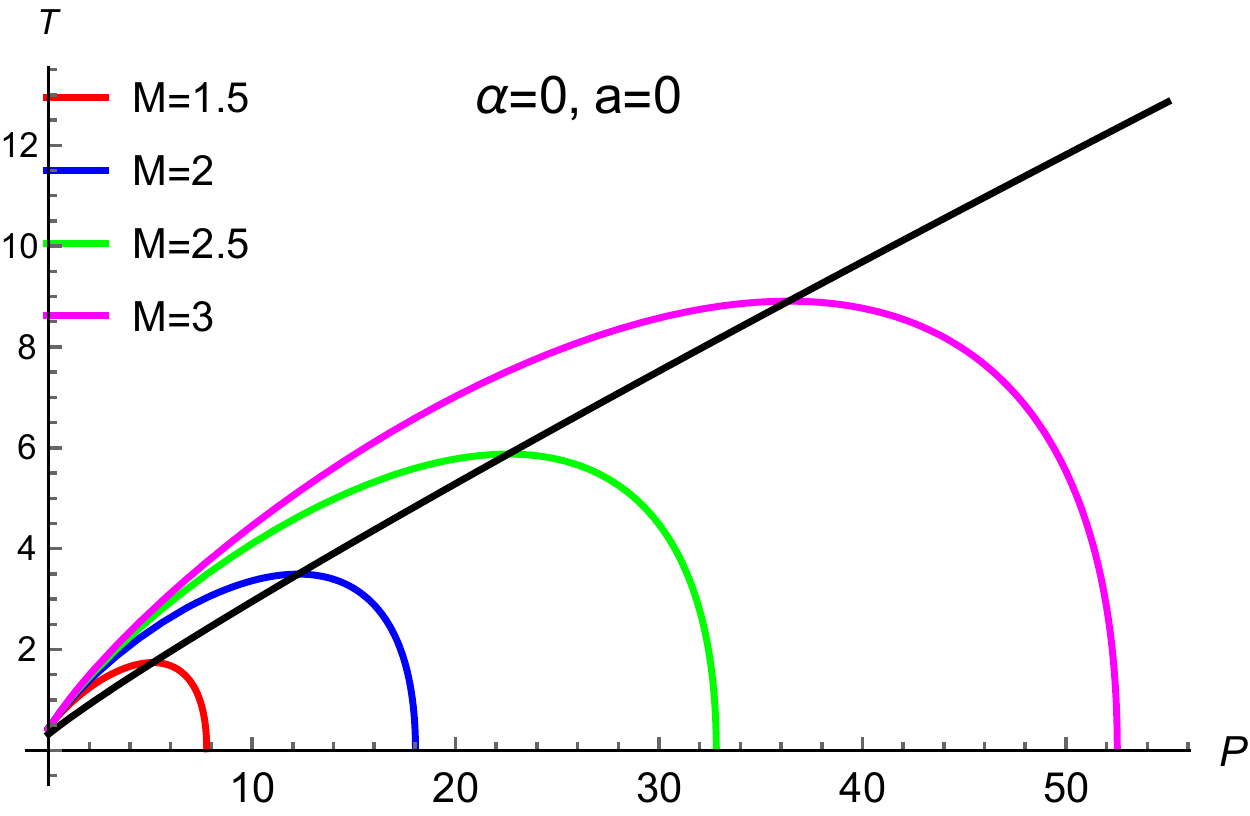}
 \label{fig:1dh800}}
   \subfigure[{$d=8$}]{
  \includegraphics[width=0.31\textwidth]{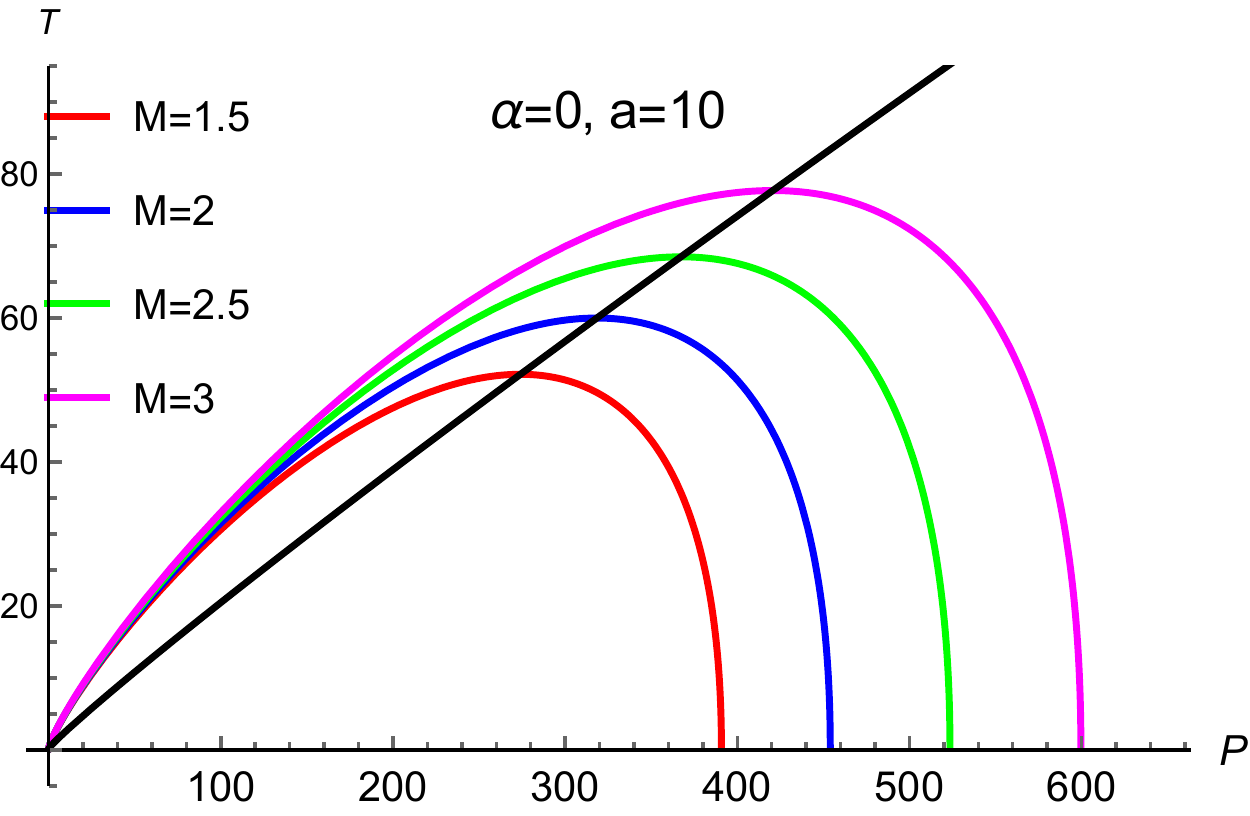}
 \label{fig:1dh8100}}
  \subfigure[{$d=8$}]{
  \includegraphics[width=0.31\textwidth]{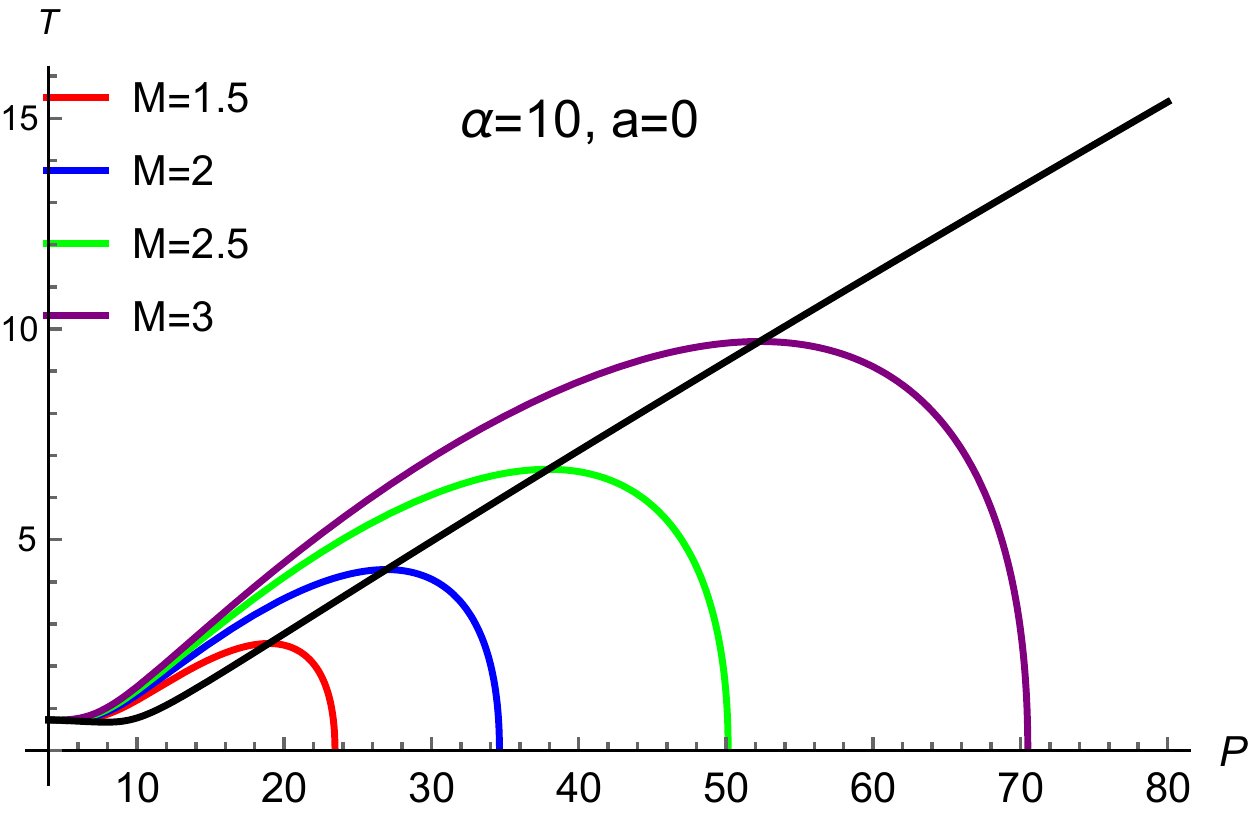}
 \label{fig:1dh8010}}
 \caption{The isenthalpic curves and inversion curves for various combinations of $d$, $a$, $\alpha$ and $M$, with $Q$=1.}\label{fig:TIC3}
\end{figure}

\begin{figure}[H]
  \centering
  \subfigure[{$d=5$}]{
  \includegraphics[width=0.31\textwidth]{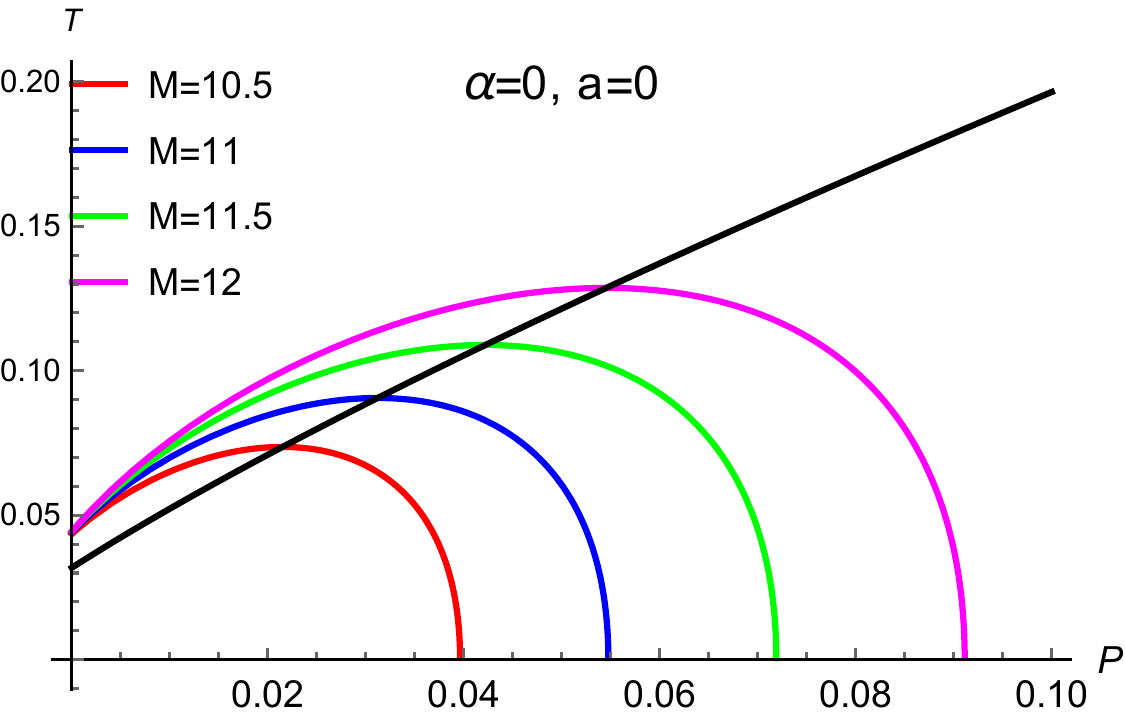}
 \label{fig:2dh500}}
   \subfigure[{$d=5$}]{
  \includegraphics[width=0.31\textwidth]{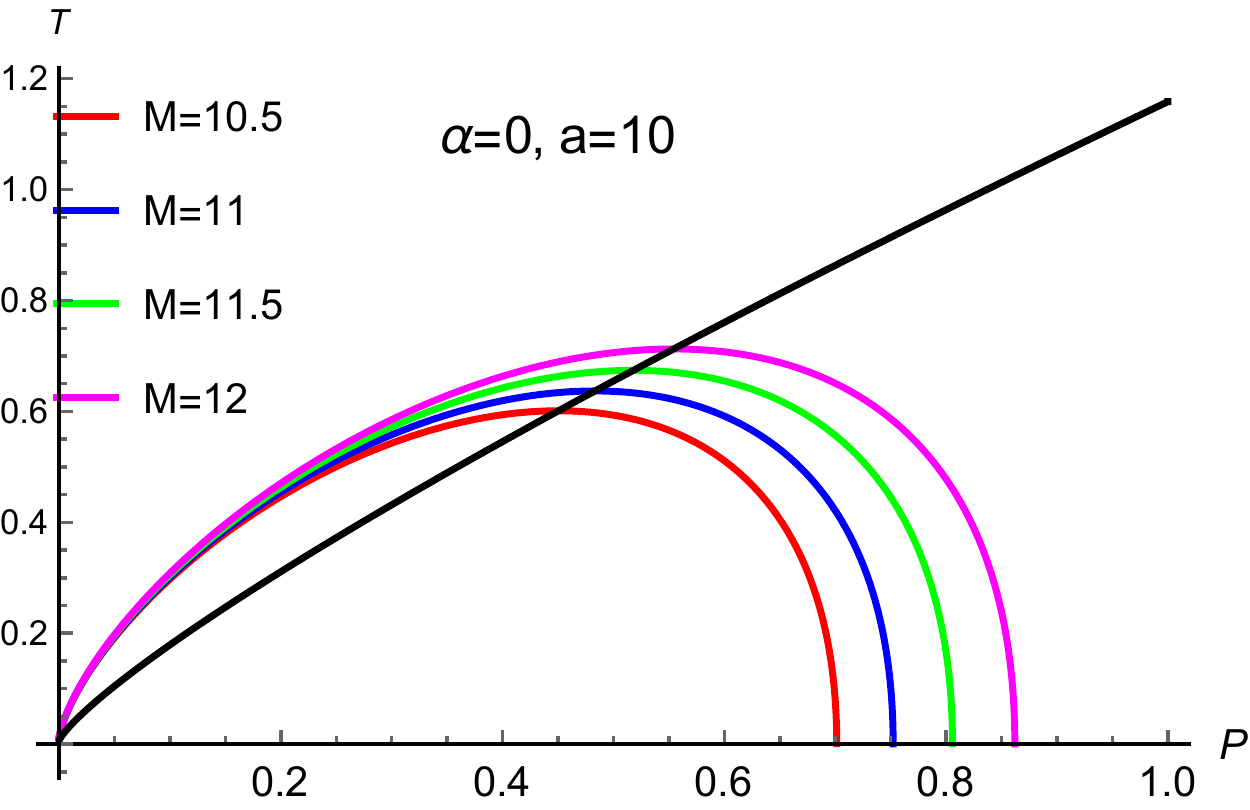}
 \label{fig:2dh5100}}
  \subfigure[{$d=5$}]{
  \includegraphics[width=0.31\textwidth]{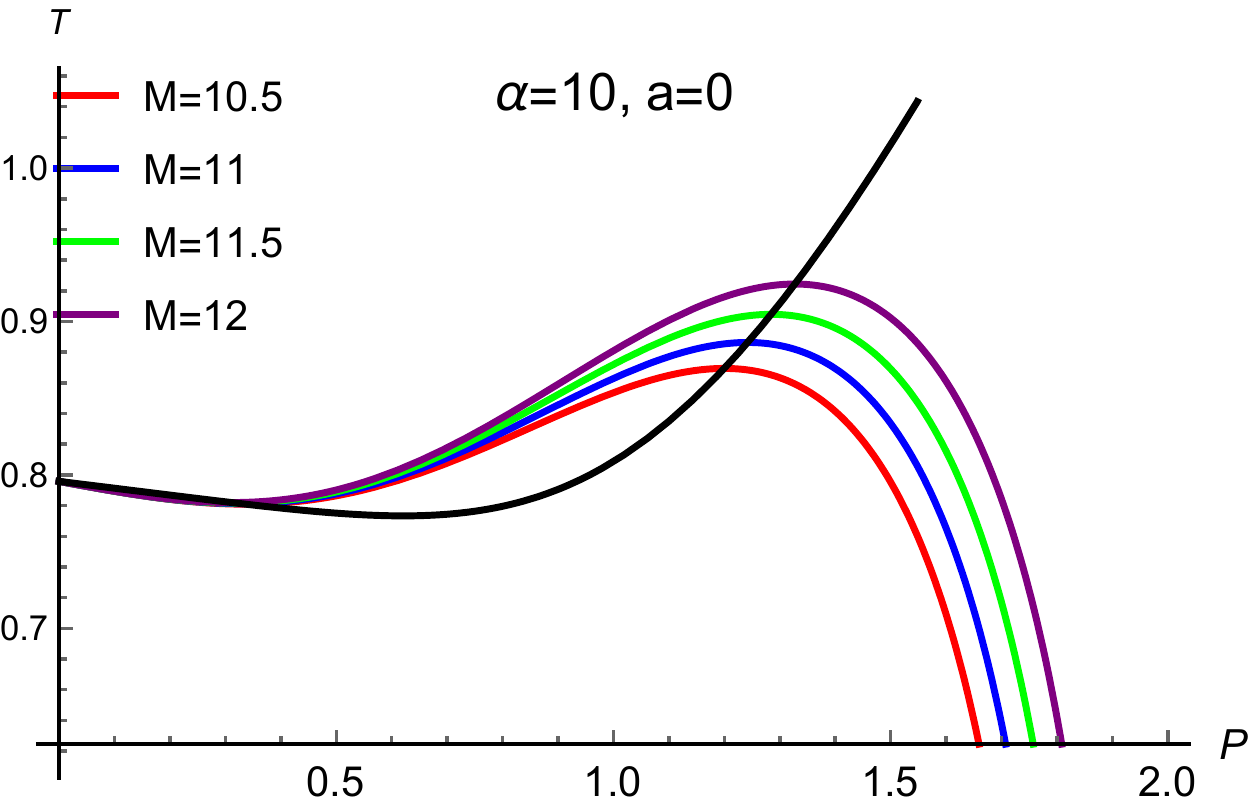}
 \label{fig:2dh5010}}
   \subfigure[{$d=6$}]{
  \includegraphics[width=0.31\textwidth]{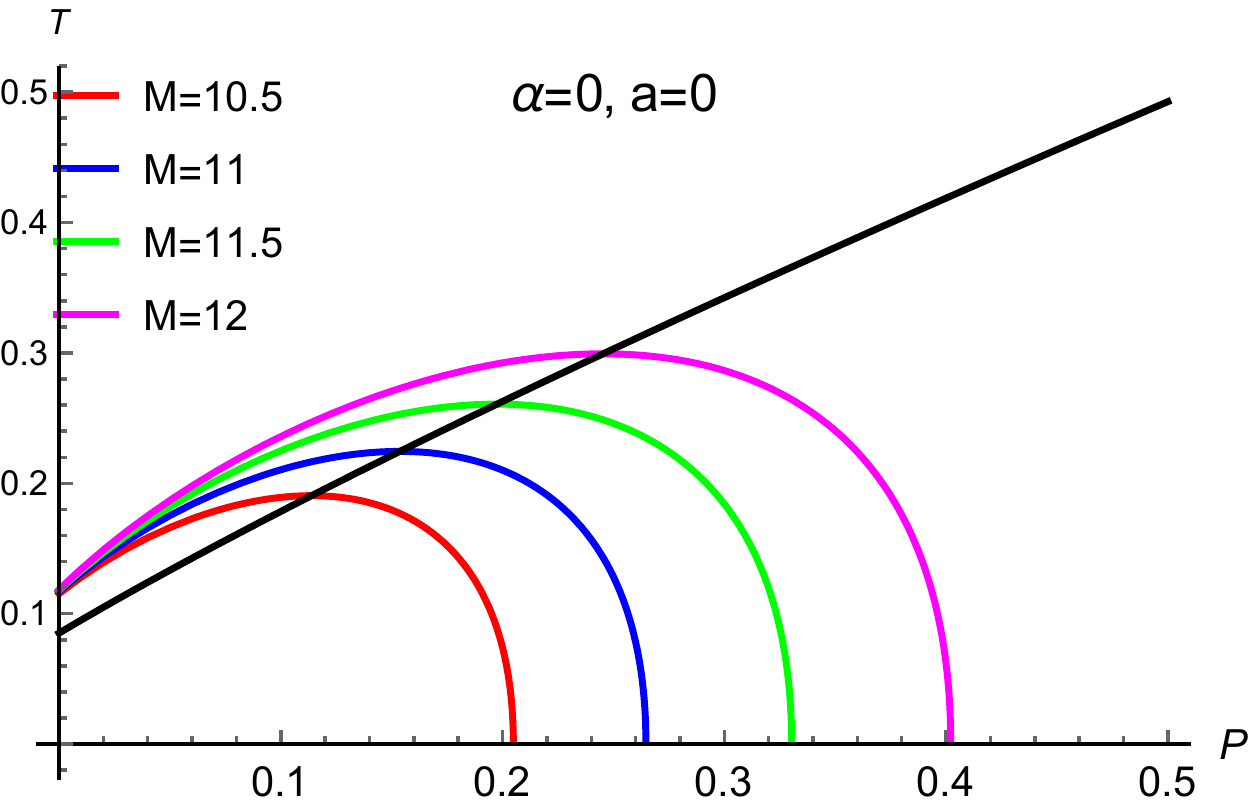}
 \label{fig:2dh600}}
   \subfigure[{$d=6$}]{
  \includegraphics[width=0.31\textwidth]{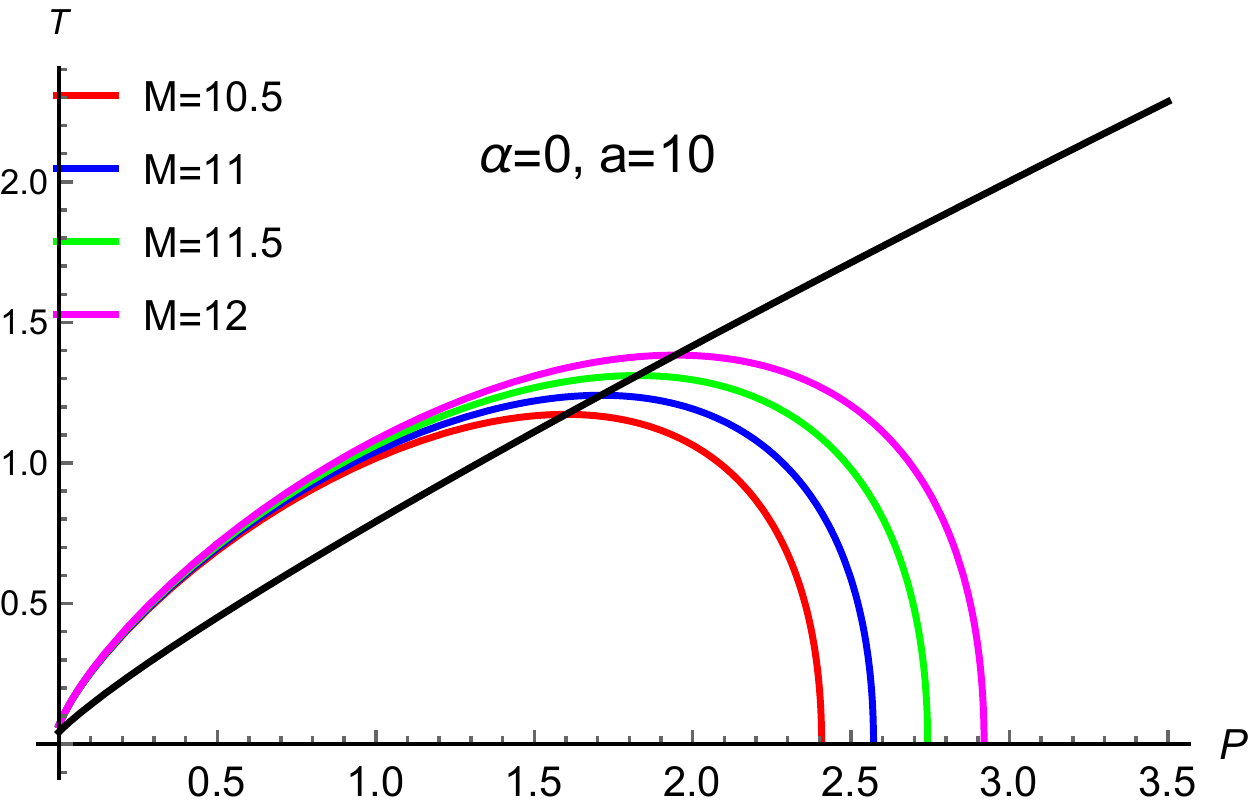}
 \label{fig:2dh6100}}
  \subfigure[{$d=6$}]{
  \includegraphics[width=0.31\textwidth]{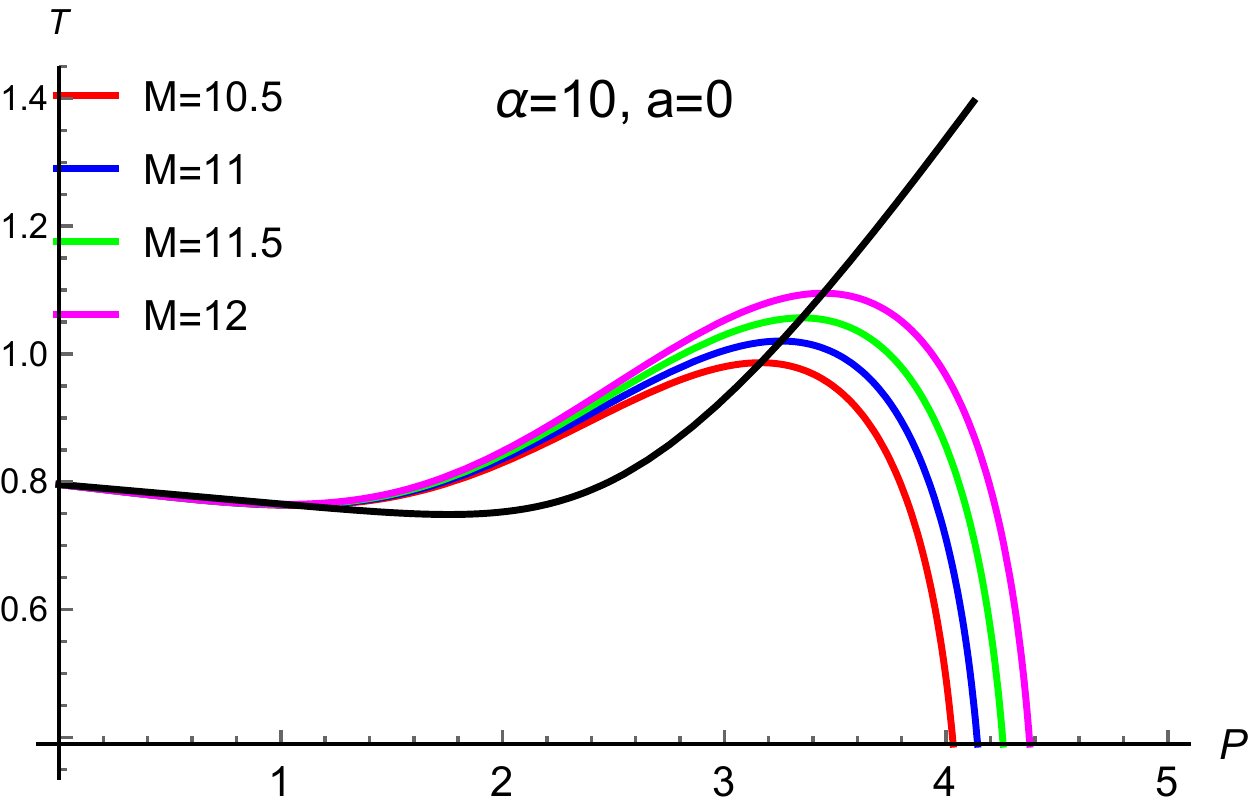}
 \label{fig:2dh6010}}
  \subfigure[{$d=7$}]{
  \includegraphics[width=0.31\textwidth]{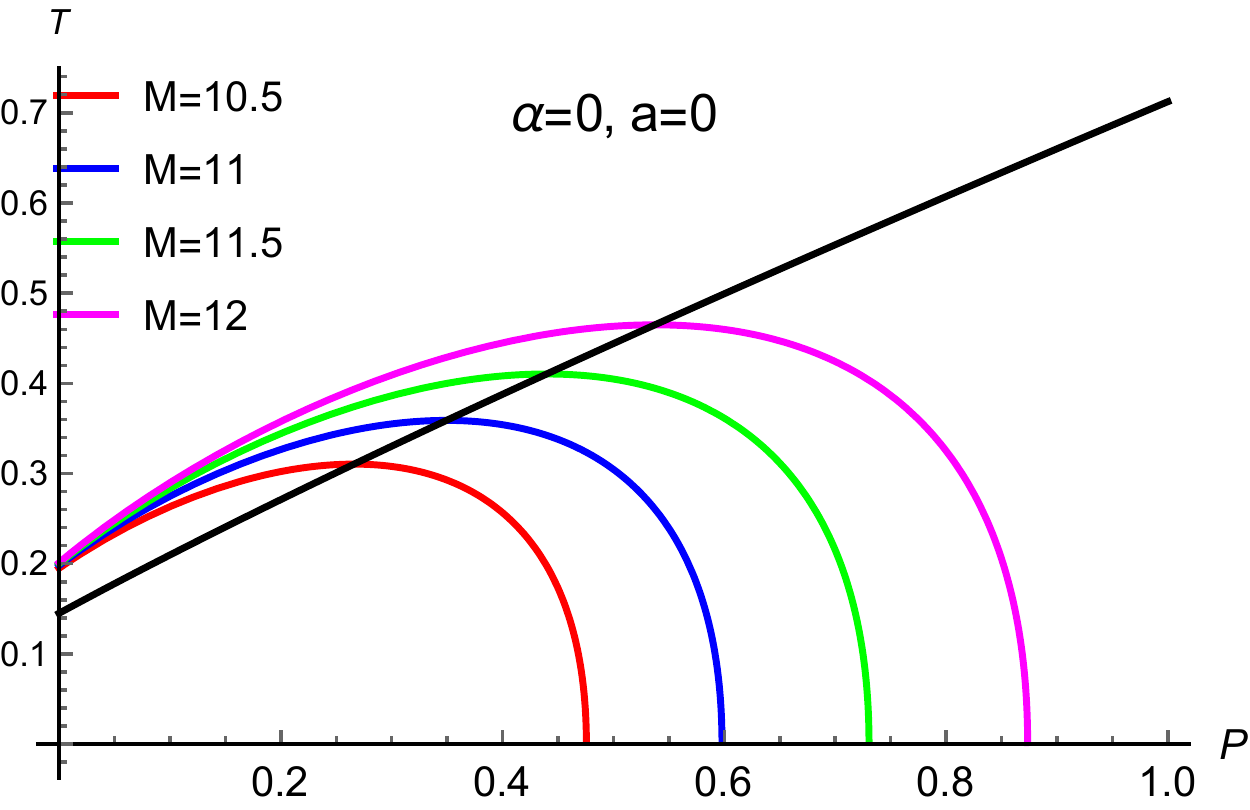}
 \label{fig:2dh700}}
   \subfigure[{$d=7$}]{
  \includegraphics[width=0.31\textwidth]{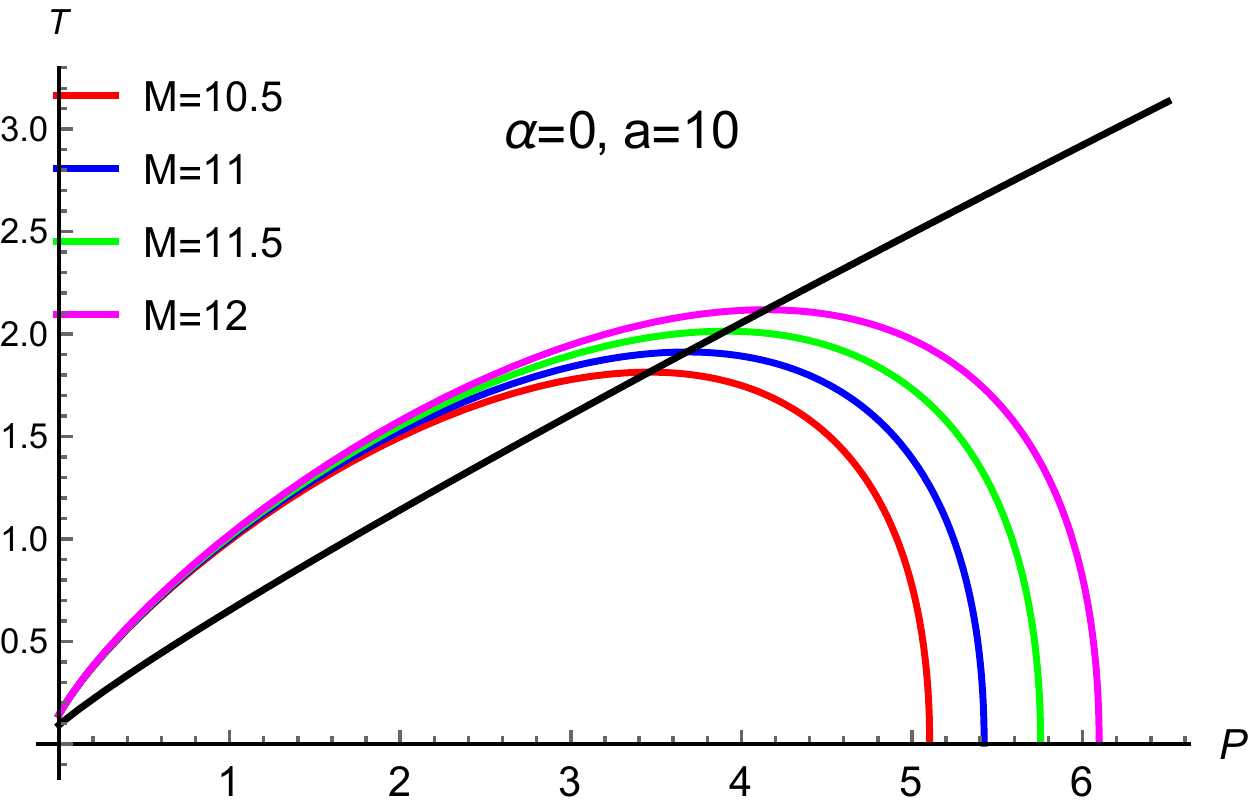}
 \label{fig:2dh7100}}
  \subfigure[{$d=7$}]{
  \includegraphics[width=0.31\textwidth]{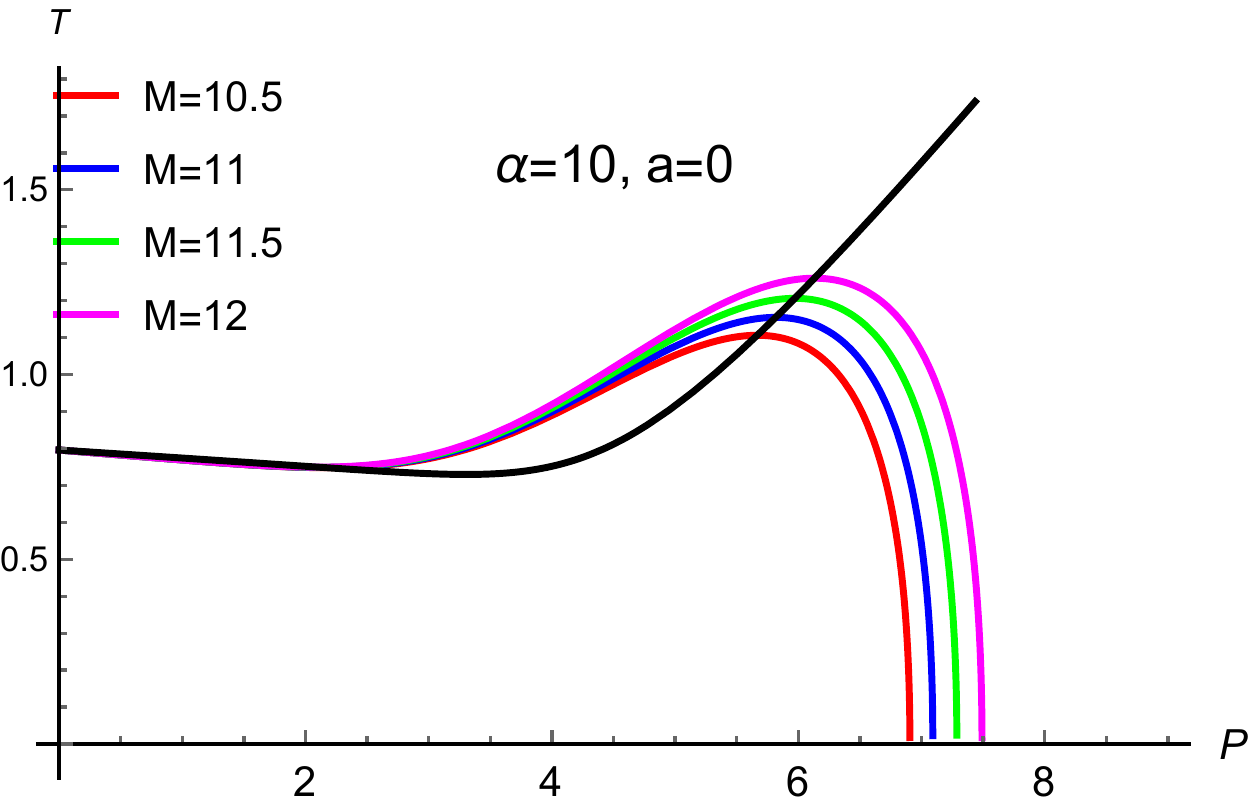}
 \label{fig:2dh7010}}
     \subfigure[{$d=8$}]{
  \includegraphics[width=0.31\textwidth]{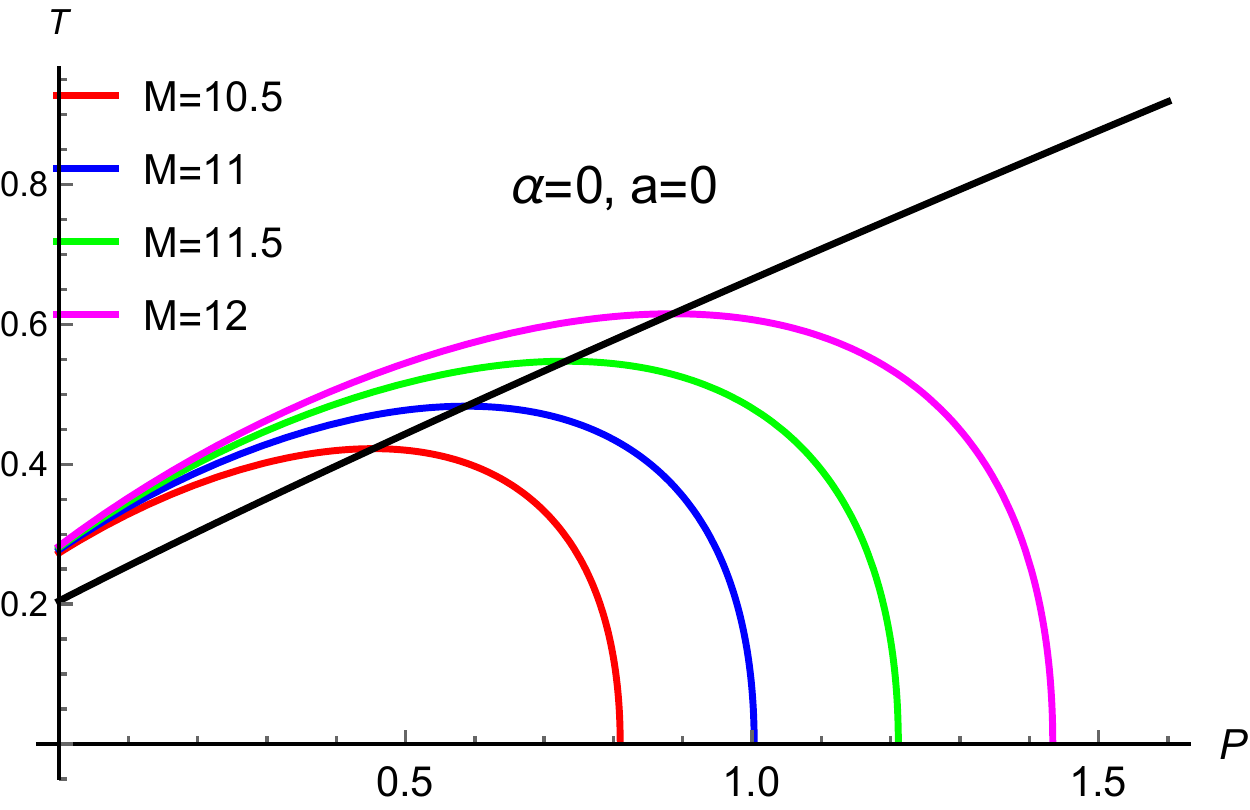}
 \label{fig:2dh800}}
   \subfigure[{$d=8$}]{
  \includegraphics[width=0.31\textwidth]{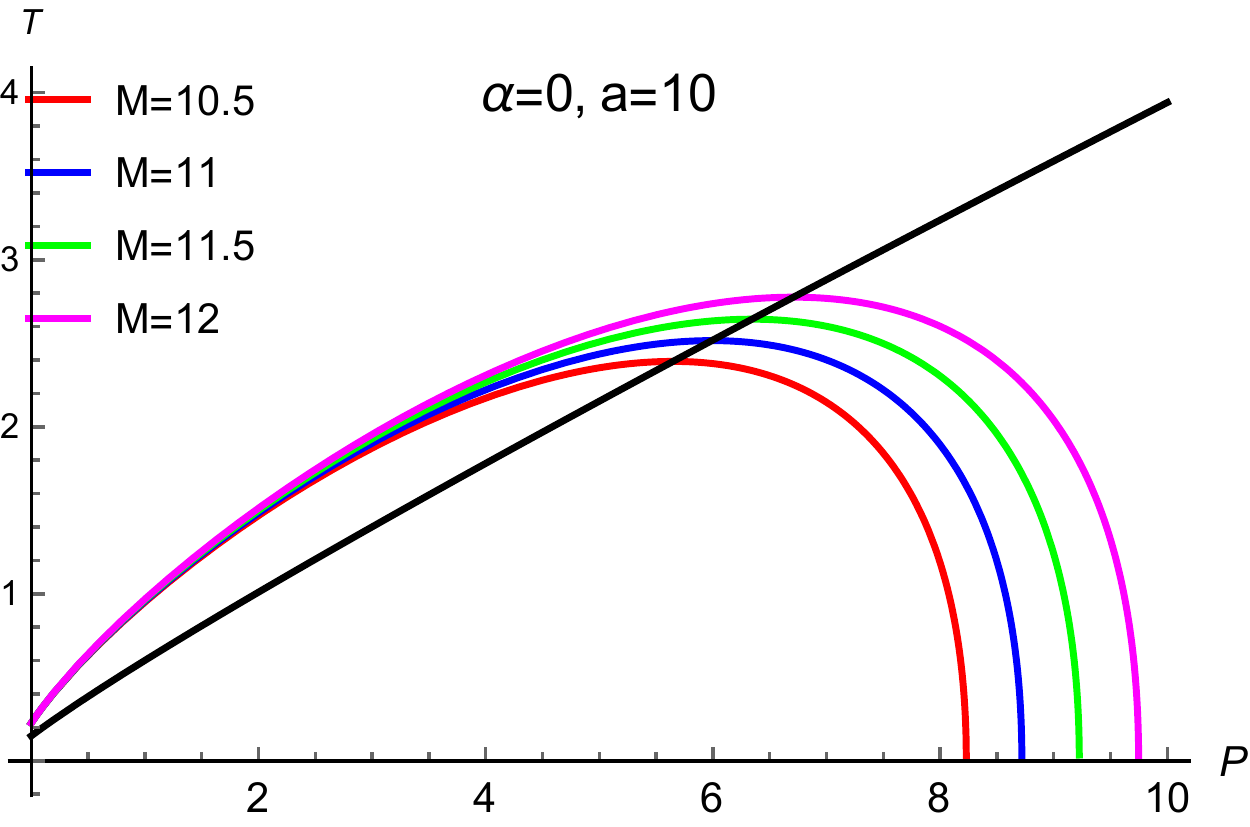}
 \label{fig:2dh8100}}
  \subfigure[{$d=8$}]{
  \includegraphics[width=0.31\textwidth]{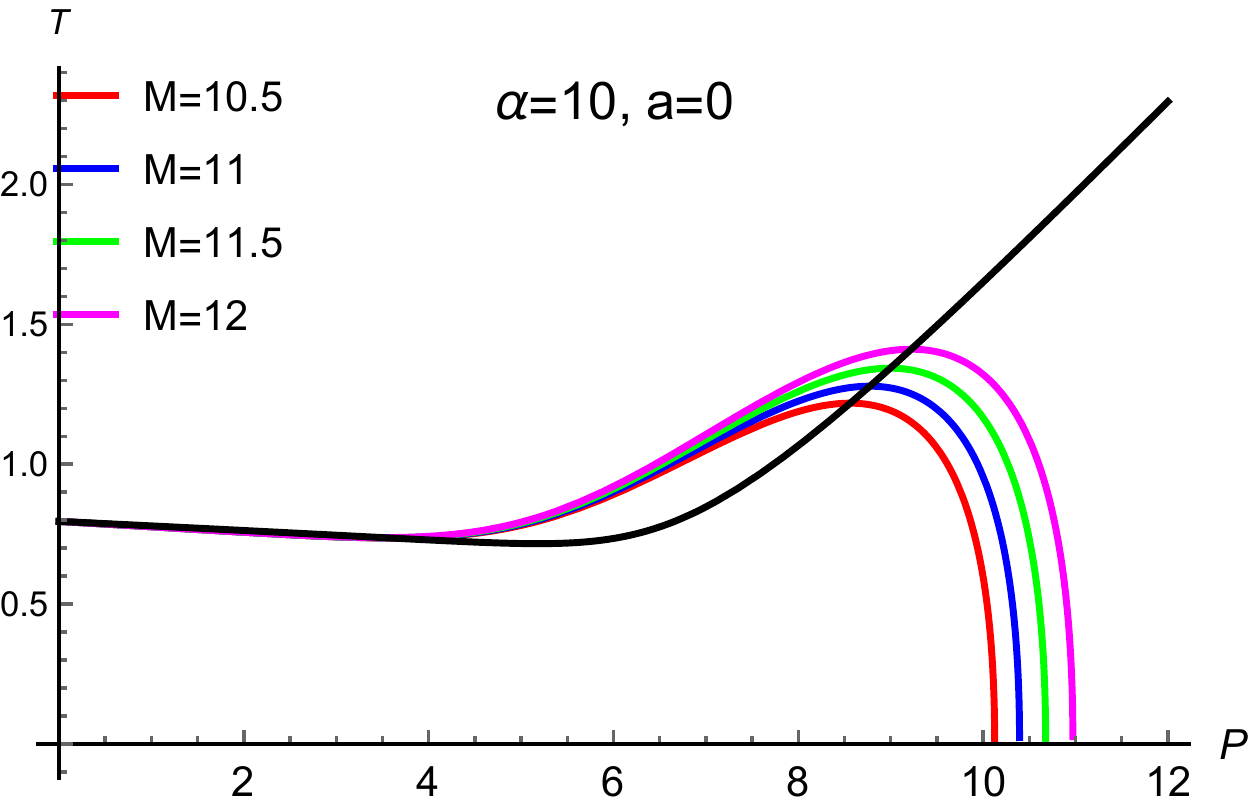}
 \label{fig:2dh8010}}
 \caption{The isenthalpic curves and inversion curves for various combinations of $d$, $a$, $\alpha$ and $M$, here $Q$=10.}\label{fig:TIC4}
\end{figure}

In the process of studying the Joule-Thomson expansion of d-dimensional charged AdS black holes with cloud of strings and quintessence, it is necessary to analyze the changes in the isenthalpic curves under the influence of different parameters. Specifically, the main purpose of this section is to disclose the fine structure of the isenthalpic curves in the case of parameters $a$, $\alpha$ and $d$ change. The solution can be solved for Eq. $\left(\ref{eqn:two6}\right)$ with a given value of $M$, and then substituting the solution into Eq. $\left(\ref{eqn:two11}\right)$ we can obtain the isenthalpic temperature. Note that $M>Q$, which is chosen to ensure that it does not appear in the $T-P$ plane as a particular hypersurfaces with naked singularity \cite{Okcu:2016tgt, Ghaffarnejad:2013cma}. The specific expressions of the solutions are not listed here.

For further study, the isenthalpic curves and the inversion curves are plotted simultaneously in Fig. \ref{fig:TIC3} and Fig. \ref{fig:TIC4} under the different parameters. As can be seen from the two figures above, both the isenthalpic curves and the inversion curves are present under the situation ($M > Q$), which shows the occurrence of the Joule-Thomson expansion. By careful observation, it is found that with $Q$ kept constant, the inversion curves all coincide with the highest point of the isenthalpic curves, and the isenthalpic curves is divided into two regions by the inversion curves. In the region above the inversion curves, the slope of the isenthalpic curves is positive, which means that cooling occurs. In the region below the inversion curves, the slope of the isenthalpic curves is negative, when heating occurs. So the inversion curves is the dividing line between the cooling and heating regions of the isenthalpic curves. By comparing Fig. \ref{fig:TIC3} and Fig. \ref{fig:TIC4}, it is evident that as $Q$ increases, the cooling-heating critical point changes and the curves tends to move to the left, which is independent of the parameters $a$, $\alpha$ and $d$. When only the effect of $M$ on the isenthalpic curve is considered, it can be seen that as $M$ increases, the curve expands to the right, correspondingly, the cooling-heating critical point also increases.

From the first and second columns of Figs. \ref{fig:TIC3} and \ref{fig:TIC4}, we can have the impression that the isenthalpic curves changes as $a$ increases, i.e., the curve tends to expand to the right, when other conditions remain constant. A similar changes in parameters $d$ and $\alpha$ are found by comparison. As $d$ increases, the curves tend to expand at higher pressures, which is independent of the other parameters and is consistent with the findings in the literature \cite{Yin:2021akt}. Comparing the first column and the third column of the two graphs, one difference is that when $\alpha$ changes, the slope variation of the isenthalpic curve may be somewhat different in the cooling region.

\subsection{The ratio between $T_{i}^{min}$ and $T_{c}$}\label{sec:CD}
This section focuses on criticality. The first step is to find the stationary inflection points in $P(r_{+})-r_{+}$ diagram. The equation of state $P = P(V,T)$ for the black hole is obtained from Eq. $\left(\ref{eqn:two11}\right)$ as
\begin{equation}\label{eqn:r1}
\begin{aligned}
&P=\frac{2ar_{+}^{4-d}+8\pi^{3-d}Q^{2}\Gamma\left(\frac{d-1}{2}\right)^{2}r_{+}^{6-2d}+(d-2)\left(\alpha(d-2)r_{+}-d+4\pi Tr_{+}+3\right)}{16\pi r_{+}^{2}},\\
\end{aligned}
\end{equation}
then, the critical point can be obtained using the following conditions \cite{Kubiznak:2012wp}
\begin{equation}\label{eqn:r2}
  \frac{\partial P}{\partial r_{+}}=0,\frac{\partial^{2}P}{\partial r_{+}^{2}}=0.
\end{equation}

Since the explicit expressions for the solution is too long to be easily expressed, the results for the critical values will be given in the Table \ref{tab:rr1} for diverse cases. It is observed that, when $d = 4$, the results for critical point are consistent with the correlation results derived in the Ref. \cite{Chabab:2020ejk}. Next, numerical calculations are performed for further study, and for the case of $Q = 1$, we present the results in Table \ref{tab:Rc4}, Table \ref{tab:Rc5}, Table \ref{tab:Rc6}, Table \ref{tab:Rc7} and Table \ref{tab:Rc8}.

Noting that $T_{i}^{min}$ can be obtained by demanding $P_{i}=0$, while $P_{i}$ can be derived from $\mu = 0$. After setting $\mu = 0$ and substituting $P_{i}=0$ into Eq. $\left(\ref{eqn:J8}\right)$, the following expression regarding $T_{i}^{min}$ can be obtained
\begin{equation}\label{eqn:tizb}
  \frac{4 r_{\min } \left(\frac{4 \pi ^{d-1} r_{\min }^{d+\frac{8}{d-1}+2} \left(4 a r_{\min }^4+r_{\min }^d \left((d-3) d \left(\alpha  r_{\min }-1\right)+2 \alpha  r_{\min }\right)\right)}{\Gamma \left(\frac{d-1}{2}\right)^2}+96 \pi ^2 Q^2 r_{\min }^{\frac{8 d}{d-1}}\right)}{(d-1) \left(\frac{4 \pi ^{d-1} r_{\min }^{d+\frac{8}{d-1}+2} \left(2 a r_{\min }^4+(d-2) r_{\min }^d \left(\alpha  (d-2) r_{\min }-d+3\right)\right)}{\Gamma \left(\frac{d-1}{2}\right)^2}+32 \pi ^2 Q^2 r_{\min }^{\frac{8 d}{d-1}}\right)}=0,
\end{equation}
and
\begin{equation}\label{eqn:tizb1}
 T_{i}^{min}=-\frac{\frac{2ar_{min}^{4-d}}{d-2}+\frac{8\pi^{3-d}Q^{2}r_{min}^{6-2d}\Gamma\left(\frac{d-1}{2}\right)^{2}}{d-2}+\alpha(d-2)r_{min}-d+3}{4\pi r_{min}}.
\end{equation}

The Tables \ref{tab:Rmin4}, \ref{tab:Rmin5}, \ref{tab:Rmin6}, \ref{tab:Rmin7} and \ref{tab:Rmin8} are listed by fixing the values of any two parameters of $d$, $a$ and $\alpha$ while changing the value of the other parameter in the case of $Q=1$. The final Table \ref{tab:Ratio1} is obtained after integration of data.

It can be observed that when the dimension is fixed, the change of $r_{c}$ is mainly related to $a$. The reason is that only $a$ but not $\alpha$ exists in the expression about $r_{c}$. For the same reason, the variation of $P_{c}$ is also mainly related to $a$. $r_{c}$ and $P_{c}$ decrease with the increase of $a$. It is obvious that $T_{c}$ decreases with the increase of $a$ or $\alpha$. Then we concentrate on observing the effect of dimension $d$. With Tables \ref{tab:Rc4}, \ref{tab:Rc5}, \ref{tab:Rc6},\ref{tab:Rc7}, and \ref{tab:Rc8}, it can be found that as $d$ increases, $r_{c}$ decreases, $P_{c}$ increases, and $T_{c}$ also increases.

By observing Tables \ref{tab:Rmin4}, \ref{tab:Rmin5}, \ref{tab:Rmin6}, \ref{tab:Rmin7} and \ref{tab:Rmin8}, changes in $d$, $a$ and $\alpha$ can also have a visible effect on $r_{min}$ and $T_{i}^{min}$. The increase of $d$, $a$ and $\alpha$ will all increase $r_{min}$ as well. $T_{i}^{min}$ increases as $d$ increases, however the increase in $a$ and $\alpha$ causes $T_{i}^{min}$ to decrease.
The analysis of the ratio of $T_{c}$ and $T_{i}^{min}$ in relation to the above observation continues as shown in the Table \ref{tab:Ratio1}. The ratio $\frac{T_{i}^{min}}{T_{c}}$ decreases with the increase of both dimension $d$ and parameter of quintessence $\alpha$. The difference is that the ratio $\frac{T_{i}^{min}}{T_{c}}$ increases with the increase of parameter of cloud of strings $a$. It is worth of attention that the ratios are all less than 1/2 and converge to 1/2, independently of $d$, $a$ and $\alpha$. It proves that the ratio turned out to be 1/2 is not a universal phenomenon but mainly exists in four-dimensional spacetime.
\begin{table}[]
\centering
\caption{The critical physical quantities ($r_{c}$, $P_{c}$ and $T_{c}$) for various dimensions.}
\begin{tabular}{p{0.5in}lp{1.0in}lp{1.9in}lp{1.9in}l}
\hline
\hline
\multicolumn{1}{l|}{$d$} & \multicolumn{1}{l|}{$r_{c}$} & \multicolumn{1}{l|}{$P_{c}$} & $T_{c}$ \\ \hline
\multicolumn{1}{l|}{4} & \multicolumn{1}{l|}{$\frac{\sqrt{6}Q}{\sqrt{1-a}}$}   & \multicolumn{1}{l|}{$\frac{(a-1)^{2}}{96\pi Q^{2}}$}   & $\frac{\sqrt{6}\text{(}1-a)^{\frac{3}{2}}-9\alpha Q}{18\pi Q}$   \\ \hline
\multicolumn{1}{l|}{5} & \multicolumn{1}{l|}{$\frac{3r_{c}^{3}(a-r_{c})+\frac{60Q^{2}}{\pi^{2}}}{4\pi r_{c}^{8}}=0$}   & \multicolumn{1}{l|}{$\frac{\pi^{2}r_{c}^{3}\left(3r_{c}-2a\right)-20Q^{2}}{8\pi^{3}r_{c}^{6}}$}   & $\frac{\pi^{2}r_{c}^{3}\left(r_{c}\left(4-3\alpha r_{c}\right)-2a\right)-16Q^{2}}{4\pi^{3}r_{c}^{5}}$   \\ \hline
\multicolumn{1}{l|}{6} & \multicolumn{1}{l|}{$\frac{6\pi^{2}r_{c}^{4}\left(a-r_{c}^{2}\right)+63Q^{2}}{4\pi^{3}r_{c}^{10}}=0$}   & \multicolumn{1}{l|}{$-\frac{3 \left(4 \pi ^2 r_c^4 \left(a-2 r_c^2\right)+21 Q^2\right)}{32 \pi ^3 r_c^8}$}   & $-\frac{2\pi^{2}r_{c}^{4}\left(a+r_{c}^{2}\left(2\alpha r_{c}-3\right)\right)+9Q^{2}}{4\pi^{3}r_{c}^{7}}$   \\ \hline
\multicolumn{1}{l|}{7} & \multicolumn{1}{l|}{$\frac{5\left(\pi^{4}r_{c}^{5}\left(a-r_{c}^{3}\right)+72Q^{2}\right)}{2\pi^{5}r_{c}^{12}}=0$}   & \multicolumn{1}{l|}{$\frac{\pi^{4}r_{c}^{5}\left(5r_{c}^{3}-2a\right)-72Q^{2}}{4\pi^{5}r_{c}^{10}}$}   & $-\frac{\pi^{4}r_{c}^{5}\left(2a+r_{c}^{3}\left(5\alpha r_{c}-8\right)\right)+64Q^{2}}{4\pi^{5}r_{c}^{9}}$   \\ \hline
\multicolumn{1}{l|}{8} & \multicolumn{1}{l|}{$\frac{15\left(8\pi^{4}r_{c}^{6}\left(a-r_{c}^{4}\right)+495Q^{2}\right)}{32\pi^{5}r_{c}^{14}}=0$}   & \multicolumn{1}{l|}{$-\frac{5\left(16\pi^{4}r_{c}^{6}\left(a-3r_{c}^{4}\right)+495Q^{2}\right)}{128\pi^{5}r_{c}^{12}}$}   & $-\frac{8\pi^{4}r_{c}^{6}\left(a+r_{c}^{4}\left(3\alpha r_{c}-5\right)\right)+225Q^{2}}{16\pi^{5}r_{c}^{11}}$   \\ \hline
\hline
\end{tabular}
\label{tab:rr1}
\end{table}

\begin{table}[]
\centering
\caption{The critical physical quantities ($r_{c}$, $P_{c}$ and $T_{c}$) for $d=4$.}
\begin{tabular}{p{1.5in}lp{2.0in}lp{2.0in}lp{2.0in}l}
\hline
\hline
\multicolumn{1}{l|}{$d=4$} & \multicolumn{1}{l|}{$r_{c}$} & \multicolumn{1}{l|}{$P_{c}$} & $T_{c}$ \\ \hline
\multicolumn{1}{l|}{$a=0,\alpha=0$}      & \multicolumn{1}{l|}{2.44949} & \multicolumn{1}{l|}{0.003316} & 0.043317 \\ \hline
\multicolumn{1}{l|}{$a=0,\alpha=0.05$}  & \multicolumn{1}{l|}{2.44949} & \multicolumn{1}{l|}{0.003316} & 0.035359 \\ \hline
\multicolumn{1}{l|}{$a=0.5,\alpha=0$}  & \multicolumn{1}{l|}{3.4641}  & \multicolumn{1}{l|}{0.000829} & 0.015315 \\ \hline
\multicolumn{1}{l|}{$a=0.5,\alpha=0.05$} & \multicolumn{1}{l|}{3.4641}  & \multicolumn{1}{l|}{0.000829} & 0.007357\\ \hline
\hline
\end{tabular}
\label{tab:Rc4}
\end{table}

\begin{table}[]
\centering
\caption{The critical physical quantities ($r_{c}$, $P_{c}$ and $T_{c}$) for $d=5$.}
\begin{tabular}{llll}
\hline
\hline
\multicolumn{1}{l|}{$d=5$} & \multicolumn{1}{l|}{$r_{c}$} & \multicolumn{1}{l|}{$P_{c}$} & $T_{c}$ \\ \hline
\multicolumn{1}{l|}{$a=0,\alpha=0$}     & \multicolumn{1}{l|}{1.19312} & \multicolumn{1}{l|}{0.055902} & 0.213431 \\ \hline
\multicolumn{1}{l|}{$a=0,\alpha=0.05$}    & \multicolumn{1}{l|}{1.19312} & \multicolumn{1}{l|}{0.055902} & 0.201494 \\ \hline
\multicolumn{1}{l|}{$a=0.5,\alpha=0$}    & \multicolumn{1}{l|}{1.34076} & \multicolumn{1}{l|}{0.036013} & 0.163367 \\ \hline
\multicolumn{1}{l|}{$a=0.5,\alpha=0.05$} & \multicolumn{1}{l|}{1.34076} & \multicolumn{1}{l|}{0.036013} & 0.15143  \\ \hline
\hline
\end{tabular}
\label{tab:Rc5}
\end{table}

\begin{table}[]
\centering
\caption{The critical physical quantities ($r_{c}$, $P_{c}$ and $T_{c}$) for $d=6$.}
\begin{tabular}{llll}
\hline
\hline
\multicolumn{1}{l|}{$d=6$} & \multicolumn{1}{l|}{$r_{c}$} & \multicolumn{1}{l|}{$P_{c}$} & $T_{c}$ \\ \hline
\multicolumn{1}{l|}{$a=0,\alpha=0$}      & \multicolumn{1}{l|}{1.01037} & \multicolumn{1}{l|}{0.175392} & 0.405054 \\ \hline
\multicolumn{1}{l|}{$a=0,\alpha=0.05$}    & \multicolumn{1}{l|}{1.01037} & \multicolumn{1}{l|}{0.175392} & 0.389139 \\ \hline
\multicolumn{1}{l|}{$a=0.5,\alpha=0$}    & \multicolumn{1}{l|}{1.10345} & \multicolumn{1}{l|}{0.126923} & 0.337043 \\ \hline
\multicolumn{1}{l|}{$a=0.5,\alpha=0.05$} & \multicolumn{1}{l|}{1.10345} & \multicolumn{1}{l|}{0.126923} & 0.321127 \\ \hline
\hline
\end{tabular}
\label{tab:Rc6}
\end{table}

\begin{table}[]
\centering
\caption{The critical physical quantities ($r_{c}$, $P_{c}$ and $T_{c}$) for $d=7$.}
\begin{tabular}{llll}
\hline
\hline
\multicolumn{1}{l|}{$d=7$} & \multicolumn{1}{l|}{$r_{c}$} & \multicolumn{1}{l|}{$P_{c}$} & $T_{c}$ \\ \hline
\multicolumn{1}{l|}{$a=0,\alpha=0$}      & \multicolumn{1}{l|}{0.962923} & \multicolumn{1}{l|}{0.343295} & 0.587673 \\ \hline
\multicolumn{1}{l|}{$a=0,\alpha=0.05$}    & \multicolumn{1}{l|}{0.962923} & \multicolumn{1}{l|}{0.343295} & 0.567779 \\ \hline
\multicolumn{1}{l|}{$a=0.5,\alpha=0$}    & \multicolumn{1}{l|}{1.0372}   & \multicolumn{1}{l|}{0.262741} & 0.50739  \\ \hline
\multicolumn{1}{l|}{$a=0.5,\alpha=0.05$} & \multicolumn{1}{l|}{1.0372}   & \multicolumn{1}{l|}{0.262741} & 0.487495 \\ \hline
\hline
\end{tabular}
\label{tab:Rc7}
\end{table}

\begin{table}[]
\centering
\caption{The critical physical quantities ($r_{c}$, $P_{c}$ and $T_{c}$) for $d=8$.}
\begin{tabular}{llll}
\hline
\hline
\multicolumn{1}{l|}{$d=8$} & \multicolumn{1}{l|}{$r_{c}$} & \multicolumn{1}{l|}{$P_{c}$} & $T_{c}$ \\ \hline
\multicolumn{1}{l|}{$a=0,\alpha=0$}      & \multicolumn{1}{l|}{0.955634} & \multicolumn{1}{l|}{0.544612} & 0.757017  \\ \hline
\multicolumn{1}{l|}{$a=0,\alpha=0.05$}    & \multicolumn{1}{l|}{0.955634} & \multicolumn{1}{l|}{0.544612} & 0.733144 \\ \hline
\multicolumn{1}{l|}{$a=0.5,\alpha=0$}    & \multicolumn{1}{l|}{1.01757}  & \multicolumn{1}{l|}{0.435529} & 0.671153  \\ \hline
\multicolumn{1}{l|}{$a=0.5,\alpha=0.05$} & \multicolumn{1}{l|}{1.01757}  & \multicolumn{1}{l|}{0.435529} & 0.64728  \\ \hline
\hline
\end{tabular}
\label{tab:Rc8}
\end{table}

%min,begining%%%%

\begin{table}[]
\centering
\caption{The minimum inversion temperature ($T_{i}^{min}$) and its corresponding root ($r_{min}$) for $d=4$.}
\begin{tabular}{lll}
\hline
\hline
\multicolumn{1}{l|}{$d=4$} & \multicolumn{1}{l|}{$r_{min}$} & $T_{i}^{min}$ \\ \hline
\multicolumn{1}{l|}{$a=0,\alpha=0$}      & \multicolumn{1}{l|}{1.22474} & 0.021658\\ \hline
\multicolumn{1}{l|}{$a=0,\alpha=0.05$}    & \multicolumn{1}{l|}{1.28859} & 0.016606 \\ \hline
\multicolumn{1}{l|}{$a=0.5,\alpha=0$}    & \multicolumn{1}{l|}{1.73205} & 0.007657   \\ \hline
\multicolumn{1}{l|}{$a=0.5,\alpha=0.05$} & \multicolumn{1}{l|}{2.09059} & 0.002365  \\ \hline
\hline
\end{tabular}
\label{tab:Rmin4}
\end{table}

\begin{table}[]
\centering
\caption{The minimum inversion temperature ($T_{i}^{min}$) and its corresponding root ($r_{min}$) for $d=5$.}
\begin{tabular}{lll}
\hline
\hline
\multicolumn{1}{l|}{$d=5$} & \multicolumn{1}{l|}{$r_{min}$} & $T_{i}^{min}$ \\ \hline
\multicolumn{1}{l|}{$a=0,\alpha=0$}      & \multicolumn{1}{l|}{0.702228} & 0.10073 \\ \hline
\multicolumn{1}{l|}{$a=0,\alpha=0.05$}    & \multicolumn{1}{l|}{0.709911} & 0.093009 \\ \hline
\multicolumn{1}{l|}{$a=0.5,\alpha=0$}    & \multicolumn{1}{l|}{0.758108} & 0.077921   \\ \hline
\multicolumn{1}{l|}{$a=0.5,\alpha=0.05$} & \multicolumn{1}{l|}{0.769404} & 0.070368  \\ \hline
\hline
\end{tabular}
\label{tab:Rmin5}
\end{table}

\begin{table}[]
\centering
\caption{The minimum inversion temperature ($T_{i}^{min}$) and its corresponding root ($r_{min}$) for $d=6$.}
\begin{tabular}{lll}
\hline
\hline
\multicolumn{1}{l|}{$d=6$} & \multicolumn{1}{l|}{$r_{min}$} & $T_{i}^{min}$ \\ \hline
\multicolumn{1}{l|}{$a=0,\alpha=0$}      & \multicolumn{1}{l|}{0.650819} & 0.183409  \\ \hline
\multicolumn{1}{l|}{$a=0,\alpha=0.05$}    & \multicolumn{1}{l|}{0.654851} & 0.172996 \\ \hline
\multicolumn{1}{l|}{$a=0.5,\alpha=0$}    & \multicolumn{1}{l|}{0.681196} & 0.1542511   \\ \hline
\multicolumn{1}{l|}{$a=0.5,\alpha=0.05$} & \multicolumn{1}{l|}{0.686482} & 0.144099  \\ \hline
\hline
\end{tabular}
\label{tab:Rmin6}
\end{table}

\begin{table}[]
\centering
\caption{The minimum inversion temperature ($T_{i}^{min}$) and its corresponding root ($r_{min}$) for $d=7$.}
\begin{tabular}{lll}
\hline
\hline
\multicolumn{1}{l|}{$d=7$} & \multicolumn{1}{l|}{$r_{min}$} & $T_{i}^{min}$ \\ \hline
\multicolumn{1}{l|}{$a=0,\alpha=0$}      & \multicolumn{1}{l|}{0.658134}            & 0.257949 \\ \hline
\multicolumn{1}{l|}{$a=0,\alpha=0.05$}    & \multicolumn{1}{l|}{0.661107}            & 0.244853 \\ \hline
\multicolumn{1}{l|}{$a=0.5,\alpha=0$}    & \multicolumn{1}{l|}{0.6797104} & 0.224907   \\ \hline
\multicolumn{1}{l|}{$a=0.5,\alpha=0.05$} & \multicolumn{1}{l|}{0.683421}            & 0.212149  \\ \hline
\hline
\end{tabular}
\label{tab:Rmin7}
\end{table}

\begin{table}[]
\centering
\caption{The minimum inversion temperature ($T_{i}^{min}$) and its corresponding root ($r_{min}$) for $d=8$.}
\begin{tabular}{lll}
\hline
\hline
\multicolumn{1}{l|}{$d=8$} & \multicolumn{1}{l|}{$r_{min}$} & $T_{i}^{min}$ \\ \hline
\multicolumn{1}{l|}{$a=0,\alpha=0$}      & \multicolumn{1}{l|}{0.681641} & 0.324289  \\ \hline
\multicolumn{1}{l|}{$a=0,\alpha=0.05$}    & \multicolumn{1}{l|}{0.684138} & 0.308516 \\ \hline
\multicolumn{1}{l|}{$a=0.5,\alpha=0$}    & \multicolumn{1}{l|}{0.69796}  & 0.290015   \\ \hline
\multicolumn{1}{l|}{$a=0.5,\alpha=0.05$} & \multicolumn{1}{l|}{0.700967} & 0.274635  \\ \hline
\hline
\end{tabular}
\label{tab:Rmin8}
\end{table}

\begin{table}[]
\centering
\caption{The ratio $\frac{T_{i}^{min}}{T_{c}}$ for various combinations of $d$, $a$ and $\alpha$.}
\begin{tabular}{lllll}
\hline
\hline
  \multicolumn{5}{c}{$\frac{T_{i}^{min}}{T_{c}}$}              \\ \hline
\multicolumn{1}{l|}{}& \multicolumn{1}{l|}{$a=0,\alpha=0$}    & \multicolumn{1}{l|}{$a=0,\alpha=0.05$}  & \multicolumn{1}{l|}{$a=0.5,\alpha=0$}   & $a=0.5,\alpha=0.05$ \\ \hline
\multicolumn{1}{l|}{$d=4$} & \multicolumn{1}{l|}{0.499999} & \multicolumn{1}{l|}{0.46965}  & \multicolumn{1}{l|}{0.5}      & 0.321498 \\ \hline
\multicolumn{1}{l|}{$d=5$ }& \multicolumn{1}{l|}{0.471956} & \multicolumn{1}{l|}{0.461594} & \multicolumn{1}{l|}{0.476968} & 0.464689 \\ \hline
\multicolumn{1}{l|}{$d=6$ }& \multicolumn{1}{l|}{0.452801} & \multicolumn{1}{l|}{0.444561} & \multicolumn{1}{l|}{0.45766}  & 0.448729 \\ \hline
\multicolumn{1}{l|}{$d=7$} & \multicolumn{1}{l|}{0.438933} & \multicolumn{1}{l|}{0.431247} & \multicolumn{1}{l|}{0.443263} & 0.435182 \\ \hline
\multicolumn{1}{l|}{$d=8$} & \multicolumn{1}{l|}{0.428377} & \multicolumn{1}{l|}{0.420812} & \multicolumn{1}{l|}{0.432115} & 0.424291 \\ \hline
\hline
\end{tabular}
\label{tab:Ratio1}
\end{table}

\section{ Conclusion and Discussion}\label{sec:D}
This paper investigated the Joule-Thomson expansion of the charged AdS black hole with cloud of strings and quintessence, which present in higher dimensional spacetime. In extended phase space, the mass of the black hole is interpreted as enthalpy, while the cosmological constant is treated as thermodynamic pressure. After a brief review of the thermodynamics of the black hole, the Joule-Thomson expansion effect of the black hole was discussed in detail from four aspects: the Joule-Thomson coefficient, the inversion curve, the isenthalpic curve and the ratio between $T_{i}^{min} $ and $T_{c}$. At the extreme values of the isenthalpic curve, the Joule-Thomson coefficient is exactly zero, and the trajectories of these points are described as inversion curve.

The Joule-Thomson effect is an irreversible adiabatic expansion, in which temperature changes with pressure and enthalpy are constant. And the variation between temperature and pressure determines the sign of the Joule-Thomson coefficient. We plotted Fig. \ref{fig:JT1} and Fig. \ref{fig:JT2} to visualize this variation when setting $P$ certain. It turns out that the existence of divergence point dividing the Joule-Thomson coefficient curve into positive and negative infinity regions, and the divergence point of Joule-Thomson coefficient corresponds to the zero point of Hawking temperature. Further, as $d$, $q$, $a$ and $\alpha$ increase, the divergence point moves to the right.

Then we investigated the inversion curve with the approach of numerical analysis. Different figures exhibit that the inversion curves change with the parameters $d$, $Q$, $a$ and $\alpha$ for each plot. The specific variations are presented in detail in the Table \ref{tab:DISS1}.
\begin{table}[H]
\centering
\caption{Findings for the effects of different parameters on the inversion curves.}
\begin{tabular}{l|lll}
\hline
\hline
parameters & \multicolumn{3}{l}{The effects of parameters on the inversion curves} \\ \hline
$d$          & \multicolumn{3}{l}{\begin{tabular}[c]{@{}l@{}}In the high-pressure case, the inversion temperature\\ for a given pressure decreases as $d$ increases;\\ In the low-pressure case, the inversion temperature\\ for a given pressure increases as $d$ increases.\end{tabular}}                               \\ \hline
$Q$          & \multicolumn{3}{l}{\begin{tabular}[c]{@{}l@{}}In the high-pressure region, the inversion temperature for\\ a given pressure increases with the increase of $Q$; \\ In the low-pressure region, the inversion temperature for\\ a given pressure decreases with the increase of $Q$.\end{tabular}}                               \\ \hline
$a$          & \multicolumn{3}{l}{\begin{tabular}[c]{@{}l@{}}The high and low pressure dividing points move to the left\\ as $a$ increases.\end{tabular}}                               \\ \hline
$\alpha$         & \multicolumn{3}{l}{\begin{tabular}[c]{@{}l@{}}The high and low pressure dividing points move to the right\\ as $\alpha$ increases.\\ The change is most pronounced in the low pressure region.\end{tabular}}                               \\ \hline
\hline
\end{tabular}
\label{tab:DISS1}
\end{table}

And thirdly, under the premise of $M > Q$, we studied the variations of the isenthalpic curve in the $T$ - $P$ plane. When combined with the inversion curve, it shows that the Joule-Thomson expansion is present in all states. Meanwhile, The inversion curve passes through the maximum point of the isenthalpic curve, thus dividing the isenthalpic curve into two regions: heating (one occurs in the region below the inversion curves) and cooling (one occurs in the region above the inversion curves). Different figures exhibit that the isenthalpic curves change with the parameters $d$, $Q$, $M$, $a$ and $\alpha$ for each plot. The specific findings are presented in detail in the Table \ref{tab:DISS2}.

\begin{center}
\begin{table}[H]
\centering
\caption{Findings for the effects of different parameters on the isenthalpic curves.}
\begin{tabular}{l|lll}
\hline
\hline
parameters & \multicolumn{3}{l}{The effects of parameters on the isenthalpic curves} \\ \hline
$d$          & \multicolumn{3}{l}{\begin{tabular}[c]{@{}l@{}}As $d$ increases, the cooling-heating critical points change,\\the curves tend to expand at higher pressures.\end{tabular}}                               \\ \hline
$M$          & \multicolumn{3}{l}{\begin{tabular}[c]{@{}l@{}}As $M$ increases, the cooling-heating critical points increase,\\and the curves expand to the right.\end{tabular}}                               \\ \hline
$Q$          & \multicolumn{3}{l}{\begin{tabular}[c]{@{}l@{}}As $Q$ increases, the cooling-heating critical points change\\ and the curves tend to move to the left\end{tabular}}                               \\ \hline
$a$          & \multicolumn{3}{l}{\begin{tabular}[c]{@{}l@{}}As $a$ increases, the cooling-heating critical points change,\\the curves tend to expand to the right.\end{tabular}}                               \\ \hline
$\alpha$         & \multicolumn{3}{l}{\begin{tabular}[c]{@{}l@{}}As $\alpha$ increases, the cooling-heating critical points change,\\the curves tend to expand to the right.\\Note that in the cooling region, when the value of $\alpha$ changes,\\ the slope of the isoenthalpy curve changes as well. \end{tabular}}                               \\ \hline
\hline
\end{tabular}
\label{tab:DISS2}
\end{table}
\end{center}

Furthermore, the critical quantities such as $r_{c}$, $P_{c}$ and $T_{c}$ were studied, where the ratio between $T_{i}^{min} $ and $T_{c}$ was the focus of the analysis. The results of the critical quantities in different dimensions are summarized in the Table \ref{tab:rr1}. It is found that even though some of the critical quantities are difficult to obtain specific expressions, the critical values all change independently of the parameter $\alpha$. This conclusion is also verified by the method of numerical calculations we subsequently employed. Next we analyzed the effects of different parameters $d$, $a$ and $\alpha$ on critical quantities, minimum inversion temperatures, and ratios, which are presented in the Table \ref{tab:DISS3}.

\begin{table}[H]
\centering
\caption{Findings for the effects of different parameters on critical quantities, minimum inversion temperatures, and ratios.}
\begin{tabular}{l|ll|ll|ll}
\hline
\hline
             & \multicolumn{2}{l|}{$d$} & \multicolumn{2}{l|}{$a$} & \multicolumn{2}{l}{$\alpha$} \\ \hline
$r_{c}$       & \multicolumn{2}{l|}{\begin{tabular}[c]{@{}l@{}}As $d$ increases,\\ $r_{c}$ decreases.\end{tabular}}    & \multicolumn{2}{l|}{\begin{tabular}[c]{@{}l@{}}As $a$ increases,\\ $r_{c}$ increases.\end{tabular}}    & \multicolumn{2}{l}{\begin{tabular}[c]{@{}l@{}}$\alpha$ has no\\ effect on $r_{c}$.\end{tabular}}        \\ \hline
$P_{c}$       & \multicolumn{2}{l|}{\begin{tabular}[c]{@{}l@{}}As $d$ increases,\\ $P_{c}$ increases.\end{tabular}}    & \multicolumn{2}{l|}{\begin{tabular}[c]{@{}l@{}}As $a$ increases,\\ $P_{c}$ decreases\end{tabular}}    & \multicolumn{2}{l}{\begin{tabular}[c]{@{}l@{}}$\alpha$ has no\\ effect on $P_{c}$.\end{tabular}}        \\ \hline
$T_{c}$       & \multicolumn{2}{l|}{\begin{tabular}[c]{@{}l@{}}As $d$ increases,\\ $T_{c}$ increases.\end{tabular}}    & \multicolumn{2}{l|}{\begin{tabular}[c]{@{}l@{}}As $a$ increases,\\ $T_{c}$ decreases\end{tabular}}    & \multicolumn{2}{l}{\begin{tabular}[c]{@{}l@{}}As $\alpha$ increases,\\ $T_{c}$ decreases\end{tabular}}        \\ \hline
$r_{min}$     & \multicolumn{2}{l|}{\begin{tabular}[c]{@{}l@{}}As $d$ increases,\\ $r_{min}$ increases.\end{tabular}}    & \multicolumn{2}{l|}{\begin{tabular}[c]{@{}l@{}}As $a$ increases,\\ $r_{min}$ increases\end{tabular}}    & \multicolumn{2}{l}{\begin{tabular}[c]{@{}l@{}}As $\alpha$ increases,\\ $r_{min}$ increases\end{tabular}}        \\ \hline
$T_{i}^{min}$ & \multicolumn{2}{l|}{\begin{tabular}[c]{@{}l@{}}As $d$ increases,\\  $T_{i}^{min}$ increases\end{tabular}}    & \multicolumn{2}{l|}{\begin{tabular}[c]{@{}l@{}}As $a$ increases,\\ $T_{i}^{min}$ decreases\end{tabular}}    & \multicolumn{2}{l}{\begin{tabular}[c]{@{}l@{}}As $\alpha$ increases,\\ $T_{i}^{min}$ decreases\end{tabular}}        \\ \hline
$\frac{T_{i}^{min}}{T_{c}}$           & \multicolumn{2}{l|}{\begin{tabular}[c]{@{}l@{}}As $d$ increases,\\ $\frac{T_{i}^{min}}{T_{c}}$ decreases.\end{tabular}}    & \multicolumn{2}{l|}{\begin{tabular}[c]{@{}l@{}}As $a$ increases,\\ $\frac{T_{i}^{min}}{T_{c}}$ increases.\end{tabular}}    & \multicolumn{2}{l}{\begin{tabular}[c]{@{}l@{}}As $\alpha$ increases,\\ $\frac{T_{i}^{min}}{T_{c}}$ decreases.\end{tabular}}        \\ \hline
\hline
\end{tabular}
\label{tab:DISS3}
\end{table}

\begin{acknowledgments}
We are grateful to Wei Hong, Peng Wang, Haitang Yang, Jun Tao, Deyou Chen and
Xiaobo Guo for useful discussions. This work is supported in part by NSFC
(Grant No. 11747171), Natural Science Foundation of Chengdu University of TCM
(Grants nos. ZRYY1729 and ZRYY1921), Discipline Talent Promotion Program of
/Xinglin Scholars(Grant no.QNXZ2018050) and the key fund project for Education
Department of Sichuan (Grantno. 18ZA0173).
\end{acknowledgments}

\end{document}